\documentclass[acmsmall,screen,nonacm]{acmart}

\newif\ifdraft\drafttrue

\usepackage{booktabs}   
\usepackage{subcaption} 

\usepackage{amsmath,bm,bbm}
\usepackage{stmaryrd}
\usepackage{mathtools}

\ifdraft
\usepackage[draft]{commenting}
\else
\usepackage[nompar]{commenting}
\fi
\usepackage{hyperref}
\usepackage{graphicx}
\usepackage{fontawesome}
\usepackage{pifont}
\usepackage[normalem]{ulem}
\usepackage{cancel}
\usepackage[capitalize]{cleveref}
\usepackage{wrapfig}
\usepackage{adjustbox}
\usepackage{longfbox}

\usepackage{float}
\usepackage{caption}

\usepackage{multirow}
\usepackage{tablefootnote}

\usepackage{tabu}
\usepackage{xcolor}

\usepackage{pgfplots}
\usepackage{listings}
\usetikzlibrary{automata, calc}

\usepackage[vlined, ruled, linesnumbered]{algorithm2e}

\newcommand{\rmblock}[1]{{\rm #1}}
\newcommand{\setdef}[2]{\left\{ #1\, \middle|\, \begin{matrix} #2 \end{matrix} \right\}}
\newcommand{\interpret}[1]{{\left\llbracket#1\right\rrbracket}}
\newcommand{\angbr}[1]{{\left\langle#1\right\rangle}}

\newfloat{Trace}{tbp}{ext}
\newfloat{Algorithm}{tbp}{ext}

\def\probbranchText{\texttt{Pb}}
\newcommand{\probbranch}[2]{{\probbranchText_{(#1, #2)}}}
\def\stskip{\mathit{skip}}
\def\assume{\texttt{assume}}

\newcommand{\nextFullProb}[2][]{^{#1}\left[#2\right]}
\newcommand{\prob}[1][]{\mathbb{P}_{#1}\nextFullProb}
\newcommand{\bs}[1]{{\left\{#1\right\}}}

\def\dom{{\text{dom}}}

\def\scriptA{\mathcal{A}}
\def\scriptB{\mathcal{B}}

\def\scriptD{\mathcal{D}}
\def\scriptE{\mathcal{E}}

\def\scriptL{\mathcal{L}}
\def\scriptM{\mathcal{M}}

\def\scriptP{\mathcal{P}}

\def\scriptS{\mathcal{S}}
\def\scriptT{\mathcal{T}}
\def\scriptU{\mathcal{U}}
\def\scriptV{\mathcal{V}}

\def\doubleP{\mathbb{P}}
\def\doubleV{\mathbb{V}}

\def\mergeable{{\texttt{Mergeable}}}

\def\lbranch{{\texttt{L}}}
\def\rbranch{{\texttt{R}}}

\newcommand{\oriLprobbranch}[1][i]{{\probbranch{#1}{\lbranch}}}
\newcommand{\oriRprobbranch}[1][i]{{\probbranch{#1}{\rbranch}}}
\newcommand{\oriDprobbranch}[1][i]{{\probbranch{#1}{d}}}
\def\lprobbranch{\oriLprobbranch}
\def\rprobbranch{\oriRprobbranch}
\def\dprobbranch{\oriDprobbranch}
\newcommand{\stwhile}[2]{\textbf{while } #1 \textbf{ do} #2 \textbf{done}}
\def\tIf{\textbf{if}}
\def\tThen{\textbf{then}}
\def\tElse{\textbf{else}}
\def\tTrue{\texttt{True}}
\def\tFalse{\texttt{False}}

\newcommand{\zap}[1]{}

\crefname{fact}{Fact}{Facts}
\crefname{remark}{Remark}{Remarks}

\crefname{trace}{trace}{traces}
\def\Undefined{\bot}
\def\doubleN{\mathbb{N}}

\newcommand{\finiteMemoryPolicy}[1][ ]{{\psi_{#1}}}

\newcommand{\applyPolicy}[2]{{{#1}^{#2}}}
\newcommand{\fnorm}[1]{{\text{Norm}\left(#1\right)}}

\def\opAssn{\mathrel{::=}}

\newcommand{\toMDP}[1]{{\scriptD\left({#1}\right)}}

\newcommand{\statementToLabel}[1]{\angbr{#1}}
\newcommand{\traceSet}[1]{\scriptL\left(#1\right)}
\newcommand{\filterViolating}[2]{{#1 \upharpoonright^\neg #2}}
\def\spec{\Psi}

\def\precond{\varphi_e}
\def\postcond{\varphi_f}

\newcommand{\drawLabelledArrow}[4][right]{\draw[-latex] (#2) -- (#3)  node[midway, #1] {\scriptsize #4};}

\SetKwFor{Match}{match}{with}{end match}
\SetKwFor{Case}{case}{$\Rightarrow$}{}

\newcommand{\determinise}[1]{\text{Det}\left(#1\right)}

\definecolor{commenting}{rgb}{0.0, 0.5, 0.0}
\definecolor{shallowred}{rgb}{0.9, 0.17, 0.31}
\definecolor{bluegray}{rgb}{0.36, 0.54, 0.66}

\def\dummyMdpState{q_{\mathit{dmy}}}
\def\setOfStatements{\mathbf{ST}}
\def\mathinit{\mathit{init}}
\def\last{\textsf{last}}

\def\Act{\mathbf{Act}}
\def\Dist{\mathit{Dist}}
\def\Supp{\mathit{Supp}}
\def\ind{\textsf{ind}}

\def\midcons{\mathop{::}}

\def\wt{\mathit{wt}}
\newcommand{\indicator}[1]{\bm{1}_{[#1]}}
\newcommand{\structBound}[1]{\prob[\scriptU]{#1}}
\def\refinement{V}
\def\mdp{\scriptD}
\newcommand{\maxReach}[3]{\textsf{MaxReach}_{#1}\left(#2,#3\right)}
\def\mc{\scriptM}
\def\trajectory{\zeta}
\def\path{\zeta}
\def\Trajectories{\mathbf{Traj}}
\def\init{\textsf{init}}
\def\reach{\textsf{Reach}}
\newcommand{\mdpProb}[3]{\prob[][\scriptD]{\not\vdash \bs{#2} \ #1 \ \bs{#3}}}
\newcommand{\traceProb}[2][ ]{\prob[#1][\scriptT]{#2}}
\def\minimise{\textsf{min}}
\def\pathCond{\textsf{PathCond}}

\newcommand{\colorblock}[2]{{\color{#1}#2}}
\newcommand{\twp}{\textsf{wp}}

\declareauthor{gl}{gl}{cyan}
\authorcommand{gl}{comment}
\declareauthor{hf}{hf}{red}
\authorcommand{hf}{comment}
\declareauthor{narrator}{narrator}{brown}
\authorcommand{narrator}{comment}
\declareauthor{todo}{todo}{red}
\authorcommand{todo}{comment}

\declareauthor{pw}{pw}{orange}
\authorcommand{pw}{comment}

\newtheorem{remark}{Remark}

\makeatletter
\newdimen\@tempdimd
\makeatother

\AtBeginDocument{%
  }

\renewcommand{\paragraph}[1]{%
  \vspace{2pt-\parskip}%
  \noindent\textbf{#1.}\hspace{1em}%
  \ignorespaces%
}

\setcopyright{acmcopyright}
\copyrightyear{2018}
\acmYear{2018}
\acmDOI{XXXXXXX.XXXXXXX}

\acmJournal{JACM}
\acmVolume{37}
\acmNumber{4}
\acmArticle{1}
\acmMonth{8}

\begin{document}

\title{Structural Abstraction and Refinement for Probabilistic Programs}

\author{Guanyan Li$^*$}
\affiliation{%
  \institution{Tsinghua University, China \& University of Oxford, UK}}
\email{guanyan.li.cs@gmail.com}

\author{Juanen Li}
\affiliation{%
  \institution{Beijing Normal University}
  \country{China}}
\email{juanen.li.se@gmail.com}

\author{Zhilei Han}
\affiliation{%
 \institution{Tsinghua University}
 \city{Beijing}
 \country{China}}
\email{hzl21@mails.tsinghua.edu.cn}

\author{Peixin Wang$^{\dagger}$}
\affiliation{%
  \institution{Shanghai Key Laboratory of Trustworthy Computing, East China Normal University,China}
  \city{Shanghai}}
\email{pxwang@sei.ecnu.edu.cn}

\author{Hongfei Fu$^{\dagger}$}
\affiliation{%
  \institution{Shanghai Jiao Tong University}
  \city{Shanghai}
  \country{China}}
\email{jt002845@sjtu.edu.cn}

\author{Fei He$^{\dagger}$}
\affiliation{%
  \institution{Tsinghua University}
  \city{Beijing}
  \country{China}}
\email{hefei@tsinghua.edu.cn}

\renewcommand{\shortauthors}{Guanyan Li et al.}

\begin{abstract}
In this paper, we present structural abstraction refinement, a novel framework for verifying the threshold problem of probabilistic programs.
Our approach represents the structure of a Probabilistic Control-Flow Automaton (PCFA) as a Markov Decision Process (MDP) by abstracting away statement semantics.
The \emph{maximum reachability} of the MDP naturally provides a proper upper bound of the violation probability, termed the \emph{structural upper bound}.
This introduces a fresh ``structural'' characterization of the relationship between PCFA and MDP,
contrasting with the traditional ``semantical'' view, where the MDP reflects semantics.
The method uniquely features a clean separation of concerns between probability and computational semantics that
the abstraction focuses solely on probabilistic computation and
the refinement handles only the semantics aspect,
where the latter allows non-random program verification techniques to be employed without modification.

Building upon this feature, we propose a general counterexample-guided abstraction refinement (CEGAR) framework,
capable of leveraging established non-probabilistic techniques for probabilistic verification.
We explore its instantiations using trace abstraction.
Our method was evaluated on a diverse set of examples against state-of-the-art tools,
and the experimental results highlight its versatility and ability to handle more flexible structures swiftly.
\end{abstract}

\renewcommand{\thefootnote}{\fnsymbol{footnote}}
\footnotetext[0]{${\dagger}$ Corresponding authors, no ordering among the corresponding authors}
\footnotetext[0]{$*$ This paper describes the work whose majority was performed while Guanyan Li was at Tsinghua University.}
\renewcommand{\thefootnote}{\arabic{footnote}}

\maketitle

\section{Introduction}
\label{sec: intro}

Since the early days of computer science, formalisms for reasoning about probability have been widely studied. 
As an extension to classical imperative programs, probabilistic programs enable efficient solutions to many algorithmic problems~\cite{efficiency_of_probabilistic_algorithm_1,efficiency_of_probabilistic_algorithm_2,efficiency_of_probabilistic_algorithm_3}. 
It has also led to new models which play key roles in cryptography~\cite{encrption_using_probability}, linguistics~\cite{plcfrs}, and especially machine learning where generic models~\cite{ppl_anglican,ppl_church,ppl_stan} are expressed by them.

In this paper, we revisit the probabilistic threshold problem.
Specifically, given a probabilistic program $P$, a pre-condition $\precond$, and a post-condition $\postcond$,
the \emph{violation probability} is defined as the probability that an input satisfying $\precond$
leads to a final result violating $\postcond$ after the execution of $P$.
This probability, represented as $\prob{\not\vdash \bs{\precond} \ P \ \bs{\postcond}}$
in the form of a Hoare triple~\cite{hoare1969axiomatic,software_foundations_1},
raises the question of whether it is bounded by a given threshold.
Such a problem is commonly encountered in practical scenarios.
For example, the accuracy of differential privacy mechanisms~\cite{dwork2006calibrating} can be analysed as threshold problems where we bound the probability that the randomly-perturbed output deviates largely from the actual value. 
The target of approximate computing~\cite{zaiser2025guaranteed,trace_abstraction_modulo_probability} can be described as the assertion violation that the faulty hardware runs into erroneous status,
and the tail bounds of randomized algorithms~\cite{exponential_analysis_of_probabilistic_program} can be described as the assertion violation that the runtime exceeds a given threshold.
In fact, the threshold problem has been extensively researched, with numerous 
approaches proposed to address the issue.
These include predicate abstraction~\cite{pcegar}, fixed-point 
computation~\cite{exponential_analysis_of_probabilistic_program,batz2023probabilistic}, 
and concentration inequalities~\cite{chakarov2013probabilistic}, etc.

We tackle the problem by proposing a novel \emph{abstraction refinement}~\cite{cegar,trace_abstraction}
framework tailored specifically for probabilistic programs, named \emph{structural abstraction refinement}.
A distinctive feature of our approach is the separation of concerns between probability and semantics,
thus enabling the direct application of existing non-probabilistic program analysis and verification techniques to probabilistic programs.

As an overview, our method exploits a distinctive yet intuitive connection between the \emph{Probabilistic Control-Flow Automaton} (PCFA) and the \emph{Markov Decision Process} (MDP).
The formalisms of MDP and \emph{Markov Chain} (MC) are widely adopted mathematical tools in the analysis of probabilistic programs,
and their relationship with program verification has also been extensively researched~\cite{principles_of_model_checking,pcegar}.
However, in existing literature, the relation between the automaton and MDP is typically \emph{semantical},
where the latter functions as the semantic representation of the PCFA by (potentially infinitely) unrolling the concrete state space of the automaton~\cite{prism_2018,bounded_model_checking_probabilistic_programs,pcegar,trace_enumeration}.
In contrast, our work seeks to establish a connection between these two formalisms in a syntactical, or rather a \emph{structural}, manner.

Unlike the semantical connection, 
the main insight of our work is that the structure of a PCFA $A$ can be viewed as an MDP by \emph{abstracting away the computational semantics} of statements in the automaton.
Following this abstraction, one obtains precisely the structure of an MDP, where the program statements serve as the \emph{action} labels.
This \emph{underlying MDP} of the PCFA $A$ provides an appropriate abstraction of the program in the sense that the violation probability is bounded by the \emph{maximum reachability} probability of the MDP from its initial to ending location,
which is termed the \emph{structural upper bound} of $A$.
In summary, we have:
\vspace{-3pt}
\begin{align}
    \text{Structural Upper Bound} := \text{Maximum MDP Probability} \ge \text{Violation Probability}
\end{align}
This intuition is visualised in~\Cref{sec: motivation example and overview} and formalised in~\Cref{subsec:pcfa-as-mdp}.

As the structural abstraction disregards semantics,
enabling us to focus solely on the probabilistic aspect,
the refinement process focuses solely on the semantic aspect,
independent of probabilities.
Specifically, we establish~\Cref{theorem: refinement enabling},
a key theoretical result allowing refinement by
constructing a \emph{refinement automaton} $V$
that over-approximates the \emph{violating traces} of the original PCFA $A$.
Here, a \emph{trace} refers to a \emph{sequence of statements} accepted by an automaton,
indicating that the construction focuses on execution sequences,
as in the non-probabilistic case,
without handling probabilities.
A rough visualisation of this procedure is provided in~\Cref{sec: motivation example and overview}
with principles detailed in~\Cref{subsec:the-refinement-of-structural-abstraction}.

These theoretical foundations thus offer a clean separation between probabilities and semantics.
Aligning with this feature, we employ a modular way of introducing the automation algorithms.
Firstly, building upon this feature, we propose a generalised probabilistic \emph{counterexample-guided abstraction refinement} (CEGAR) framework.
This framework is uniquely capable of directly automating various \emph{non-random} analysis and verification methodologies for probabilistic program verification.
Second, we instantiate the framework with \emph{trace abstraction}~\cite{trace_abstraction,trace_abstraction_incremental}.
Thanks to the clean separation between probabilities and semantics --- with the probabilistic aspect handled entirely by our framework --- the non-probabilistic refinement technique can be applied \emph{directly} to the semantic component exactly as in the deterministic case, \emph{without requiring any modifications}.
This serves as an example of how non-random refinement techniques can be utilised to verify probabilistic threshold problems, empowered by our framework.
Further, beyond such a direct instantiation, we observed that the integration could be further optimised,
allowing it to retain the \emph{refutational completeness} characteristic of the original trace abstraction technique,
which is a property rarely seen in probabilistic verification beyond the finite-state (incl.~bounded) techniques~\cite{prism,hensel2022probabilistic,bounded_model_checking_probabilistic_programs}.

Finally, we implemented our framework and conducted a comparative evaluation with state-of-the-art verification tools on a diverse and arguably fair set of benchmark examples collected from a wide range of literature.
Our method shows evident advantages in better handling a broader range of examples as compared to the existing tools,
including those with a large or even unbounded state space and
more tricky control-flow structures.
It successfully handles over 2$\times$ more examples than the state-of-the-art tools for the benchmarks,
and demonstrates better efficiency in more than 73\% of the examples.

\paragraph{Contributions}
In summary, our contributions are as follows:
\begin{enumerate}
  \item We introduced structural abstraction and its refinement principle for probabilistic program verification, leveraging a novel structural link between PCFA and MDP.\@
  This method distinctly separates probability from semantics.
  \item To automate the theories developed in (1), we developed a generalised CEGAR framework that can integrate directly with non-random techniques.
  We instantiated this framework with trace abstraction and optimise it to retain refutational completeness.
  \item We implemented the optimised algorithm in (2) and benchmarked it against state-of-the-art tools across a diverse set of examples from the literature, demonstrating the method's versatility.
  The experimental result strongly backs our claim on the strength of our method.
\end{enumerate}

\paragraph{Outline}
The remainder of this paper is organised as follows.
We begin with a motivating example to illustrate our approach in more detail in~\Cref{sec: motivation example and overview}.
We then formally define the relevant concepts and the problem we aim to address in~\Cref{sec: preliminaries}.
Next, we explore the theoretical foundations of our framework in~\Cref{sec: PCFA as MDP},
establishing the validity of structural abstraction and principles of refinement (i.e., \Cref{theorem: refinement enabling}).
To automate this theoretical principle, we discuss a general CEGAR framework utilising structural abstraction and refinement, and the instantiations of the framework in~\Cref{sec: refinement}.
Finally, we present our experimental results in~\Cref{sec: experiments} and conclude with a discussion of related work in~\Cref{sec: related work}.

\section{Motivating Example}
\label{sec: motivation example and overview}

To better present the key idea of our approach from a high-level perspective, we illustrate it through a motivating example.

\begin{example}[Limit]
	\label{example: motivation}
    Consider the Hoare-style program~\cite{hoare1969axiomatic,software_foundations_1} shown in 
		\cref{figure: motivating example program}, where 
		$X$ and $C$ are of integer value and 
		the operator $\oplus$ is a fair binary probabilistic choice.
\end{example}

\begin{figure}[t]
    \vspace{-5pt}
	\centering
	\begin{minipage}{0.3\textwidth}
		\begin{align*}
			1\quad & \bs{\texttt{True}} \\
			2\quad & X ::= 0; \\
			3\quad & C ::= 0 \oplus \stskip; \\
			4\quad & \stwhile{C > 0}{\\
			5\quad & \qquad X ::= X + 1 \oplus \stskip; \\
			6\quad & \qquad C ::= C - 1 \\
			7\quad & } \\
			8\quad & \bs{X = 0}
		\end{align*}
    	\caption{Program of \cref{example: motivation}}
    	\label{figure: motivating example program}
	\end{minipage}
	\hfill
	\begin{minipage}{0.3\textwidth}
		\centering
		\resizebox{!}{4.8cm}{
			\begin{tikzpicture}[yscale=0.8]
    \node at (0, 1) (N0) {};
    \node[draw, circle] at (0, 0) (N1) {};
    \node[draw, circle] at (0, -1) (N2) {};
    \node[draw, circle] at (0.5, -2) (N3) {};
    \node[draw, circle] at (-0.5, -2) (N4) {};
    \node[draw, circle] at (0, -3) (N5) {};
    \node[draw, circle] at (0, -4) (N6) {};
    \node[draw, circle] at (-0.5, -5) (N7) {};
    \node[draw, circle] at (0.5, -5) (N8) {};
    \node[draw, circle] at (0, -6) (N9) {};
    \node[draw, circle, accepting] at (0, -7) (NE) {};

    \draw[-latex] (N0) -- (N1);
    \draw[-latex] (N1) -- (N2)  node[midway, right] {\scriptsize $X::=0$};
    \draw[-latex] (N2) -- (N3)  node[midway, right] {\scriptsize $\rprobbranch[0]$};
    \draw[-latex] (N2) -- (N4)  node[midway, left] {\scriptsize $\lprobbranch[0]$};
    \draw[-latex] (N3) -- (N5)  node[midway, right] {\scriptsize $\stskip$};
    \draw[-latex] (N4) -- (N5)  node[midway, left] {\scriptsize $C::=0$};
    \draw[-latex] (N5) -- (N6)  node[midway, left] {\scriptsize $\assume\ C>0$};
    \draw[-latex] (N6) -- (N7)  node[midway, left] {\scriptsize $\lprobbranch[1]$};
    \draw[-latex] (N6) -- (N8)  node[midway, right] {\scriptsize $\rprobbranch[1]$};
    \draw[-latex] (N7) -- (N9)  node[midway, left] {\scriptsize $X \opAssn X+1$};
    \draw[-latex] (N8) -- (N9)  node[midway, right] {\scriptsize $\stskip$};
    \draw[-latex] (N9) -- (0, -6.5) -- (2, -6.5) node[midway, below] {\scriptsize $C ::= C - 1$} -- (2, -3) -- (N5);
    \draw[-latex] (N5) -- (-2, -3) -- (-2, -7) -- (NE) node[midway, below] {\scriptsize $\assume\ \neg(C>0)$};
\end{tikzpicture}
		}
        \vspace{-6pt}
		\caption{PCFA $P$ of \cref{example: motivation}}
		\label{figure: motivation example cfa}
	\end{minipage}
    \hfill
	\begin{minipage}{0.35\textwidth}
		\centering
		\resizebox{!}{4.8cm}{
			\begin{tikzpicture}[yscale=0.7]
    \node at (0, 1) (N0) {};
    \node[draw, circle] at (0, 0) (N1) {};
    \node[circle, fill, inner sep = 1pt] at (0, -0.7) (D12) {};
    \node[draw, circle] at (0, -1.5) (N2) {};
    \node[circle, fill, inner sep = 1pt] at (0, -2.2) (D234) {};
    \node[draw, circle] at (0.5, -3) (N3) {};
    \node[circle, fill, inner sep = 1pt] at (8/30, -3.7) (D35) {};
    \node[draw, circle] at (-0.5, -3) (N4) {};
    \node[circle, fill, inner sep = 1pt] at (-8/30, -3.7) (D45) {};
    \node[draw, circle] at (0, -4.5) (N5) {};
    \node[circle, fill, inner sep = 1pt] at (0, -5.2) (D56) {};
    \node[draw, circle] at (0, -6) (N6) {};
    \node[circle, fill, inner sep = 1pt] at (0, -6.7) (D678) {};
    \node[draw, circle] at (-0.5, -7.5) (N7) {};
    \node[circle, fill, inner sep = 1pt] at (-8/30, -8.2) (D79) {};
    \node[draw, circle] at (0.5, -7.5) (N8) {};
    \node[circle, fill, inner sep = 1pt] at (8/30, -8.2) (D89) {};
    \node[draw, circle] at (0, -9) (N9) {};
    \node[circle, fill, inner sep = 1pt] at (2, -7) (D95) {};
    \node[circle, fill, inner sep = 1pt] at (-2.5, -7) (D5E) {};
    \node[draw, circle, accepting] at (0, -10.5) (NE) {};

    \draw[-latex] (N0) -- (N1);
    \draw (N1) -- (D12)  node[midway, right] {\scriptsize $\statementToLabel{X::=0}$};
    \draw[-latex] (D12) -- (N2) node[midway, right] {1};
    \draw (N2) -- (D234)  node[midway, right] {\scriptsize $\statementToLabel{0}$};
    \draw[-latex] (D234) -- (N3) node[midway, right] {\scriptsize $1/2$};
    \draw[-latex] (D234) -- (N4) node[midway, left] {\scriptsize $1/2$};
    \draw (N3) -- (D35)  node[midway, right] {\scriptsize $\statementToLabel{\stskip}$};
    \draw[-latex] (D35) -- (N5) node[midway, right] {\scriptsize $1$};
    \draw (N4) -- (D45)  node[midway, left] {\scriptsize $\statementToLabel{C::=0}$};
    \draw[-latex] (D45) -- (N5) node[midway, left] {\scriptsize $1$};
    \draw (N5) -- (D56)  node[midway, right] {\scriptsize $\statementToLabel{\assume\ C>0}$};
    \draw[-latex] (D56) -- (N6) node[midway, left] {\scriptsize $1$};
    \draw (N6) -- (D678)  node[midway, left] {\scriptsize $\statementToLabel{1}$};
    \draw[-latex] (D678) -- (N7) node[midway, left] {\scriptsize $1/2$};
    \draw[-latex] (D678) -- (N8) node[midway, right] {\scriptsize $1/2$};
    \draw (N7) -- (D79)  node[midway, left] {\scriptsize $\statementToLabel{X::=X+1}$};
    \draw[-latex] (D79) -- (N9) node[midway, left] {\scriptsize $1$};
    \draw (N8) -- (D89)  node[midway, right] {\scriptsize $\statementToLabel{\stskip}$};
    \draw[-latex] (D89) -- (N9) node[midway, right] {\scriptsize $1$};
    
    \draw (N9) -- (0, -9.5) -- (2, -9.5) node[midway, below]  {\scriptsize $\statementToLabel{C ::= C - 1}$} -- (D95);
    \draw (N5) -- (-2.5, -4.5) node[pos=0.7, above]  {\scriptsize $\statementToLabel{\assume\ \neg(C>0)}$} -- (D5E);

    \draw[-latex] (D95) -- (2, -4.5) node[midway, right]  {\scriptsize $1$} -- (N5);
    \draw[-latex] (D5E) -- (-2.5, -10.5) node[midway, left]  {\scriptsize $1$} -- (NE);

    \draw[-latex] (NE) -- (0, -11) node[midway, left] {\scriptsize $a_{\mathit{dmy}}$} -- (2, -11) node[draw, circle, fill, inner sep = 1pt, midway] {} -- (2, -10.5) -- (NE) node[midway, above] {\scriptsize $1$};
\end{tikzpicture} 
		}
        \vspace{-6pt}
		\captionof{figure}{The MDP $\toMDP{P}$ Underlying $P$}
		\label{figure: mdp of motivating example}
	\end{minipage}
    \vspace{-8pt}
\end{figure}

The program can be interpreted as a coin-flipping game between two players, say, Alice and Bob. 
They make an initial toss (line 3); if heads-down, Alice wins immediately. 
Otherwise, Bob chooses a number of rounds (variable $C$) to continue.
Alice then guesses that no heads-up will occur during these rounds.
The variable $X$ (line 5) tracks heads-up events, and Alice wins if the postcondition $X = 0$ is met.
Notably, the probability of Bob winning, i.e., violating the postcondition, increases as $C$ grows.
However, this probability is capped at $0.5$ as $C \to \infty$.

We proceed to demonstrate how our approach can establish the \emph{exact} bound of $0.5$ on the violation probability.
First, the program is transformed into a Probabilistic Control-Flow Automaton (PCFA), shown in~\cref{figure: motivation example cfa}.
In this transformation, each binary probabilistic choice is assigned a unique \emph{distribution tag} ($0$ and $1$ here).
Additionally, each choice is tagged as left ($\lbranch$) or right ($\rbranch$) for clarity.
Thus, probabilistic statements appear in the form $\probbranch{i}{d}$, where $i$ is the distribution tag, and $d$ is either $\lbranch$ or $\rbranch$.
The resulting PCFA $P$ will serve as the primary focus of our analysis.

\paragraph{Structural Abstraction}
Given the PCFA, one may observe that the formalism's structure resembles a Markov decision process (MDP),
where~\cref{figure: mdp of motivating example} visualised this intuition, in which distributions are marked by hyper-edges with a central $\bullet$ to indicate branching points.
This MDP is called the underlying MDP of $P$, denoted by $\toMDP{P}$.
In the MDP, statements are treated purely as labels without semantic functions, noted in $\statementToLabel{-}$.
Tags $0$ and $1$ appear as action names, $\statementToLabel{0}$ and $\statementToLabel{1}$, representing fair Bernoulli distributions, while other actions denote Dirac distributions.
Finally, to match the definition of MDP, we add a dummy action $a_{\mathit{dmy}}$ inducing a self-looping Dirac distribution at the ending location.

Notably, unlike the usual semantic approach seen in the literature~\cite{pcegar,bounded_model_checking_probabilistic_programs},
the MDP is \emph{not} an unrolling of the state space of the program.
In our method, \emph{structural abstraction}, 
the semantics of the statements are completely omitted,
so that disjoint assumptions like $\statementToLabel{\assume\ \neg(C>0)}$  
and $\statementToLabel{\assume\ (C>0)}$  
now become simply the usual \emph{non-deterministic} choices (actions) in the MDP.\@

With this MDP, the probability of violating the property 
(i.e., Bob winning) is surely bounded by the maximum reachability probability 
from the start to the end location.
Namely, the violation probability $\prob{\not\vdash \bs{\tTrue} \ P \ \bs{X = 0}}$ is bounded by the \emph{maximum reachability probability} in $\toMDP{P}$, termed the \emph{structural upper bound}, $\structBound{P}$.

However, $\toMDP{P}$ over-approximates $P$, yielding $\structBound{P} = 1$, a trivial upper bound.
Thus, this abstraction necessitates \emph{refinement}.
Examining the trivial bound reveals that both violating and non-violating traces contribute to this over-estimation.
For example, \cref{eq:trace-example-1} shows a safe trace that is not violating,
while \cref{eq:trace-example-2} illustrates an \emph{infeasible} trace that is impossible to occur in actual execution.
In~\cref{eq:trace-example-2}, the small $*$ marks the starting location of the loop in~\cref{figure: motivation example cfa}.

\begin{figure} [t]
    \vspace{-12pt}
	\centering
	\begin{minipage}{0.12\textwidth}
	\centering
        \vspace{5pt}
		\resizebox{!}{6cm}{
		\begin{tikzpicture}[yscale=0.7]
    \node at (0, 1) (N0) {};
    \node[draw, circle] at (0, 0) (N1) {};
    \node[draw, circle] at (0, -1.5) (N2) {};
    \node[draw, circle] at (-0.5, -3) (N4) {};
    \node[draw, circle] at (0, -4.5) (N5) {};
    \node[draw, circle, accepting] at (0, -7.5) (NE) {};

    \draw[-latex] (N0) -- (N1);
    \draw[-latex] (N1) -- (N2)  node[midway, right] {\scriptsize $X::=0$};
    \draw[-latex] (N2) -- (N4)  node[midway, left] {\scriptsize $\lprobbranch[0]$};
    \draw[-latex] (N4) -- (N5)  node[midway, left] {\scriptsize $C::=0$};
    \draw[-latex] (N5) -- (-0.5, -4.5) -- (-0.5, -7.5) node[midway, right]  {\scriptsize $\assume\ \neg(C>0)$} -- (NE);
\end{tikzpicture}
		}
            \vspace{-17pt}
		\captionof{Trace}{Safe}
		\label[trace]{eq:trace-example-1}
	\end{minipage}
	\hfill
    \hspace{-5pt}
	\begin{minipage}{0.17\textwidth}
	\centering
		\resizebox{!}{6cm}{
		\begin{tikzpicture}[yscale=0.7]
    \node at (0, 1) (N0) {};
    \node[draw, circle] at (0, 0) (N1) {};
    \node[draw, circle] at (0, -1) (N2) {};
    \node[draw, circle] at (-0.5, -2) (N4) {};
    \node[draw, circle] at (0, -3) (N5) {\scriptsize $*$};
    \node[draw, circle] at (0, -4) (N6) {};
    \node[draw, circle] at (0.5, -5) (N7) {};
    \node[draw, circle] at (0, -6) (N9) {};
    \node[draw, circle] at (0, -7) (N10) {\scriptsize $*$};
    \node[draw, circle, accepting] at (0, -8) (NE) {};

    \draw[-latex] (N0) -- (N1);
    \draw[-latex] (N1) -- (N2)  node[midway, right] {\scriptsize $X::=0$};
    \draw[-latex] (N2) -- (N4)  node[midway, left] {\scriptsize $\lprobbranch[0]$};
    \draw[-latex] (N4) -- (N5)  node[midway, left] {\scriptsize $C::=0$};
    \draw[-latex] (N5) -- (N6)  node[midway, left] {\scriptsize $\assume\ C>0$};
    \draw[-latex] (N6) -- (N7)  node[midway, left] {\scriptsize $\rprobbranch[1]$};
    \draw[-latex] (N7) -- (N9)  node[midway, left] {\scriptsize $\stskip$};
    \draw[-latex] (N9) -- (1, -6) -- (1, -7) node[midway, left]  {\scriptsize $C ::= C - 1$} -- (N10);
    \draw[-latex] (N10) -- (-1, -7) -- (-1, -8) node[midway, right]  {\scriptsize $\assume\ \neg(C>0)$} -- (NE);
\end{tikzpicture}
		}
            \vspace{-13pt}
		\captionof{Trace}{Infeasible}
		\label[trace]{eq:trace-example-2}
	\end{minipage}
	\hfill
    \hspace{-5pt}
	\begin{minipage}{0.3\textwidth}
	\centering
		\resizebox{!}{6cm}{
			\begin{tikzpicture}[yscale=0.7]
    \node at (0, 1) (HiddenStart) {};
    \node[draw, circle] at (0, 0) (Assn0) {};
    \node[draw, circle] at (0, -1) (PreProbChoice) {};
    \node[draw, circle] at (0.5, -2) (PreSkip) {};
    \node[draw, circle] at (0, -3) (FirstIf) {};
    \node[draw, circle] at (0, -4) (FirstProbChoice) {};
    \node[draw, circle] at (1, -4) (FirstSkip) {};
    \node[draw, circle] at (-1, -4) (FirstPlus) {};
    \node[draw, circle] at (2, -4) (FistSkipMinus) {};
    \node[draw, circle] at (-2, -4) (FirstPlusMinus) {};
    \node[draw, circle] at (0, -5) (SecondIf) {};
    \node[draw, circle] at (0, -6) (SecondProbChoice) {};
    \node[draw, circle] at (0.5, -7) (SecondSkip) {};
    \node[draw, circle] at (-0.5, -7) (SecondPlus) {};
    \node[draw, circle] at (0, -8) (SecondMinus) {};
    \node[draw, circle, accepting] at (0, -9) (NE) {};

    \draw[-latex] (HiddenStart) -- (Assn0);
    \drawLabelledArrow{Assn0}{PreProbChoice}{$X \opAssn 0$}
    \drawLabelledArrow{PreProbChoice}{PreSkip}{$\rprobbranch[0]$}
    \drawLabelledArrow{PreSkip}{FirstIf}{$\stskip$}
    \drawLabelledArrow{FirstIf}{FirstProbChoice}{$\assume\ C>0$}
    \drawLabelledArrow[below]{FirstProbChoice}{FirstSkip}{$\rprobbranch[1]$}
    \drawLabelledArrow[below]{FirstSkip}{FistSkipMinus}{$\stskip$}
    \draw[-latex] (FistSkipMinus) -- (2, -3) -- (FirstIf) node[pos=0.3, above] {\scriptsize $C \opAssn C - 1$};
    \drawLabelledArrow[above]{FirstProbChoice}{FirstPlus}{$\lprobbranch[1]$}
    \drawLabelledArrow[above=2]{FirstPlus}{FirstPlusMinus}{$X \opAssn X + 1$}
    \draw[-latex] (FirstPlusMinus) -- (-2, -4.5) -- (0, -4.5) -- (SecondIf) node[midway, right] {\scriptsize $C \opAssn C - 1$};
    \draw[-latex] (SecondIf) -- (-2, -5) -- (-2, -9) -- (NE) node[midway, above=0.1] {\scriptsize $\assume~\neg (C > 0)$};
    \draw[-latex] (SecondMinus) -- (2, -8) node[midway, below] {\scriptsize $C \opAssn C - 1$} -- (2, -5) -- (SecondIf);
    \drawLabelledArrow{SecondIf}{SecondProbChoice}{$\assume\ C>0$}
    \drawLabelledArrow{SecondProbChoice}{SecondSkip}{$\rprobbranch[1]$}
    \drawLabelledArrow[left]{SecondProbChoice}{SecondPlus}{$\lprobbranch[1]$}
    \drawLabelledArrow{SecondSkip}{SecondMinus}{$\stskip$}
    \drawLabelledArrow[left]{SecondPlus}{SecondMinus}{$X \opAssn X + 1$}
\end{tikzpicture}
		}
        \vspace{-1pt}
		\captionof{figure}{The Refined PCFA $P'$}
		\label[figure]{figure: motivating example violating automaton}
	\end{minipage}
    \hfill
    \hspace{-20pt}
	\begin{minipage}{0.3\textwidth}
	\centering
		\resizebox{!}{6cm}{
			\begin{tikzpicture}[yscale=0.7]
    \node at (0, 1) (HiddenStart) {};
    \node[draw, circle] at (0, 0) (Assn0) {};
    \node[circle, fill, inner sep = 1pt] at (0, -0.7) (D01) {};
    \node[draw, circle] at (0, -1.5) (PreProbChoice) {};
    \node[circle, fill, inner sep = 1pt] at (0, -2.2) (DPreChoice) {};
    \node[draw, circle, fill=gray!30] at (-0.5, -3) (Empty) {};
    \draw[-latex] (DPreChoice) -- (Empty) node[midway, left] {\scriptsize $1/2$};

    \draw[-latex] (Empty) -- (-0.5, -5) node[midway, left] {\scriptsize $a_{\mathit{dmy}}$} -- (-1.5, -5) node[draw, circle, fill, inner sep = 1pt, midway] {} -- (-1.5, -3) -- (Empty) node[midway, above] {\scriptsize $1$};
    
    \node[draw, circle] at (0.5, -3) (PreSkip) {};
    \node[circle, fill, inner sep = 1pt] at (8/30, -3.7) (DPreSkip) {};
    \node[draw, circle] at (0, -4.5) (FirstIf) {};
    \node[circle, fill, inner sep = 1pt] at (0, -5.2) (DFirstIf) {};
    \node[draw, circle] at (0, -6) (FirstProbChoice) {};
    \node[circle, fill, inner sep = 1pt] at (0, -7) (DFirstChoice) {};
    \node[draw, circle] at (1, -7) (FirstSkip) {};
    \node[circle, fill, inner sep = 1pt] at (2, -7) (DFirstSkip) {};
    \node[draw, circle] at (-1, -7) (FirstPlus) {};
    \node[circle, fill, inner sep = 1pt] at (-2, -7) (DFirstPlus) {};
    \node[draw, circle] at (3, -7) (FirstSkipMinus) {};
    \node[circle, fill, inner sep = 1pt] at (3, -5) (DFirstSkipMinus) {};
    \node[draw, circle] at (-2.8, -7) (FirstPlusMinus) {};
    \node[circle, fill, inner sep = 1pt] at (-2.8, -8.2) (DFirstPlusMinus) {};
    \node[draw, circle] at (0, -9) (SecondIf) {};
    \node[circle, fill, inner sep = 1pt] at (0, -9.7) (DSecondIf) {};
    \node[draw, circle] at (0, -10.5) (SecondProbChoice) {};
    \node[circle, fill, inner sep = 1pt] at (0, -11.2) (DSecondChoice) {};
    \node[draw, circle] at (0.5, -12) (SecondSkip) {};
    \node[circle, fill, inner sep = 1pt] at (8/30, -12.7) (DSecondSkip) {};
    \node[draw, circle] at (-0.5, -12) (SecondPlus) {};
    \node[circle, fill, inner sep = 1pt] at (-8/30, -12.7) (DSecondPlus) {};
    \node[draw, circle] at (0, -13.5) (SecondMinus) {};
    \node[circle, fill, inner sep = 1pt] at (3, -11) (DSecondMinusLoop) {};
    \node[circle, fill, inner sep = 1pt] at (-2.8, -9) (DSecondIfEnd) {};
    \node[draw, circle, accepting] at (0, -14.3) (NE) {};
    
    \draw[-latex] (HiddenStart) -- (Assn0);
    
    \draw (Assn0) -- (D01) node[midway, right] {\scriptsize $\statementToLabel{X \opAssn 0}$};
    \draw[-latex] (D01) -- (PreProbChoice) node[midway, right] {1};
    
    \draw (PreProbChoice) -- (DPreChoice) node[midway, right] {\scriptsize $\statementToLabel{0}$};
    \draw[-latex] (DPreChoice) -- (PreSkip) node[midway, right] {\scriptsize $1/2$};
    
    \draw (PreSkip) -- (DPreSkip) node[midway, right] {\scriptsize $\statementToLabel{\stskip}$};
    \draw[-latex] (DPreSkip) -- (FirstIf) node[midway, right] {\scriptsize $1$};
    
    \draw (FirstIf) -- (DFirstIf) node[midway, right] {\scriptsize $\statementToLabel{\assume\ C>0}$};
    \draw[-latex] (DFirstIf) -- (FirstProbChoice) node[midway, left] {\scriptsize $1$};
    
    \draw (FirstProbChoice) -- (DFirstChoice) node[midway, left] {\scriptsize $\statementToLabel{1}$};
    
    \draw[-latex] (DFirstChoice) -- (FirstSkip) node[midway, above] {\scriptsize $1/2$};
    \draw[-latex] (DFirstChoice) -- (FirstPlus) node[midway, below] {\scriptsize $1/2$};
    \draw (FirstSkip) -- (DFirstSkip) node[midway, above] {\scriptsize $\statementToLabel{\stskip}$};
    \draw[-latex] (DFirstSkip) -- (FirstSkipMinus) node[midway, below] {\scriptsize $1$};
    
    \draw (FirstPlus) -- (DFirstPlus) node[midway, above] {\scriptsize $\statementToLabel{X \opAssn X + 1}$};
    \draw[-latex] (DFirstPlus) -- (FirstPlusMinus) node[midway, below] {\scriptsize $1$};
    
    \draw (FirstSkipMinus) -- (DFirstSkipMinus) node[midway, left] {\scriptsize $\statementToLabel{C \opAssn C - 1}$};
    \draw[-latex] (DFirstSkipMinus) -- (3, -4.5) -- (FirstIf) node[pos=0.3, above] {\scriptsize $1$};
    
    \draw (FirstPlusMinus) -- (DFirstPlusMinus) node[midway, right] {\scriptsize $\statementToLabel{C \opAssn C - 1}$};
    \draw[-latex] (DFirstPlusMinus) -- (-2.8, -8.2) -- (0, -8.2) -- (SecondIf) node[pos=0.7, left] {\scriptsize $1$};
    
    \draw (SecondIf) -- (DSecondIf) node[midway, right] {\scriptsize $\statementToLabel{\assume\ C>0}$};
    \draw[-latex] (DSecondIf) -- (SecondProbChoice) node[midway, left] {\scriptsize $1$};
    
    \draw (SecondIf) -- (DSecondIfEnd) node[midway, below] {\scriptsize $\statementToLabel{\assume~\neg (C > 0)}$};
    \draw[-latex] (DSecondIfEnd) -- (-2.8, -14.3) -- (NE) node[midway, above] {\scriptsize $1$};
    
    \draw (SecondProbChoice) -- (DSecondChoice) node[midway, left] {\scriptsize $\statementToLabel{1}$};
    \draw[-latex] (DSecondChoice) -- (SecondSkip) node[midway, right] {\scriptsize $1/2$};
    \draw[-latex] (DSecondChoice) -- (SecondPlus) node[midway, left] {\scriptsize $1/2$};
    
    \draw (SecondSkip) -- (DSecondSkip) node[midway, right] {\scriptsize $\statementToLabel{\stskip}$};
    \draw[-latex] (DSecondSkip) -- (SecondMinus) node[midway, right] {\scriptsize $1$};
    
    \draw (SecondPlus) -- (DSecondPlus) node[midway, left] {\scriptsize $\statementToLabel{X \opAssn X + 1}$};
    \draw[-latex] (DSecondPlus) -- (SecondMinus) node[midway, left] {\scriptsize $1$};
    
    \draw (SecondMinus) -- (3, -13.5) -- (DSecondMinusLoop) node[midway, left] {\scriptsize $\statementToLabel{C \opAssn C - 1}$};
    \draw[-latex] (DSecondMinusLoop) -- (3, -9) -- (SecondIf) node[midway, above] {\scriptsize $1$};

    \draw[-latex] (NE) -- (0, -14.8) node[midway, left] {\scriptsize $a_{\mathit{dmy}}$} -- (3, -14.8) node[draw, circle, fill, inner sep = 1pt, midway] {} -- (3, -14.3) -- (NE) node[midway, above] {\scriptsize $1$};
    
\end{tikzpicture}
		}
        \vspace{-2pt}
		\captionof{figure}{The Refined MDP $P'$}
		\label[figure]{figure: motivating example violating automaton mdp}
	\end{minipage}
    \vspace{-12pt}
\end{figure}

\paragraph{Structural Refinement}
The refinement process reduces the structural probability of the MDP derived from the PCFA by \emph{eliminating} non-violating traces. 
To make this process operational, we show (in~\cref{theorem: refinement enabling}) that it suffices to construct a \emph{refinement automaton} $\refinement$ that over-approximates the \emph{violating traces} of $P$.
With this condition, the following equality holds:
$\prob{\not\vdash \bs{\precond} \ P \ \bs{\postcond}} = \prob{\not\vdash \bs{\precond} \ \minimise (P \cap \refinement) \ \bs{\postcond}}$, 
where $\minimise$ and $\cap$ represent the standard deterministic minimisation and intersection operations~\cite{principles_of_model_checking} for finite automata.
This implies that $\structBound{\minimise (P \cap \refinement)}$ remains an upper bound for $\prob{\not\vdash \bs{\precond} \ P \ \bs{\postcond}}$, referred to as the \emph{refined (structural) upper bound}.

Notably, constructing $\refinement$ depends solely on the computational semantics of statements, without involving any probabilistic considerations.
Specifically, as shown in traces like~\cref{eq:trace-example-1,eq:trace-example-2},
the probabilistic statements ($\probbranch{i}{d}$) do not affect whether a trace is violating.
In essence, these statements act as neutral $\stskip$ statements in computation.
This enables the use of traditional non-probabilistic program verification and analysis techniques for constructing $\refinement$.
We illustrate this approach through \emph{trace abstraction}~\cite{trace_abstraction, trace_abstraction_incremental},
the renowned technique in non-probabilistic verification, for this example.

\paragraph{Refinement with Trace Abstraction}
Using trace abstraction, we first \emph{generalise} the two identified non-violating traces into automata $Q_1$ and $Q_2$, shown in~\cref{fig: generalised automaton for trace 1} and~\cref{fig: generalised automaton for trace 2}.
The generalisation intuitively yields automata that encompass non-violating traces 
for the ``same reasons''~\footnote{Usually captured by the technique of ``interpolation''~\cite{craig_interpolation,trace_abstraction,trace_abstraction_modulo_probability}, to be detailed in~\Cref{subsec: refine by trace abstraction}.} as the provided traces.
For instance, \Cref{fig: generalised automaton for trace 1} generalises 
\Cref{eq:trace-example-1} by including all traces where $X$ does not 
increase after being assigned to $0$,
which must be safe as they always satisfy $X = 0$.
The complete rationale for this 
construction and more details on the techniques of trace abstraction are elaborated subsequently in~\Cref{subsec: refine by trace abstraction}.
Following the principles of trace abstraction,
the refinement automaton $\refinement$ is defined as $\refinement := \overline{Q_1 \cup Q_2}$, where the overline denotes the standard complement operation for finite automata;
since $Q_1$ and $Q_2$ contain only non-violating traces, the complement of their union, $\refinement$, over-approximates the violating traces in $P$.

The refined PCFA $P'$ is then constructed as: $P' := \min(P \cap \refinement)$, as illustrated in~\cref{figure: motivating example violating automaton}, with its underlying MDP depicted in~\cref{figure: motivating example violating automaton mdp}.
Intuitively, this PCFA includes \emph{only} traces that could lead to violation.
Specifically, it avoids the $\lprobbranch[0]$ branch, which resets $C$ and prevents loop entry.
To align with the definition of MDP, when reflected in the underlying MDP in~\Cref{figure: motivating example violating automaton mdp}, this becomes a transition with probability $1/2$ to go to a dummy, self-absorbing MDP node, as to be detailed in~\Cref{subsec:pcfa-as-mdp}.
Further, for a trace to be accepted, the loop must iterate at least once through the $\lprobbranch[1]$ branch before termination.
The refined upper bound $\structBound{\min(P \cap \refinement)}$ is then $0.5$, implying that the probability of Bob winning is also bounded by $0.5$, as expected.
It is straightforward to observe that the refinement process, 
including the generation of $\refinement$ and the computation of the 
refined PCFA $\min(P \cap \refinement)$, does not entail any 
probabilistic computation. 
Further, as will be further elaborated below,
the refinement above amounts to applying well-established 
non-probabilistic techniques.
This demonstrates the clear separation 
between probability and semantics in our approach.

\begin{figure} [t]
	\begin{minipage}{0.5\textwidth}
        \hspace{-8pt}
		\resizebox{!}{2.5cm}{
			\begin{tikzpicture}[xscale=0.75]
    \node[draw, circle, minimum size=20pt] at (0, 0) (N0) {\scriptsize \tTrue};
    \node[draw, circle, accepting, minimum size=20pt] at (3, 0) (N1) {\scriptsize $X=0$};



    \draw[-latex] (N0) edge [bend left=20] node[above] {\scriptsize $X ::= 0$} (N1);
    \draw[-latex] (N1) edge [bend left=20] node[below] {\scriptsize $\Sigma$} (N0);
    \draw[-latex] (N0) edge [out=200, in=160,looseness=12] node[above=5] {\scriptsize $\Sigma$} (N0); 
    \draw[-latex] (N1) edge [out=20, in=-20,looseness=12] node[above=8] {\scriptsize $\Sigma \setminus \bs{X ::= X + 1}$} (N1);
    \draw[-latex] (0, 1.2) -- (N0);
\end{tikzpicture}
		}
        \vspace{-23pt}
		\captionof{figure}{Generalised Automaton $Q_1$ of~\cref{eq:trace-example-1}}
		\label[figure]{fig: generalised automaton for trace 1}
	\end{minipage}
	\hfill
	\begin{minipage}{0.47\textwidth}
		\resizebox{!}{2.5cm}{
			\begin{tikzpicture}[xscale=0.75]
    \node[draw, circle, minimum size=20pt] at (0, 0) (N0) {\scriptsize \tTrue};
    \node[draw, circle, accepting, minimum size=20pt ] at (6, 0) (N1) {\scriptsize \tFalse};
    \node[draw, circle, minimum size=20pt] at (3, 0) (N2) {\scriptsize $C\le 0$};
    \draw[-latex] (N1) edge [out=-90, in=-90, looseness=0.6] node[above] {\scriptsize $\Sigma$} (N0);
    \draw[-latex] (N2) edge [out=-45, in=-135, looseness=3] node[above] {\scriptsize $\Sigma$} (N2);
    \draw[-latex] (N0) edge [bend left=20] node[above] {\tiny $\bs{\begin{matrix}\assume\ \neg (C > 0), \\ C ::= 0\end{matrix}}$} (N2);
    \draw[-latex] (N2) edge [bend left=20] node[below] {\scriptsize $\Sigma$} (N0);
    \draw[-latex] (N2) edge [bend left=20] node[above] {\scriptsize $\assume\ C > 0$} (N1);
    \draw[-latex] (N1) edge [bend left=20] node[below] {\scriptsize $\Sigma$} (N2);
    \draw[-latex] (N0) edge [out=200, in=160,looseness=12] node[above=5] {\scriptsize $\Sigma$} (N0);
    \draw[-latex] (N1) edge [out=20, in=-20,looseness=12] node[above=5] {\scriptsize $\Sigma$} (N1);
    \draw[-latex] (0, 1.2) -- (N0);
    
\end{tikzpicture}
		}
        \vspace{-13pt}
		\captionof{figure}{Generalised Automaton $Q_2$ of~\cref{eq:trace-example-2}}
		\label[figure]{fig: generalised automaton for trace 2}
	\end{minipage}
    \vspace{-10pt}
\end{figure}

\section{Preliminaries}
\label{sec: preliminaries}

Following functional conventions, we assume the data structure of lists to be linked lists with constructor ``$\midcons$''.
We use the function $\init$ to take the whole list except the last element.
We denote the update of a function $f$ with a new mapping $X \mapsto V$ by $f[X \mapsto V]$.
By slightly abusing symbols, we will use $\in$ and $\notin$ for lists.

\paragraph{Probabilistic Theory}
A distribution over a countable set $S$, $\doubleP \in \Dist(S)$, is a function that assigns to each subset~\footnote{
  A simplified version of $\sigma$-algebra from~\cite{mciver2005abstraction}.
} of $S$ a number between $0$ and $1$ such that: $\prob{\emptyset} = 0$, $\prob{S} = 1$ and for any countable sequence of pairwise disjoint subsets of $S$, $S_1, S_2, \dots$, $\prob{\uplus_{i = 1}^\infty S_i} = \sum_{i = 1}^\infty \prob{S_i}$.
For single element $s$, if it is clear from the context, we denote $\prob{\bs{s}}$ by just $\prob{s}$.
We use $\Supp(\doubleP)$ to denote exactly all the non-$0$ elements of $S$, that:
$\Supp(\doubleP) = \bs{s \in S \ | \ \prob{s} \neq 0}$.
A sub-distribution $\doubleP$ over set $S$ is almost the same as a distribution except that
$0 \le \prob{S} \le 1$.

A \emph{Markov chain} (MC) $\mc$ is a pair $(Q, \scriptP)$, where
$Q$ is a set of nodes~\footnote{
  We use the name \emph{MDP nodes} to distinguish from \emph{program states} below.
}, and
$\scriptP : Q \rightharpoonup \Dist(Q)$ is a partial function that assigns nodes 
a distribution.
A trajectory $\trajectory$ of an MC $\mc (Q, \scriptP)$ is a finite sequence of nodes $q_1, \dots, q_n$, where for each $i \in \bs{1, \dots, n - 1}$, $q_{i + 1} \in \Supp(\scriptP(q_i))$.
The set of trajectories starting from a nodes $q$ is denoted by $\Trajectories_\mc(q)$.
Then, the \emph{reachability probability} from a node $q$ to a node $q'$ is
\(
  \reach_\mc(q, q') := \prob{\setdef{
  \trajectory \in \Trajectories_\mc(q)
  }{
  \last(\trajectory) = q', q' \notin \init(\trajectory)}
  }
\),
where $\last$ denotes the final element of the sequence and $\init$ denotes the sequence with the last element removed.

A \emph{Markov decision process} (MDP) $\mdp$ is a tuple $(Q, \Act, \scriptP)$, where
$Q$ is a set of nodes;
$\Act$ is a set of actions;
$\scriptP : Q \times \Act \rightharpoonup \Dist(Q)$ is a partial function that 
assigns a node-action pair a distribution on nodes, where for a given node $q$, an action $a$ is said to be \emph{enabled} by the node, if \rmblock{$(q, a) \in \dom(\scriptP)$}.
Here $\dom$ means the domain of a (partial) function.
A (deterministic) policy $\psi$ of an MDP is a tuple $(S, \delta, s_0)$, where $S$ is a (possibly infinite) set of policy states;
$\delta : Q \times S \rightharpoonup \Act \times S$ selects an enabled action and returns the next policy state given a policy state and MDP node; $s_0$ is the initial policy state.
The policy set of an MDP $\mdp$ is denoted by $\scriptS(\mdp)$.
A policy with $|S| = 1$ is a \emph{simple policy}.
Applying a policy $\psi (S, \delta, s_0)$ to an MDP $\mdp (Q, \Act, \scriptP)$, denoted by $\applyPolicy{\mdp}{\psi}$, produces an MC, $(Q \times S, \scriptP^\psi)$, where
$\delta(q, s) = (a, s')$ implies $\scriptP^\psi(q, s)(q', s') = \scriptP(q, a)(q')$, with $\scriptP^\psi(q, s)(q', s'') = 0$ for $s'' \neq s'$.
The maximum reachability probability from $q$ to $q'$ in $\mdp$ is:
\(
  \maxReach{\mdp}{q}{q'} := 
  \sup_{\psi \in \scriptS(\mdp)} 
  \reach_{\applyPolicy{\mdp}{\psi}}(q, q')
\).

\paragraph{Statements and PCFA}
Throughout this paper, we fix a finite set of variables $\scriptV$ and define $\scriptE$ and $\scriptB$ as the sets of \emph{arithmetic} and \emph{Boolean} expressions over $\scriptV$, respectively.
We denote variables by upper case letters $X, Y, \dots$.
For simplicity, we assume each variable in $\scriptV$ takes values in $\doubleN$, the natural numbers.
A \emph{valuation} $v$ is a function $\scriptV \to \doubleN \uplus \{\Undefined\}$, where $\Undefined$ intuitively represents a ``non-sense'' value;
the set of all valuations is denoted $\doubleV$.
We write $\interpret{E}_v$ and $\interpret{B}_v$ for the \emph{evaluation} of expressions $E$ and $B$ under valuation $v$, yielding values other than $\Undefined$.
A predicate $\varphi$ is simply a Boolean expression, and we use $\Phi$ to denote the set of all predicates.
We write $v \models \varphi$ to represent $\interpret{\varphi}_v = \tTrue$.
Notably, $\interpret{\varphi}_\bot=\tFalse$ for any predicate $\varphi$.
The satisfiability and unsatisfiability of predicates follow standard definitions.
The program $P$ studied in our paper is essentially standard, 
which is recursively given by the following production rules:
\begin{align*}
 P := \ & X \opAssn E
	\ | \ P_1 \oplus P_2
	\ | \ P_1 \circledast P_2
	\ | \ \stskip
	\ | \ P_1; P_2
	\ | \ \texttt{if} \ B \ \texttt{then} \ P_1 \ \texttt{else} \ P_2 
	\ | \ \texttt{while} \ B \ \texttt{do} \ P' \ \texttt{done}
\end{align*}
Here, $E \in \scriptE$ and $B \in \scriptB$, the operator $\oplus$ denotes the fair binary probabilistic choice~\footnote{
  For simplicity, we introduce only the fair binary choice,
  but it can be easily generalised to work with any number.
  And fair binary choice itself does not hinder theoretical generality~\cite{prob_turing_complete}.
} and $\circledast$ represents the non-deterministic choice.
As in the rest of this work, programs are represented as \emph{probabilistic control-flow automata} (PCFA) in the standard way~\cite{wang2024static,exponential_analysis_of_probabilistic_program}, for which, the details are presented in the appendix.

We then delve into the introduction of PCFA and its related concepts.
A \emph{statement} (to appear in a PCFA) $\sigma$ over $\scriptV$ is given by:
\[
  \sigma := 
  \stskip \ | \ 
  X \opAssn E \ | \ 
  \assume~B \ | \ 
  \lprobbranch \ | \ \rprobbranch \ | \ 
  *_i
\]
where, again, 
$E \in \scriptE$ and $B \in \scriptB$ range over arithmetic and Boolean expressions respectively;
the statements $\lprobbranch$ and $\rprobbranch$ are the (fair) probabilistic statements, and $*_i$ is the non-deterministic statement, 
where $i$ ranges over identifiers~\footnote{
  The identifiers can be any set of symbols, e.g., natural numbers in the motivating example.
}, which is called a \emph{distribution tag} when appearing in a probabilistic statement and non-deterministic tag when in a non-deterministic statement.
The set of all statements is denoted $\setOfStatements$.
The \emph{evaluation} of a statement $\sigma$ over a valuation $v$, denoted by $\interpret{\sigma}_v$, is 
defined in the standard way, 
that
$\interpret{\assume~B}_v := \Undefined$ if $\interpret{B}_v = \tFalse$,
$\interpret{\sigma}_\Undefined := \Undefined$,
and $\interpret{X\opAssn~E}_v := v[X \mapsto \interpret{E}_v]$,
with all other cases having $\interpret{\sigma}_v := v$.

\begin{definition}[Probabilistic Control-Flow Automaton]
  \label{def: pcfa}
  A probabilistic control-flow automaton (PCFA) $A$ is a tuple $(L, \Sigma, \Delta, \ell_0, \ell_e)$, where 
  $L$ is a \emph{finite} set of locations;
  $\Sigma \subseteq \setOfStatements$ is a \emph{finite} set of statements;
  $\Delta : L \times \Sigma \rightharpoonup L$ is a \emph{partial} transition function, and we write $\ell \xrightarrow{\sigma} \ell'$ to denote that 
  $(\ell, \sigma) \in \dom(\Delta)$ and $\Delta(\ell, \sigma) = \ell'$;
  $\ell_0 \in L$ is the initial; and, $\ell_e \in L$ is the ending location.
  Specifically, no transition starts from the ending location $\ell_e$, that is: \rmblock{$\forall \sigma.(\ell_e, \sigma) \notin \dom(\Delta)$}.
\end{definition}

Notably, from an automata-theoretic perspective, PCFA is inherently \emph{restricted}:  
it is essentially a \emph{Deterministic Finite Automaton} (DFA)~\cite{principles_of_model_checking}  
with a \emph{single} ending location,
which has \emph{no out-transitions}.
These restrictions are removed later in the concept of \emph{general PCFA} in~\Cref{subsec:the-refinement-of-structural-abstraction}.

A \emph{program state} $C$ of a PCFA $A$ is a pair $(\ell, v)$, where
$\ell$ is a location of $A$ and $v$ is a valuation.
A transition $\ell \xrightarrow{\sigma} \ell'$ is said to be enabled by program state $(\ell, v)$ if either: $\sigma = \assume~B$ is an assumption statement with $\interpret{B}_v = \tTrue$, or $\sigma$ is another type of statement.
A program state $(\ell', v')$ is the \emph{successor program state} of $C = (\ell, v)$ under transition $r := \ell \xrightarrow{\sigma} \ell'$ if:
$r$ is an enabled transition by $C$, and 
either $v' = v[X \mapsto \interpret{E}_v]$ for an assignment $\sigma = X \opAssn E$; or $v' = v$ otherwise.
A \emph{computation} $\pi$ is a finite sequence of program states with transitions:
$(\ell_0, v_0) \xrightarrow{\sigma_1} (\ell_1, v_1) \cdots \xrightarrow{\sigma_n} (\ell_n, v_n)$, where:
$\ell_0$ is the initial location of $A$;
each $\ell_i$ for $i \in \bs{0,\dots,n}$ is a location of $A$;
and, for each $(\ell_i, v_i) \xrightarrow{\sigma_{i+1}} (\ell_{i+1}, v_{i+1})$, the program state $(\ell_{i+1}, v_{i+1})$ is a successor of $(\ell_i, v_i)$ under transition $\ell_i \xrightarrow{\sigma_{i+1}} \ell_{i+1}$.
A computation with the final program state at location $\ell_e$ is termed a \emph{terminating computation}.
The weight of a computation $\pi$ is given by 
$wt(\pi) = {(1/2)}^n$, where $n$ is the count of probabilistic transitions in $\pi$.
Given a finite-length computation $\pi$ starting from the initial location,
a \emph{trace} $\tau$ \emph{induced} by it, denoted $\ind(\pi)$, refers to the sequence of labels involved.
It is termed \emph{accepting} if $\pi$ ends in $\ell_e$ (i.e., is terminating).
Additionally, we require traces to be non-empty.
The set of accepting traces of a PCFA $A$ is denoted $\scriptL(A)$.
Notably, a trace $\tau$ uniquely determines a computation and the definition of the \emph{weight} of $\tau$, follows naturally.

Multiple non-probabilistic transitions may be enabled by a given program state simultaneously, introducing non-determinism. 
To address this, we define an \emph{automaton scheduler}.
An automaton scheduler $\xi$ for a PCFA $A$ is a function that assigns each non-terminating computation $\pi$ of $A$ a unique (sub-)probabilistic distribution over the transitions enabled by the final program state of $\pi$.
Specifically, $\xi(\pi)$ is:
(a) a Dirac distribution on a transition with a non-probabilistic statement;
(b) a fair Bernoulli distribution over two probabilistic transitions, \rmblock{$\lprobbranch$} and \rmblock{$\rprobbranch$}, sharing a distribution tag $i$; or, 
(c) a sub-distribution on a single probabilistic transition, \rmblock{$\lprobbranch$} or \rmblock{$\rprobbranch$}, tagged by $i$, with probability $1/2$.
The set of schedulers for $A$ is denoted by $\Xi_A$.
A computation $\pi := C_0 \xrightarrow{\sigma_1} C_1 \cdots \xrightarrow{\sigma_n} C_n$ is \emph{guided} by an automaton scheduler $\xi$ if, for every prefix $C_0 \xrightarrow{\sigma_1} \cdots \xrightarrow{\sigma_i} (\ell_i, v_i) \xrightarrow{\sigma_{i+1}} (\ell_{i+1}, v_{i+1})$ of $\pi$, the transition $\ell_i \xrightarrow{\sigma_{i+1}} \ell_{i+1}$ belongs to $\Supp(\xi(C_0 \xrightarrow{\sigma_1} \cdots \xrightarrow{\sigma_i} C_i))$.

Given an initial valuation $v_\mathinit$ and a scheduler $\xi$, a PCFA $A$ defines a probabilistic distribution over the final valuation at the ending location.
Formally, let the set of terminating computations starting from $(\ell_0, v_\mathinit)$ and being guided by $\xi$ to be $\Pi_{v_\mathinit}^\xi$, which is also called a \emph{run} of $A$, the (sub)distribution~\footnote{
  Depending on whether the program is almost surely terminating (AST).
} over a set of valuations at the ending location is:
\[
  \prob[A][(v_\mathinit, \xi)]{V} := 
  \sum_{\setdef{\pi \in \Pi_{v_\mathinit}^\xi}{\last(\pi) = (\ell_e, v), v \in V}} wt(\pi)
\]
Intuitively, given a fixed initial value and a scheduler, a run of a PCFA represents the Markov chain that encompasses all the computations that may be executed due to probabilistic choices.
See, e.g.,~\Cref{fig: example of a run} for an example.

\paragraph{Problem Statement}
Hence, quantifying both the initial valuation and the scheduler, {\bf \textit{violation probability}} of a PCFA $A$ against a pre-condition $\varphi_e$ and a post-condition $\varphi_f$ is:
\begin{align}
  \prob{\not\vdash \bs{\varphi_e} \ A \ \bs{\varphi_f}} := 
  \sup_{v_\mathinit \models \varphi_e} \sup_{\xi \in \Xi_A} 
  \prob[A][(v_\mathinit, \xi)]{\bs{v \in \doubleV \ | \ v \models \neg \varphi_f}}
  \label{eq:std-prob-def}
\end{align}
The pair of pre- and post-conditions $(\precond, \postcond)$ is called the specification or property of the program.
Further, given a threshold $\beta$, the verification problem is then to answer whether $\prob{\not\vdash \bs{\precond} \ A \ \bs{\postcond}} \le \beta$.
In the rest of the paper, we fix a PCFA $A$ and a specification $(\precond, \postcond)$.

\begin{figure}
  \begin{tikzpicture}[yscale=0.6]
    \node at (-1, 0) (N0) {};
    \node[draw, circle] at (0, 0) (N1) {};
    \node[draw, circle] at (2, 0) (N2) {};
    \node[draw, circle] at (4, 0.5) (N3) {};
    \node[draw, circle] at (4, -0.5) (N4) {};
    \node[draw, circle] at (6, 0) (N5) {};
    \node[draw, circle, accepting] at (9, 0) (NE) {};

    \draw[-latex] (N0) -- (N1);
    \draw[-latex] (N1) -- (N2)  node[midway, above] {\scriptsize $X::=0$};
    \draw[-latex] (N2) -- (N3)  node[midway, above] {\scriptsize $\rprobbranch[0]$};
    \draw[-latex] (N2) -- (N4)  node[midway, below] {\scriptsize $\lprobbranch[0]$};
    \draw[-latex] (N3) -- (N5)  node[midway, above] {\scriptsize $\stskip$};
    \draw[-latex] (N4) -- (N5)  node[midway, below] {\scriptsize $C::=0$};
    \draw[-latex] (N5) -- (NE) node[midway, above] {\scriptsize $\assume\ \neg(C>0)$};
\end{tikzpicture}
  \vspace{-3pt}
  \captionof{figure}{Example of a Run in~\cref{example: motivation}}
  \label[figure]{fig: example of a run}
  \vspace{-7pt}
\end{figure}

\section{Structural Abstraction and Refinement}
\label{sec: PCFA as MDP}

In this section, we present the theoretical foundations of our method, 
namely, structural abstraction and its refinement.
These established 
principles form the theoretical basis for the algorithms to be 
introduced in the subsequent section.
In~\Cref{subsec:pcfa-as-mdp}, we present a formal treatment of PCFA as an MDP,
which lays down a solid foundation for the central concept of this paper: \emph{structural abstraction},
introduced in~\Cref{subsec:structural-abs}.
We complement the abstraction with its corresponding refinement principle in~\Cref{subsec:structural-refinement}.

\subsection{PCFA as MDP}
\label{subsec:pcfa-as-mdp}

This part delves into the key insights of our method -- ``\emph{structurally} viewing PCFA as MDP''.
Formalising this insight necessitates a \emph{new definition} of the violation probability.
As a result, an equality between it and the standard definition will be established in~\Cref{theorem: two defs equal}.
As briefly overviewed in~\Cref{figure: motivation example cfa}  
and~\Cref{figure: mdp of motivating example}, a PCFA $A$ 
can be viewed as an MDP.\@
This is achieved by mapping  
automata locations to MDP nodes, statements (or distribution tags)  
to actions, and transitions to probabilistic distributions.  
Notably, our method differs significantly from the \emph{semantical}  
approaches established in the literature~\cite{pcegar,cegar}.  
In our approach, the nodes in the MDP correspond to \emph{locations},
rather than to \emph{program states} of the original PCFA.\@
More specifically:

\begin{definition}[Underlying MDP]
  For PCFA $A (L, \Sigma, \Delta, \ell_0, \ell_e)$, its underlying MDP $\toMDP{A}$ is $(L \cup \{\dummyMdpState\}, \Act(A), \scriptP)$ where:
  \begin{itemize}
    \item \textbf{Nodes}: $A$'s locations plus dummy state $\dummyMdpState$
    \item \textbf{Actions} $\Act(A)$ contains:  
    (1) $\angbr{\sigma}$ for non-probabilistic $\sigma \in \Sigma$;  
    (2) $\angbr{i}$ for each $\probbranch{i}{d} \in \Sigma$; and,
    (3) $a_{\mathit{dmy}}$ the dummy action for $\dummyMdpState$ and the ending location $\ell_e$.
    \item \textbf{MDP Transitions} $\scriptP$:  
    \begin{enumerate}
      \item Every transition $\ell \xrightarrow{\sigma} \ell' \in \Delta$ with non-probabilistic $\sigma$ triggers a Dirac distribution on $\ell'$,
      namely, $\scriptP(\ell, \angbr{\sigma})(\ell') = 1$.
      \item For probabilistic transitions $\ell \xrightarrow{\probbranch{i}{d}} \ell_1$,
      $P := \scriptP(\ell, \angbr{i})$ depends on whether the complementary transition $\ell \xrightarrow{\probbranch{i}{\overline{d}}} \ell_2$ exists
      (where $\overline{\lbranch} = \rbranch$ and $\overline{\rbranch} = \lbranch$):
      (1) if it exists, $P$ is a fair Bernoulli distribution on $\ell_1$ and $\ell_2$ if $\ell_1 \neq \ell_2$;
      (2) if it exists with $\ell_1 = \ell_2$ then $P(\ell_1) = 1$ is Dirac again; and,
      (3) if it does not exist, $P$ is a fair Bernoulli distribution on $\ell_1$ and $\dummyMdpState$.
      \item To match the definition of MDP, $\dummyMdpState$ and the ending location $\ell_e$ have a unique dummy action $a_{\mathit{dmy}}$, which yield Dirac self-loop for both of them.
    \end{enumerate}
  \end{itemize}
  \vspace{-6pt}
\end{definition}

Similar to MDP and Markov Chains (MC), a PCFA $A$ where each location has at most one enabled action is effectively an MC.\@
Thus, we define \emph{policies for PCFA}, also denoted $\psi$, analogous to MDP~\cite{principles_of_model_checking},
as a triple $(S, \delta, s_0)$, where
$S$ is a set of policy nodes;
$\delta : L \times S \rightharpoonup \Act(A) \times S$ 
selects an action
from the current node and PCFA location;
$s_0$ is the initial policy node.

Inheriting from MDP, a policy with only one node, $|S| = 1$, is called a \emph{simple policy}.
The set of PCFA policies is denoted by $\scriptS(A)$.
Clearly, each policy $\psi \in \scriptS(A)$
corresponds uniquely to a policy in $\scriptS(\toMDP{A})$.
Applying a policy $\psi = (S, \delta, s_0)$ to a PCFA $A = (L, \Sigma, \Delta, \ell_0, \ell_e)$ induces another PCFA, denoted $\applyPolicy{A}{\psi} = (L^\psi, \Sigma, \Delta^\psi, (\ell_0, s_0), \ell_e)$, where:
the new location set $L^\psi := ((L \setminus \bs{\ell_e}) \times S) \uplus \bs{\ell_e}$ isolates the ending location $\ell_e$, making other locations the product with $S$;
the new transition function $\Delta^\psi$ selects transitions per the policy, so $(\ell, s) \xrightarrow{\sigma} (\ell', s')$ in $\Delta^\psi$ iff:
there exists $\ell \xrightarrow{\sigma} \ell'$ in $\Delta$, and either
(1) $\sigma$ is not probabilistic and $\delta(\ell, s) = (\sigma, s')$, or
(2) $\sigma$ is probabilistic with distribution tag $i$, and $\delta(\ell, s) = (i, s')$;
also, $(\ell, s) \xrightarrow{\sigma} \ell_e$ exists in $\Delta^\psi$ iff the above holds for some $s'$.
This application induces a PCFA interpretable as an MC.\@

Each run $\Pi$ of a PCFA corresponds to a specific policy.
Intuitively, policy states correspond to program states, and actions to the scheduler's resolution of non-determinism.
Applying a policy produces a PCFA where the set of accepting traces matches the run's induced traces $\setdef{\ind(\pi)}{\pi \in \Pi}$.
Conversely, not every policy represents a PCFA run.
Some policies produce ``non-sense'' traces not induced by computations.
For instance, in~\cref{eq:trace-example-2}, assigning $C = 0$ followed by $\assume~C > 0$ creates a contradiction.
Such traces are \emph{infeasible}; others are \emph{feasible}.

We define the \emph{evaluation of a trace}, $\interpret{\tau}$, as chaining its elements' computations:
$\interpret{[\sigma]}(v) := \interpret{\sigma}_v$ for a singleton trace,
and $\interpret{\sigma :: \tau}(v) := \interpret{\tau}(\interpret{\sigma}_v)$.
Given an initial valuation $v_\mathinit$ and a post-condition $\postcond$, a trace $\tau$ is a \emph{violating trace} if $\interpret{\tau}(v_\mathinit) \models \neg \postcond$; otherwise, it is \emph{non-violating}.
These concepts generalise to a pre-condition $\precond$ by existential quantification: $\exists v_\mathinit \models \precond.\interpret{\tau}(v_\mathinit) \models \neg \postcond$.
The \emph{path condition} of a trace $\tau$, denoted $\pathCond(\tau)$, is the \emph{weakest} predicate indicating $\precond$ and satisfying $\forall v \models \pathCond(\tau).\interpret{\tau}(v) \models \neg \postcond$.
For a trace set $\Theta$, $\pathCond(\Theta) := \bigwedge_{\tau\in\Theta} \pathCond(\tau)$.

Thus, we observe: the probability of a PCFA $A$ violating the pre- and post-condition $\precond$ and $\postcond$ is derived by traversing all possible policies and, for each, accumulating all violating traces, formally expressed as:
\begin{align}
  \mdpProb{A}{\precond}{\postcond} := 
  \sup_{v \models \varphi_e} \sup_{\psi \in \scriptS(A)} 
  \sum_{\tau \in \scriptL(\applyPolicy{A}{\psi})} 
  \wt(\tau) \cdot \indicator{\interpret{\tau}(v) \models \neg \varphi_f}
  \label{eq: prob def}
\end{align}
where $\indicator{-}$ is the indicator function that returns $1$ if the given value equals to $\tTrue$ and $0$ otherwise.
We call the formula $\indicator{\interpret{\tau}(v) \models \neg \postcond}$ the \emph{semantic check} of trace $\tau$.
Comparing this definition with the standard definition in~\Cref{eq:std-prob-def}, 
although both policies and schedulers resolve non-determinism, 
schedulers account for the concrete valuation, whereas policies 
can be arbitrary and do not consider semantics.
This definition hence intuitively ``delays'' the step-wise valuation transition within 
a computation to the final semantic check,
allowing the selection 
of the next transition to be \emph{arbitrary}.
More specifically, in the standard definition, for any computation 
$(\ell_0, v_0) \xrightarrow{\sigma_1} \cdots \xrightarrow{\sigma_n} (\ell_n, v_n)$, 
one must ensure the transition between each $v_i$ and $v_{i+1}$ is valid. 
The new definition, however, delays all step-wise obligations to an 
overall semantic check, i.e., $\indicator{\interpret{\sigma_1\dots\sigma_n}(v_0) \models \neg \postcond}$. 
This new definition (\Cref{eq: prob def}) thus distinguishes itself by separating semantic 
checks from probability computation, enabling a potential division 
between semantic verification and policy selection.

This approach, however, inevitably introduces nonsensical traces, which 
lack corresponding computations and are evaluated to $\bot$.
Nevertheless, we argue that by considering \emph{all} possible policies, 
the overall probabilities are the same as the standard definition.
Formally, we have the following theorem, with a detailed proof provided in the appendix.

\begin{theorem}
  \label{theorem: two defs equal}
  The two definitions in~\Cref{eq:std-prob-def,eq: prob def} are equivalent for any $A$, $\precond$ and $\postcond$:
  \begin{align}
  \mdpProb{A}{\precond}{\postcond} =
  \prob{\not\vdash \bs{\precond} \ A \ \bs{\postcond}}
  \label{eq: two defs equal}
  \end{align}
\end{theorem}

The theorem then concludes the validity of viewing a PCFA as an MDP,
laying down a solid foundation for the structural abstraction below.

As a brief recapitulation, to introduce our new definition, we defined 
several key concepts, e.g., the policies, especially the simple ones,
the evaluation of traces, feasibility and the path conditions.
They all play significant roles in the below introduction as we will focus on the new definition.

\subsection{Structural Abstraction}
\label{subsec:structural-abs}

Based on the aforementioned definition, the concept of \emph{structural abstraction} emerges naturally;
this approach disregards computational semantics and utilises the underlying MDP of the PCFA to abstract the probabilistic program.
Formally, in~\cref{eq: prob def}, the semantic check $\indicator{\interpret{\tau}(v) \models \neg \postcond}$ is ignored, and is consistently treated as $1$.
This therefore yields an upper bound of the violation probability:

\begin{definition}[Structural Upper Bound]
  The \emph{structural upper bound} of a PCFA $A$ with initial and ending locations $\ell_0$ and $\ell_e$ is given by:
  \begin{align}
    \structBound{A} :=
    \maxReach{\toMDP{A}}{\ell_0}{\ell_e} =
    \sup_{\psi \in \scriptS(A)}
    \sum_{\tau \in \scriptL(\applyPolicy{A}{\psi})}
    \wt(\tau)
    \label{eq: structural bound}
  \end{align}
\end{definition}

By definition and~\Cref{eq: two defs equal}, we instantly have:
\begin{align}
  \prob{\not\vdash \bs{\precond} \ A \ \bs{\postcond}} =
  \mdpProb{A}{\precond}{\postcond} \le \structBound{A}
  \label{eq: struct bound is an upper bound}
\end{align}

Thus, while calculating the exact value in~\eqref{eq: prob def} is challenging,
the structural upper bound can be efficiently computed with standard algorithms~\cite{principles_of_model_checking}
commonly used to determine maximum reachability probabilities in MDP.

Besides the principles of abstraction above,
we also developed the following result for the new definition as the principle behind \emph{refinement},
which essentially exclude non-violating traces while retaining the violating ones.
A proof is provided using the new definition in the appendix.

\begin{lemma}
  \label{theorem: cross-structural equality}
  For two PCFA $A_1$ and $A_2$, when they have the same set of 
  violating traces against the given pre- and post-conditions $\precond$ and $\postcond$, 
  then, we have:
  \[
    \prob{\not\vdash \bs{\precond} \ A_1 \ \bs{\postcond}} 
    = 
    \mdpProb{A_1}{\precond}{\postcond}
    = 
    \mdpProb{A_2}{\precond}{\postcond}
    = 
    \prob{\not\vdash \bs{\precond} \ A_2 \ \bs{\postcond}}
  \]
\end{lemma}

Notably, the lemma only requires the two PCFA to have the same set of violating traces,
and has no requirement on the non-violating traces:
$A_1$ and $A_2$ are even allowed to have different sets of statements $\Sigma$.

As the refinement procedure involves removal of the non-violating traces from $A$,
which essentially yields another PCFA $A'$,
the result essentially enables the \emph{cross-structural} comparison of the violating probability between different PCFA.

\subsection{The Refinement of Structural Abstraction}
\label{subsec:the-refinement-of-structural-abstraction}

\paragraph{General PCFA}
Prior to exploring the details of refinement, let us first introduce the concept of \emph{general PCFA}.
As discussed in~\Cref{sec: preliminaries}, the notion of PCFA is essentially restricted.
We then introduce the concept of a general PCFA by easing these restrictions, so that in such a notion:\@
(1) the transition function $\Delta$ is now a transition relation $\Delta : L \times \Sigma \times L$,
hence $\ell \xrightarrow{\sigma} \ell'$ then denotes $(\ell, \sigma, \ell') \in \Delta$;
(2) the ending location is not unique, and now there is a set $L_e \subseteq L$ called 
the ending location set;
(3) transitions are allowed to start from an ending location.
The terminologies for general PCFA are inherited directly from PCFA.\@
This definition essentially renders general PCFA simply any \emph{Non-deterministic Finite Automata} (NFA) without restriction.

\begin{example}[General PCFA]
  The generalised automata in~\cref{fig: generalised automaton for trace 1,fig: generalised automaton for trace 2} for~\cref{eq:trace-example-1,eq:trace-example-2} are general PCFA but NOT PCFA,
  as they are non-deterministic, also, there are transitions starting from the ending location.\qed
\end{example}

Building on~\Cref{theorem: cross-structural equality},
consider a PCFA $A$ and a general PCFA $\refinement$ that over-approximates the violating traces of $A$ against the given pre- and post-conditions $\precond$ and $\postcond$.
One might naturally attempt to derive a refined bound as $\structBound{A \cap \refinement}$ by directly applying~\Cref{theorem: cross-structural equality}.
However, a technical obstacle arises:
to apply~\Cref{theorem: cross-structural equality},
both automata $A_1$ and $A_2$ must be PCFA conforming to~\Cref{def: pcfa}.
Recall that a PCFA imposes three structural constraints:
\hypertarget{cond-a}{(a)} determinism, \hypertarget{cond-b}{(b)} a unique ending location, and \hypertarget{cond-c}{(c)} no out-transitions from the ending location.
Since $\refinement$ is a general PCFA, the intersection $A \cap \refinement$ may violate these constraints and thus fail to be a PCFA.

To resolve this issue, we apply the \emph{deterministic minimisation} operation $\minimise(-)$ to $A \cap \refinement$,
which restores the PCFA structure.
The following lemma formalises this transformation:

\begin{lemma}
  \label{thm:min(intersection)-is-pcfa}
  For any PCFA $A$ and general PCFA $\refinement$, \rmblock{$\minimise(A \cap \refinement)$} is a PCFA.
\end{lemma}

\begin{proof}
To show this, observe the following straightforward properties~\footnote{
  To save space, the detailed proofs are postponed to the appendix.
}:
(1) all PCFA (hence $A$) satisfy the property that:
(**) \emph{every accepting trace of the automaton is not a prefix of another accepting trace};
(2) any deterministically minimal automaton satisfies Property \hyperlink{cond-b}{(b)} and \hyperlink{cond-c}{(c)} (i.e., a single ending location without out-edge) iff it satisfies (**);
(3) as $\scriptL(A \cap \refinement)$ is a subset of $\scriptL(A)$, $A \cap \refinement$ also satisfies (**).
So that, by (2), we can conclude that $\minimise(A \cap \refinement)$ is a PCFA, as deterministic minimisation already guarantees determinism (i.e., Property \hyperlink{cond-a}{(a)}).
Notably, pure determinisation is not sufficient here, as it cannot guarantee \hyperlink{cond-b}{(b)} uniqueness of the ending location.
\end{proof}

By the rationale above, we can now apply~\Cref{theorem: cross-structural equality} to conclude the following theorem:

\begin{theorem}
  \label{theorem: refinement enabling}
  Consider a PCFA $A$ and a (general) PCFA $\refinement$ that over-approximates 
  the violation traces of $A$ against the given 
  pre- and post-conditions $\precond$ and $\postcond$.
  Let the deterministic automata minimisation operation be denoted by \rmblock{$\minimise(-)$}.
  The following equality holds: \rmblock{
  \[ 
    \prob{\not\vdash \bs{\precond} \ A \ \bs{\postcond}} 
    = 
    \prob{\not\vdash \bs{\precond} \ \minimise(A \cap \refinement) \ \bs{\postcond}} 
  \]}
\end{theorem}
\begin{proof}
  Let $A_1 = A$ and $A_2 = \minimise(A \cap \refinement)$.
  By~\Cref{thm:min(intersection)-is-pcfa}, $A_2$ is a PCFA.
  Since $\refinement$ over-approximates the violating traces of $A$,
  $A_1$ and $A_2$ have the same set of violating traces.
  The equality follows from~\Cref{theorem: cross-structural equality}.
\end{proof}

With 
the theorem established, 
refining structural abstraction fundamentally thus entails identifying an appropriate \emph{general PCFA},
termed the \emph{refinement automaton} $\refinement$.
The new structural upper bound $\structBound{\minimise(A \cap \refinement)}$ serves as a \emph{refined structural upper bound} for the violating probability $\prob{\not\vdash \bs{\precond} \ A \ \bs{\postcond}}$.

\paragraph{Separating Probability from Semantics}
Notably, \Cref{theorem: refinement enabling} facilitates a clear separation 
between probability and semantics.
Since the construction of $\refinement$ 
must exclusively include violating traces, and given the definition of the 
evaluation of probabilistic statements as $\interpret{\dprobbranch}_v = v$ 
--- identical to $\stskip$ and $*_i$ statements --- probabilistic statements 
do not influence whether a trace is violating.
This implies that non-random 
techniques are \emph{directly} applicable in constructing $\refinement$.
This principle will be fully leveraged and demonstrated in the subsequent section.

\section{Automating Refinement of Structural Abstraction}
\label{subsec:structural-refinement}
\label{sec: refinement}

This section introduces algorithms for automatically verifying the thresholds problem,
building on the introduced structural abstraction (\Cref{eq: struct bound is an upper bound}), and the principles of the refinement (\Cref{theorem: refinement enabling}).
In the following, we first propose a general CEGAR framework capable of integrating non-random techniques, 
as detailed in~\Cref{subsec: general cegar framework},
exploiting the separation of probability and semantics afforded by these foundations.
Within the general framework, we present concrete instantiations through trace abstraction in~\Cref{subsec: refine by trace abstraction,subsec: rc refine by trace abstraction}.
The former provides a direct instantiation,
while the latter introduces a further optimisation to achieve refutational completeness.
We argue that this demonstration offers a modular and well-structured description,  
where our framework handles the probabilistic aspects of verification,
while leaving the semantic aspects to be addressed by established non-random techniques.

\subsection{A General CEGAR Framework}
\label{subsec: general cegar framework}


This part presents a general CEGAR-based automatic verification framework for probabilistic programs,
compatible with non-probabilistic verification and analysis techniques.
We begin by introducing the concept of counterexamples,
followed by a formal presentation of the key framework.
This then leads to two subtle technical challenges:
first, verifying whether a counterexample has indeed been identified,
and second, resolving the incompatibility of path conditions between violating traces.
The two points arise to complement the overall framework and present unique challenges as compared to the non-random verification.

\begin{figure} [t]
  \resizebox{\textwidth}{!}{
    \begin{tikzpicture}[yscale=0.4]
    \node at (-3, 0) (NS) {};
    \node[draw, circle] at (-2, 0) (N0) {};
    \node[draw, circle] at (2, 0) (N1) {};
    \node[draw, circle] at (3, 1) (N2) {};
    \node[draw, circle] at (7, 1) (N4) {};
    \node[draw, circle] at (8.5, 1) (N6) {};
    \node[draw, circle] at (3, -1) (N3) {};
    \node[draw, circle] at (7, -1) (N5) {};
    \node[draw, circle] at (8.5, -1) (N7) {};
    \node[draw, circle] at (9.5, 0) (N8) {};
    \node[draw, accepting, circle] at (12, 0) (N9) {};

    \draw[-latex] (NS) -- (N0);
    \draw[-latex] (N0) -- (N1)  node[midway, above] {\scriptsize $X::=0; \rprobbranch[0];\stskip;\assume\ C>0$};
    \draw[-latex] (N1) --  node[pos=0.9, left=0.2] {\scriptsize $\lprobbranch[1]$} (N2);
    \draw[-latex] (N1) --  node[pos=0.9, left=0.2] {\scriptsize $\rprobbranch[1]$} (N3);
    \draw[-latex] (N2) -- node[above=2] {\scriptsize $X::=X+1;C::=C-1;\assume\ C>0$} (N4);
    \draw[-latex] (N3) -- node[below] {\scriptsize $\stskip;C::=C-1;\assume\ C>0$} (N5);
    \draw[-latex] (N4) -- node[above] {\scriptsize $\rprobbranch[1]$} (N6);
    \draw[-latex] (N5) -- node[below] {\scriptsize $\lprobbranch[1]$} (N7);
    \draw[-latex] (N4) -- node[left] {\scriptsize $\lprobbranch[1]$} (N7);
    \draw[-latex] (N6) -- node[pos=0.1, right=0.2] {\scriptsize $\stskip$} (N8);
    \draw[-latex] (N7) -- node[pos=0.1, right=0.2] {\scriptsize $X::=X+1$} (N8);
    \draw[-latex] (N8) -- node[above=2] {\scriptsize $C::=C-1;\assume\ C\le 0$} (N9);
\end{tikzpicture}
  }
  \captionof{figure}{Example of A Counterexample of~\cref{example: motivation} when $\beta = 0.3$}
  \label{fig: example of a counterexample}
  \vspace{-5pt}
\end{figure}

\paragraph{Counterexamples}
Following common CEGAR principles~\cite{cegar,pcegar,trace_abstraction},
our procedure either verifies the threshold or identifies a \emph{counterexample} (CE) disproving it.
Drawing inspiration from~\cite{pcegar,trace_enumeration},
a \emph{counterexample} in this framework is a finite set of \emph{compatible} violating traces,
certifying the existence of a run exceeding the violation probability threshold.
Formally, a set of violating traces $\Theta$ is compatible if:
(1) the path condition of $\Theta$ is satisfiable; and
(2) any two distinct traces $\tau_1$ and $\tau_2$ share a prefix (possibly $\varepsilon$) after which they diverge with probabilistic statements $\lprobbranch$ and $\rprobbranch$ sharing the same distribution tag $i$.
The latter ensures that the traces form a PCFA that is effectively an MC,
and hence the overall set $\Theta$ forms a subset of violating computations of a potential violating run.
Finally, a lemma affirms the validity of these counterexamples~\footnote{
  More details are in the appendix.
}.
\begin{lemma}
  \label{theorem: valid counterexample}
  Consider a PCFA $A$ with pre- and post-conditions $\precond$ and $\postcond$, and a threshold $\beta \in [0, 1]$.
  The violation probability $\prob{\not\vdash \bs{\precond} \ A \ \bs{\postcond}}$ exceeds $\beta$, 
  iff there is a violating run of $A$, 
  iff there is a finite set of compatible and violating traces of $A$.
\end{lemma}

\begin{example}[Counterexample]
  Referring to~\cref{example: motivation}, consider the threshold $\beta = 0.3$ (instead of $0.5$ in~\cref{sec: motivation example and overview}).
  A counterexample with three traces is shown in~\cref{fig: example of a counterexample}.
  Some trivial locations are abbreviated using semicolons ``;'' for brevity.
  All three traces are violating with a total probability of $0.375$ when the initial valuation is $C = 2$,
  indicating that the violating probability of~\cref{example: motivation} exceeds the threshold $\beta = 0.3$.
  Notably, this set is compatible as they have a path condition $C = 2$.
\end{example}

\begin{figure}[t]
  \includegraphics[width=\textwidth]{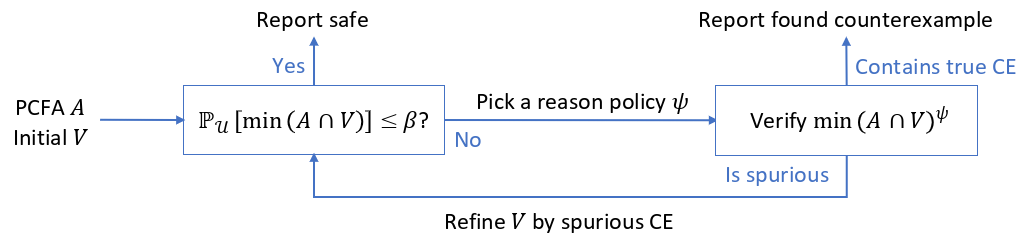}
  \vspace{-22pt}
  \caption{General CEGAR Framework}
  \label{fig: general cegar framework}
  \vspace{-10pt}
\end{figure}

\paragraph{Framework Overview}
With the concept of counterexample established, we are ready to present our framework,
as shown in~\cref{fig: general cegar framework}.
Initially, the process is provided with a PCFA $A$ and an initial refinement automaton $\refinement$,
which over-approximates the violating traces of $A$.
Then, the process computes the refined structural bound based on the current $\refinement$.
Should the bound be $\le \beta$, the process deems it safe, for which the soundness is indicated by~\cref{eq: struct bound is an upper bound}.
Otherwise, in accordance with well-established conclusions for MDP~\cite{principles_of_model_checking}, there must be a \emph{simple} policy $\psi$ such that
\(
  \sum_{
  \tau \in \scriptL(\applyPolicy{\minimise(A\cap\refinement)}{\psi})
  } \wt(\tau) = 
  \structBound{\minimise(A \cap \refinement)} > \beta
\).
We call such a policy $\psi$ a \emph{reason policy} (that induces the bound).
Referring to the previous discussion on counterexamples, the traces of $\applyPolicy{\minimise(A\cap\refinement)}{\psi}$ that are accepted already meet the second condition.
Hence, we call the automaton and the accepting traces of $\applyPolicy{\minimise(A\cap\refinement)}{\psi}$ a \emph{candidate counterexample}.
With such a reason policy, the process proceeds to ascertain whether there indeed exists a counterexample within $\applyPolicy{\minimise(A\cap\refinement)}{\psi}$ -- 
that a finite set of traces satisfying the first condition of counterexamples.
Should such a set exist, by~\cref{theorem: valid counterexample}, it is justifiable to report a violation with the discovered counterexample.
Otherwise, this intuitively suggests that the current $\refinement$ remains too coarse and necessitates refinement of $\refinement$ 
with the knowledge that the current candidate counterexample, i.e., $\applyPolicy{\minimise(A\cap\refinement)}{\psi}$, is \emph{spurious}.

\begin{example}[Reason Policy and Candidate Counterexample]
  \label[example]{example: reason policy and candidate counterexample}
  Casting our minds back to~\cref{example: motivation}, suppose the initial refinement automaton $\refinement$ is the trivial automaton that permits all traces.
  As a consequence, the initial PCFA $\minimise(P \cap \refinement)$ retains its original structure $P$ (\cref{figure: motivation example cfa}).

  Within this figure, the majority of the locations are elementary in that they permit at most one action,
  with the exception of the loop's starting location, which possesses two actions.
  In this scenario, the sole reason policy $\psi$ is determined by opting for the action $\angbr{\assume~\neg (C > 0)}$
  at this particular location and defaulting to the only action available for the remainder of the locations.

  By applying this reason policy $\psi$ to $P$, the resulting PCFA $\applyPolicy{P}{\psi}$ (the candidate counterexample) is exactly the structure portrayed in~\cref{fig: example of a run}.
  \qed
\end{example}

\paragraph{Verify Candidate Counterexample by Enumeration}
To verify the candidate counterexample, inspired by~\cite{pcegar,trace_enumeration}, we enumerate traces from $\scriptL(\applyPolicy{\minimise(A\cap\refinement)}{\psi})$ to find a finite subset of violating traces with satisfiable path conditions.
During the process, we maintain and enlarge a set $\Theta$ of violating traces in an on-the-fly manner.

The enumeration proceeds in two phases:
\begin{itemize}
\item \emph{Explicit enumeration:} We iteratively enumerate traces from the (potentially infinite) candidate counterexample $\applyPolicy{\minimise(A\cap\refinement)}{\psi}$ in descending order wrt.~the weights of the traces.
For each trace, we verify its violation status using SMT solvers~\cite{smt_interpol,de2008z3}; only violating traces are retained in the accumulating set $\Theta$. 
\item \emph{Symbolic subset search:} For the current finite set $\Theta$ of enumerated violating traces, we search for the maximum-weight \emph{compatible} subset. 
This constitutes a combinatorial optimisation problem: given $\Theta$'s weighted path conditions, find the subset with satisfiable conjunction and maximum total weight.
Formally, we are to solve the following formula (let $\mathbb{I}[P]$ denotes the Iverson bracket that if $P$ is true, it returns $1$, otherwise $0$):
\[\textstyle
  \mathop{\mathsf{MaxArg}}_{\Theta' \subseteq \Theta} \left(\mathbb{I}[\pathCond(\Theta') \text{ is satisfiable}] \cdot \wt(\Theta') \right)
\]
This problem can be solved using the MAX-SMT approach~\cite{maxsmt}.
\end{itemize}

The enumeration terminates when either:
(1) The maximum compatible subset weight exceeds threshold $\beta$, yielding a valid counterexample; or
(2) The sum of the current maximum and remaining unenumerated weights cannot exceed $\beta$, confirming spuriousness.

\begin{example}[Trace Enumeration]
  Continuing the story in~\cref{example: reason policy and candidate counterexample}, let us assume again $\beta = 0.3$.
  By employing the enumeration algorithm as delineated in~\cite{trace_enumeration} to the candidate counterexample depicted in~\cref{fig: example of a run}, both traces have an equal weight of $0.5$, and thus, either could be enumerated first.
  When both traces are enumerated,
  since both of them are safe, the process terminates upon enumerating the two traces, as the residual probability becomes $0 \leq \beta = 0.3$, confirming the spuriousness of the counterexample.\qed
\end{example}

Before introducing the customisable refinement procedure,
there is one more subtlety to address:
the instant divergence that may occur from incompatible path conditions of violating traces.
We address this by \emph{pre-condition splitting}.

\paragraph{Splitting the Pre-conditions}
Notice that the existence of violating yet incompatible traces potentially leads to instant divergence of the CEGAR process.
Consider the following example:

\begin{figure}
  \begin{minipage}{0.4\textwidth}
    \resizebox{!}{3cm}{
      \begin{tikzpicture}[yscale=0.4,xscale=0.9]
    \node at (2, 0) (N0) {};
    \node[draw, circle, green] at (3, 0) (N2) {};
    \node[draw, circle] at (4, 2) (N3) {};
    \node[draw, circle] at (4, -2) (N4) {};
    \node[draw, circle] at (6, 3) (N5) {};
    \node[draw, circle] at (6, -3) (N6) {};
    \node[draw, circle] at (6, 1) (N7) {};
    \node[draw, circle] at (6, -1) (N8) {};
    \node[draw, circle, accepting] at (8, 0) (NE) {};

    \draw[-latex] (N0) -- (N2);
    \draw[-latex, blue] (N2) -- (N3)  node[midway, left] {\scriptsize $\lprobbranch[0]$};
    \draw[-latex, blue] (N2) -- (N4)  node[midway, left] {\scriptsize $\rprobbranch[0]$};
    \draw[-latex, white!40!black] (N3) -- (N5)  node[pos=0.3, above=2] {\scriptsize $\assume~X > 0$};
    \draw[-latex, blue] (N3) -- (N7)  node[pos=0.3, below=3] {\scriptsize $\assume~\neg (X > 0)$};
    \draw[-latex, blue] (N4) -- (N8)  node[pos=0.3, above=2] {\scriptsize $\assume~X > 0$};
    \draw[-latex, white!30!black] (N4) -- (N6)  node[pos=0.3, below=3] {\scriptsize $\assume~\neg (X > 0)$};
    \draw[-latex, white!30!black] (N5) -- (NE)  node[midway, above=3] {\scriptsize $Y \opAssn 0$};
    \draw[-latex, white!30!black] (N6) -- (NE)  node[midway, below=3] {\scriptsize $Y \opAssn 0$};
    \draw[-latex, blue] (N7) -- (NE)  node[pos=0.3, above] {\scriptsize $Y \opAssn 1$};
    \draw[-latex, blue] (N8) -- (NE)  node[pos=0.3, above] {\scriptsize $Y \opAssn 1$};
\end{tikzpicture}
    }
    \captionof{figure}{PCFA of~\cref{example: splitting condition}}
    \label[figure]{fig: example split condition cfa}
  \end{minipage}
  \hfill
  \begin{minipage}{0.58\textwidth}
    \resizebox{!}{3cm}{
      \begin{tikzpicture}[yscale=0.4,xscale=0.9]
    \node at (0, 0) (N0) {};
    \node[draw, circle] at (1, 0) (N1) {};
    \node[draw, circle, green] at (3, 0) (N2) {};
    \node[draw, circle] at (4, 2) (N3) {};
    \node[draw, circle] at (4, -2) (N4) {};
    \node[draw, circle] at (6, 3) (N5) {};
    \node[draw, circle] at (6, -3) (N6) {};
    \node[draw, circle] at (6, 1) (N7) {};
    \node[draw, circle] at (6, -1) (N8) {};
    \node[draw, circle, accepting] at (8, 0) (NE) {};

    \draw[-latex] (N0) -- (N1);
    \draw[-latex] (N1) edge [out=45, in=135,looseness=1, red!70!blue] node[above] {\scriptsize $\assume~X > 0$} (N2);
    \draw[-latex] (N1) edge [out=-45, in=-135,looseness=1, orange] node[below] {\scriptsize $\assume~\neg (X > 0)$} (N2);
    \draw[-latex, red!70!blue] (N2) -- (N3)  node[pos=0.9, left] {\scriptsize $\lprobbranch[0]$};
    \draw[-latex, orange] (N2) -- (N4)  node[pos=1, left] {\scriptsize $\rprobbranch[0]$};
    \draw[-latex, white!40!black] (N3) -- (N5)  node[pos=0.3, above=2] {\scriptsize $\assume~X > 0$};
    \draw[-latex, red!70!blue] (N3) -- (N7)  node[pos=0.3, below=3] {\scriptsize $\assume~\neg (X > 0)$};
    \draw[-latex, orange] (N4) -- (N8)  node[pos=0.3, above=2] {\scriptsize $\assume~X > 0$};
    \draw[-latex, white!30!black] (N4) -- (N6)  node[pos=0.3, below=3] {\scriptsize $\assume~\neg (X > 0)$};
    \draw[-latex, white!30!black] (N5) -- (NE)  node[midway, above=3] {\scriptsize $Y \opAssn 0$};
    \draw[-latex, white!30!black] (N6) -- (NE)  node[midway, below=3] {\scriptsize $Y \opAssn 0$};
    \draw[-latex, red!70!blue] (N7) -- (NE)  node[pos=0.3, above] {\scriptsize $Y \opAssn 1$};
    \draw[-latex, orange] (N8) -- (NE)  node[pos=0.3, above] {\scriptsize $Y \opAssn 1$};
\end{tikzpicture}
    }
    \captionof{figure}{PCFA of~\Cref{example: splitting condition} with Split Conditions}
    \label[figure]{fig: example added split condition cfa}
  \end{minipage}
  \vspace{-5pt}
\end{figure}

\begin{figure}
  \begin{tikzpicture}[yscale=0.4,xscale=0.9]
    \node at (0, 0) (N0) {};
    \node[draw, circle] at (1, 0) (N1) {};
    \node[draw, circle, green] at (3, 2) (N2) {};
    \node[draw, circle, green] at (3, -2) (N2') {};
    \node[draw, circle] at (4, 2) (N3) {};
    \node[draw, circle] at (4, -2) (N4) {};
    \node[draw, circle] at (6, 1) (N7) {};
    \node[draw, circle] at (6, -1) (N8) {};
    \node[draw, circle, accepting] at (8, 0) (NE) {};

    \draw[-latex] (N0) -- (N1);
    \draw[-latex] (N1) edge [red!70!blue] node[pos=1,left=3] {\scriptsize $\assume~X > 0$} (N2');
    \draw[-latex] (N1) edge [orange] node[pos=1,left=3] {\scriptsize $\assume~\neg (X > 0)$} (N2);
    \draw[-latex, red!70!blue] (N2) -- (N3)  node[pos=0.5,above=2] {\scriptsize $\lprobbranch[0]$};
    \draw[-latex, orange] (N2') -- (N4)  node[pos=0.5,below=2] {\scriptsize $\rprobbranch[0]$};
    \draw[-latex, red!70!blue] (N3) -- (N7)  node[pos=0.3, below=3] {\scriptsize $\assume~\neg (X > 0)$};
    \draw[-latex, orange] (N4) -- (N8)  node[pos=0.3, above=2] {\scriptsize $\assume~X > 0$};
    \draw[-latex, red!70!blue] (N7) -- (NE)  node[pos=0.3, above] {\scriptsize $Y \opAssn 1$};
    \draw[-latex, orange] (N8) -- (NE)  node[pos=0.3, above] {\scriptsize $Y \opAssn 1$};
\end{tikzpicture}
  \caption{Refined PCFA of~\Cref{example: splitting condition} with Split Conditions}
  \label{fig:example-split-condition-cfa-refined}
  \vspace{-10pt}
\end{figure}

\begin{example}[Divergence from Incompatibility]
  \label{example: splitting condition}
  Consider the following Hoare style program whose PCFA is depicted in~\cref{fig: example split condition cfa}, let the threshold $\beta$ be $0.75$.
  \[
    \bs{\tTrue} \
    (
      \tIf \ X > 0 \ \tThen \ Y \opAssn 0 \ \tElse \ Y \opAssn 1
    )
    \oplus
    (
      \tIf \ X > 0 \ \tThen \ Y \opAssn 1 \ \tElse \ Y \opAssn 0
    ) \
    \bs{Y = 0}
  \]

In this example, consider the candidate counterexample comprising the two blue traces with total weights $1$.
After enumeration, both traces are found violating but incompatible.
However, standard refinement cannot distinguish them:
since $\refinement$ must contain \emph{all} violating traces,
no refinement automaton can separate these two traces.
Hence, the CEGAR process diverges by repeatedly picking the blue candidate counterexample as it has structural upper bound $1 > 0.75 = \beta$.
This issue also occurs when distinguishing traces with different values of $C$ in~\cref{example: motivation}.
\qed
\end{example}

To resolve the kind of divergence as illustrated in the previous example, we propose ``pre-condition splitting'' to ensure traces with incompatible path conditions cannot appear in the same candidate counterexample again.
Specifically, consider the previous example, we append the conflicting path conditions $\assume~X > 0$ and $\assume~\neg(X > 0)$ before the traces in~\Cref{fig: example split condition cfa}, yielding~\Cref{fig: example added split condition cfa}.
Consequently, both the red and orange traces now contain contradictory assumption conditions $\assume~X > 0$ and $\assume~\neg(X > 0)$, rendering them infeasible rather than violating.
We call the above prepended conditions \emph{split conditions}.
These infeasible traces are subsequently eliminated during refinement, eventually producing the refined PCFA as shown in~\Cref{fig:example-split-condition-cfa-refined}.
The resulting PCFA contains only two traces under different reason policies, yielding an exact structural upper bound of $0.5 \leq 0.75 = \beta$ and confirming safety. 
We generalise the above idea as follows.

\vspace{3pt}
\noindent{\it PCFA with Split Conditions.}\quad
Formally, a set of \emph{split conditions} $\{\precond^1, \ldots, \precond^n\}$ partitions the pre-condition $\precond$ such that $\bigvee_{i=1}^n \precond^i = \precond$ and $\precond^i \wedge \precond^j$ is unsatisfiable for $i \neq j$.
A PCFA with split conditions enforces that edges from the initial location $\ell_0$ are labeled by $\assume~\precond^i$ for all split conditions $\precond^i$ and do not return to $\ell_0$.
This structure is used for the refinement automaton $\refinement$ in our framework, with the set of split conditions maintained on-the-fly during CEGAR.
Initially, the set is the singleton $\{\precond\}$.
Specifically,
when instantiating the framework with an initial refinement automaton $\refinement$, we construct the actual automaton $\refinement'$ by introducing a fresh initial location $\ell_0'$ with edge $\assume~\precond$ leading to the original initial location.

This resembles introducing a fresh initial location with split condition edges as shown in~\Cref{fig: example added split condition cfa} versus~\Cref{fig: example split condition cfa}.
By further refinement,
the original location may be partitioned as in~\Cref{fig:example-split-condition-cfa-refined}, where green locations correspond to the original initial location in~\Cref{fig: example split condition cfa,fig: example added split condition cfa}.

\vspace{3pt}
\noindent{\it Modified Intersection Operation.}\quad
Incorporating split conditions into the refinement automaton $\refinement$ necessitates modifications to the intersection operation.
Since $\refinement$ now contains split conditions while the original PCFA $A$ does not, we adapt the intersection $A \cap \refinement$ as follows:
First, we synchronise $A$ with the split conditions by constructing $A' := (L \uplus \{\ell_0'\}, \Sigma, \Delta', \ell_0', \ell_e)$ where $\ell_0'$ is a fresh initial location and $\Delta'$ is additionally enlarged with the transitions $\ell_0' \xrightarrow{\assume~\precond^i} \ell_0$, where $\ell_0$ is the original initial location of $A$, for each split condition $\precond^i$ currently in $\refinement$.
The intersection is then computed as $A' \cap \refinement$.
This renewed method is employed in computing the refined upper bound as in the left blue box of~\Cref{fig: general cegar framework}.

\vspace{3pt}
\noindent{\it Deriving Split Conditions.}\quad
Split conditions are derived on-the-fly during the CEGAR process, tightly integrated with the MAX-SMT and enumeration procedures.
Notably, as shown above, now with the split conditions, every candidate counterexample must have a unique split condition --- 
by the definition of policies (in~\Cref{subsec:pcfa-as-mdp}), and every $\assume$ statement must be an independent action.
When a candidate counterexample is deemed spurious, let its current split condition be $\precond^i$ and the enumerated set of violating traces be $\Theta$,
we extract the path condition $E$ of the maximum-weight compatible trace set of $\Theta$ from the MAX-SMT analysis.
We then replace the original split condition $\precond^i$ with two refined conditions: $\precond^i \wedge E$ and $\precond^i \wedge \neg E$.

The above ideas are integrated into the refinement procedure below.

\vspace{3pt}
\noindent{\bf Refinement Procedure.}\quad
With split condition derivation established, the refinement procedure in~\Cref{fig: general cegar framework} proceeds through three phases. Below let $E$ denote the identified path condition from the MAX-SMT analysis as mentioned previously. 
\begin{enumerate}
  \item \emph{Split condition update:} Replace the transition $\smash{\ell_0 \xrightarrow{\assume~\precond^i} \ell}$ in both the current $\refinement$ and the spurious candidate counterexample with two transitions: $\ell_0 \xrightarrow{\assume~(\precond^i \wedge E)} \ell$ and $\ell_0 \xrightarrow{\assume~(\precond^i \wedge \neg E)} \ell$, where $\ell_0$ is the initial location and $\ell$ is the destination location (intuitively representing the potentially partitioned original initial location).
    
  \item \emph{Trace relabeling:} For refinement techniques requiring enumerated traces $\Theta$, we update trace labels accordingly: 
  traces in the maximum-weight set have their first statement $\assume~\precond^i$ replaced by $\assume~(\precond^i \wedge E)$,
  while the remaining traces are updated with $\assume~(\precond^i \wedge \neg E)$.
  
  \item \emph{Refinement:} The updated objects --- including $\refinement$, the candidate counterexample, and $\Theta$ --- 
  are processed by the customisable refinement method to be introduced below to compute a new general PCFA $\refinement$ with split conditions.

  Crucially, split conditions do not alter behaviors of refinement techniques adopted.
  Specifically,
  the refinement procedure can be intuitively described as:
  given a control-flow automaton ($\refinement$ in our case) and its subset of identified non-violating traces
  (from the candidate counterexample),
  refine the automaton to eliminate as much as possible more non-violating traces from $V$.
  This intuition is generally shared among standard non-probabilistic refinement techniques,
  e.g., predicate abstraction~\cite{cegar} and
  trace abstraction~\cite{trace_abstraction} to be exemplified below in~\Cref{subsec: refine by trace abstraction}.
  By substituting new split conditions ($\assume~(\precond^i \wedge E)$ and $\assume~(\precond^i \wedge \neg E)$) for $\assume~\precond^i$,
  both the modified candidate counterexample and $\Theta$ remain subsets of $\scriptL(\refinement)$, allowing 
  refinement methods to proceed unaware of split conditions.
\end{enumerate}

Split conditions prevent re-enumeration of identical candidate counterexamples, as traces with path conditions contradicting the split condition become infeasible rather than violating,
hence effectively reducing divergence.
For finite-state programs like~\Cref{example: splitting condition}, since the number of potential split conditions is finite, this approach guarantees convergence of the CEGAR process.

\begin{remark}[Methodology of Adapting the Framework]
  Within the framework, most components are predetermined, leaving only \emph{two} customisable elements:  
  the construction of the initial $\refinement$ and the refinement method.
  As demonstrated in the principle of refinement in~\Cref{theorem: refinement enabling},
  the construction and refinement of $\refinement$ does not require handling probabilities as the probabilistic statements will not affect whether a trace is violating or not.
  This renders typically straightforward instantiations.
\end{remark}

To show our claim of the straightforwardness of the adaptation, in the following,
we exemplify the instantiation of our framework with trace abstraction.
We begin by a direct instantiation with the non-random technique briefly introduced and straightforwardly applied to adapt to the above methodology.
Further, we explore a potential optimisation based on the direct instantiation, which further retains the \emph{refutational completeness} property of trace abstraction,
which is a guarantee hardly seen in probabilistic verification beyond finite-state (incl.~bounded) techniques.
In fact, we also investigated more instances of the instantiations, e.g., predicate abstraction and value analysis.
As they are all direct applications of the well-established techniques, in the interests of space, we postponed the demonstration to the appendix.

\subsection{Direct Instantiation with Trace Abstraction}
\label{subsec: refine by trace abstraction}


Next, we shall adapt the framework by \emph{trace abstraction}~\cite{trace_abstraction}, a successful non-probabilistic verification technique.
In this part, again, we first present a brief account of the approach of trace abstraction
and then shift to a discussion on the direct instantiation of the general CEGAR framework with the renowned technique.

\paragraph{Trace Abstraction}
In the non-probabilistic context, trace abstraction over-approximates the non-violating traces of the original control-flow automaton (CFA)~\footnote{
  Simply given by (general) PCFA without probabilistic statements.
} $A$ by a series of certified CFA $Q_1, \dots, Q_n$ with $\scriptL(A) \subseteq \bigcup_{i = 1}^n \scriptL(Q_i)$, where a certified CFA contains only the non-violating traces.

The CEGAR-driven refinement algorithm for trace abstraction entails constructing a series of \emph{Floyd-Hoare automata}, which serve as the certified CFA that contains no violating trace.
A Floyd-Hoare automaton is a (general) CFA $(L, \Sigma, \Delta, \ell_0, L_e)$, augmented with a predicate assignment function $\lambda : L \to \Phi$.
For each transition $\ell \xrightarrow{\sigma} \ell'$, the condition represented by a Hoare triple~\cite{software_foundations_1} $\bs{\lambda(\ell)} \ \sigma \ \bs{\lambda(\ell')}$ must hold.
Although previously mentioned, we here formally present an account for this concept.
The validity of a Hoare triple $\bs{\varphi_1} \ \sigma \ \bs{\varphi_2}$ is given by that for all valuations $v \models \varphi_1$, the valuation $\interpret{\sigma}_v$ either is $\Undefined$ or satisfies $\varphi_2$.

In the $(n + 1)$-th iteration of trace abstraction refinement, a trace (candidate counterexample in the non-probabilistic scenario) is selected from the rest of the automaton $A \setminus \bigcup_{i = 1}^n \scriptL(Q_i)$, which is followed by an attempt to verify it.
Should the trace indeed be violating, the counterexample is reported;
otherwise, the trace is generalised into a Floyd-Hoare automaton $Q_{n + 1}$.
The iteration goes until the $m$-th round when $A \setminus \bigcup_{i = 1}^m \scriptL(Q_i) = \emptyset$.

In the above process, the generalisation process encompasses two steps: \emph{interpolation}~\cite{smt_interpol,craig_interpolation} and the construction of a Floyd-Hoare automaton.
More specifically, given a non-violating trace $\sigma_1 \dots \sigma_n$, the interpolation of the trace produces a \emph{tagged trace}, of form: $\bs{\varphi_0} \ \sigma_1 \ \bs{\varphi_1} \ \dots \ \bs{\varphi_{n-1}} \ \sigma_n \ \bs{\varphi_{n}}$, where the propositions $\varphi_i$ are called \emph{interpolants} and the Hoare triple $\bs{\varphi_{i - 1}} \ \sigma_i \ \bs{\varphi_{i}}$ holds for every segment.
At the same time, $\precond$ implies $\varphi_0$ and $\varphi_{n}$ implies $\postcond$.
Utilising the tagged trace, a Floyd-Hoare automaton is constructed by establishing a location set corresponding to each generated predicates $\bs{\varphi_0, \dots, \varphi_{n}}$~\footnote{
  Duplicate predicates are naturally removed.
}.
In what follows, the transitions $\ell \xrightarrow{\sigma} \ell'$ for $\sigma$ in the statement set of the program CFA $A$ are added, whenever the Hoare triple $\bs{\lambda(\ell)} \ \sigma \ \bs{\lambda(\ell')}$ holds.

\begin{example}[Generalisation]
  Notably, this generalisation is directly applicable to our context.
  Referring back to~\cref{eq:trace-example-1,eq:trace-example-2}, the interpolation process by the established tool SMTInterpol~\cite{smt_interpol} returns the following tagged traces for~\Cref{eq:trace-example-1,eq:trace-example-2}:
  \begin{align}
    \bs{\tTrue} \
    X \opAssn 0 \
    \bs{X = 0} \
    \lprobbranch[0] \
    \bs{X = 0} \
    C \opAssn 0 \
    \bs{X = 0} \
    \assume~\neg (C > 0) \
    \bs{X = 0}
  \end{align}
  \begin{align}
    \begin{matrix}
      \bs{\tTrue} \
      X \opAssn 0 \
      \bs{\tTrue} \
      \lprobbranch[0] \
      \bs{\tTrue} \
      C \opAssn 0 \
      \bs{C \le 0} \
      \assume~C > 0 \
      \bs{\tFalse} \\
      \rprobbranch[1] \
      \bs{\tFalse} \
      \stskip \
      \bs{\tFalse} \
      C \opAssn C - 1 \
      \bs{\tFalse} \
      \assume~\neg (C > 0) \
      \bs{\tFalse}
    \end{matrix}
  \end{align}
  
  Then, by introducing the location set for the interpolants for each tagged trace and subsequently adding the transitions,
  the generalised certified general PCFA~\cref{fig: generalised automaton for trace 1,fig: generalised automaton for trace 2} are obtained.\qed
\end{example}

\paragraph{Probabilistic Refinement with Trace Abstraction}
The theoretical aim of the instantiation involves to construct a $\refinement = \overline{\bigcup_{i = 1}^n Q_i}$ where each $Q_i$ is a certified (general) PCFA.\@
Specifically:
\begin{itemize}
  \item the initial $\refinement$ is the trivial PCFA that accepts all traces $\Sigma^*$; and,
  \item when a spurious counterexample is found, generalise the enumerated non-violating traces and union all the Floyd-Hoare automata obtained by generalisation from each trace into a PCFA $Q$.
  Next, update $V$ in an incremental manner (so there is no need to keep every generated $Q$) by: $V \leftarrow \overline{\overline{V} \cup Q}$.
\end{itemize}

\subsection{Optimised Instantiation with Trace Abstraction for Refutational Completeness}
\label{subsec: rc refine by trace abstraction}


When we probe closer to the flexibility on examining each traces provided by trace abstraction,
we discern that there is room for further optimisation.
In this section, we shall dig deeper into the integration with trace abstraction to explore potential optimisations and their impact.

\begin{figure}
  \begin{minipage}{0.7\textwidth}
    \includegraphics[width=\textwidth]{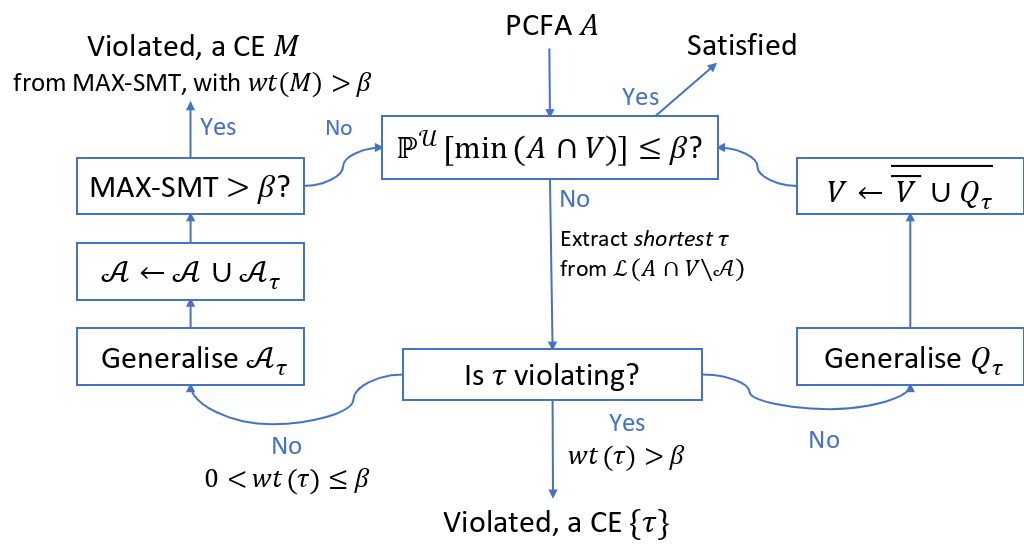}
    \vspace{-15pt}
    \captionof{figure}{Optimised CEGAR with Trace Abstraction}
    \label{fig: optimised trace abs cegar}
  \end{minipage}
  \vspace{-5pt}
\end{figure}

In general, the verification problem is undecidable, which, in light of our CEGAR-driven algorithm, implies that the refinement loop might never terminate.
Nevertheless, we can still modify the algorithm to ensure termination and a valid counterexample when the problem is genuinely unsatisfiable; this is referred to as \emph{refutational completeness}.
It is worth noting that, this property suggests a theoretical guarantee on automatically proving an \emph{arbitrarily close} lower bound.

To achieve this effect, it is worth noting the following trivial facts:
\begin{enumerate}
  \item The main objects we are dealing with are now single traces (rather than states), allowing us to consider each trace separately;
  \item \label[fact]{item: finite trace counterexample} The counterexample consists of only finite traces as declared in~\cref{theorem: valid counterexample};
  \item \label[fact]{item: trace length finiteness} There are only a finite number of traces of a certain length in any PCFA $A$; and, 
  \item \label[fact]{item: finite decidability} It is possible to decide the existence of a counterexample in a finite trace set.
\end{enumerate}

Based on the facts, to achieve refutational completeness, two principles can be derived:
(1) storing the enumerated violating traces in a specific storage, from and only from which counterexamples are sought; and,
(2) preventing the re-enumeration of traces within the storage.
As there are only finite traces, the structure of the storage can be either a set of separated traces or a general PCFA, which does not hinder the operations below.

\paragraph{Optimised CEGAR Framework}
Incarnating the principles, the direct instantiation can be optimised to form the algorithmic procedure depicted in~\cref{fig: optimised trace abs cegar}.
Compared to the direct instantiation, this new algorithm possesses some subtle differences.

The algorithm consists of two loops --
the left loop modifies and examines the mentioned storage $\scriptA$, while the right loop eliminates discovered non-violating traces.
Initially, rather than extracting a reason policy directly, the process picks a \emph{shortest} trace $\tau$ from $A \cap \refinement \setminus \scriptA$, thereby avoiding re-enumerating traces in the storage.
Next, verify this single trace $\tau$, from which three kinds of results may be yielded.
If the trace is found to be non-violating, the standard generalisation process is employed to construct a certified general PCFA $Q_\tau$. After this, the aforementioned updating procedure is applied to $\refinement$.
When $\tau$ is violating, a simple scenario occurs if its weight surpasses the threshold independently, for which we report a counterexample with this single trace.

A more intriguing scenario arises otherwise.
As the examination of $\scriptA$ is rather computationally intensive, a specialised generalisation procedure is also executed to add traces with ``similar reasons for violation'' to the discovered $\tau$.
This process, unlike the aforementioned generalisation for non-violating traces, faces two main hurdles.
First, the interpolation process cannot be applied, which we'll explain in more detail below.
Second, the process must avoid introducing loops to make use of~\cref{item: finite decidability} mentioned earlier.
In response to these challenges, a novel finite generalisation process for violating traces will be introduced below.
For now, one just needs to know that the finite generalisation results in a (general) PCFA, denoted by $\scriptA_\tau$,
such that $\tau \in \scriptL(\scriptA_\tau)$ and there are only finite traces in $\scriptL(\scriptA_\tau)$.

With the finitely generalised automaton $\scriptA_\tau$ for the picked trace $\tau$, the traces are then integrated into the storage $\scriptA$.
Subsequently, the storage $\scriptA$ retains only finite traces.
Recall (in~\cref{subsec: general cegar framework}) that the existence of a potential counterexample for finite traces can be decided by encoding the problem in MAX-SMT~\cite{maxsmt,pcegar}.
Also, one may notice that the direct instantiation of trace abstraction~\cref{subsec: refine by trace abstraction} is also complete for finite traces.
Hence, adopting the analysis process may also be a possible choice for examining $\scriptA$.
As a result from the examinations, if a counterexample is identified, i.e., MAX-SMT produces a result $> \beta$, the subset identified by MAX-SMT is reported;
otherwise, the process recommences from the beginning for the next iteration.

\paragraph{Finite Generalisation}
We now turn our attention to the finite generalisation process for violating traces.
The fundamental steps for this process resemble those for non-violating traces:
one first computes a set of \emph{tagged traces},
from which an automaton is subsequently built.
However, the construction methods for these formalisms significantly \emph{differ} from those used for generalising non-violating traces.
Let us delve deeper into the details.

\paragraph{Tagging Traces by Weakest Pre-condition}
The interpolation process used for non-violating traces cannot be applied when tagging violating traces. 
To understand why, consider the details of interpolation~\cite{henzinger2004abstractions,craig_interpolation}:
for an \emph{unsatisfiable} proposition $\varphi \wedge \varphi'$, comprising sub-propositions $\varphi$ and $\varphi'$, a proposition $\psi$ exists such that:
(1) $\varphi$ implies $\psi$; 
(2) $\psi \wedge \varphi'$ is \emph{unsatisfiable}; and 
(3) $\psi$ contains only variables present in both $\varphi$ and $\varphi'$.
Given the \emph{unsatisfiability} requirement, it is evident why this cannot be applied to violating traces, where reaching $\postcond$ is, by definition, \emph{satisfiable}.

To address this challenge, we use the \emph{weakest pre-condition} technique, introduced by Dijkstra~\cite{dijkstra1975guarded}.
The tagging procedure involves computing backwardly from $\postcond$. 
The weakest pre-condition function $\twp(\sigma, \varphi)$ yields the weakest proposition that ensures $\varphi$ after executing statement $\sigma$, defined as:
\[
  \twp(X \opAssn E, \varphi) := \varphi[X / E]
  \quad
  \twp(\assume~\varphi', \varphi) := \varphi' \to \varphi
  \quad
  \twp(\sigma, \varphi) := \varphi
\]
Here, $\varphi[X / E]$ substitutes every occurrence of $X$ in $\varphi$ with $E$, and $\to$ denotes logical implication.

Given a violating trace $\sigma_1 \dots \sigma_n$ with pre- and post-conditions $\precond$ and $\postcond$, one can compute a tagged trace that:
\(
  \bs{\varphi_0} \ \sigma_1 \ \bs{\varphi_1} \ \dots \ \bs{\varphi_{n-1}} \ \sigma_n \ \bs{\varphi_{n}}
\), where:
$\varphi_i := \twp(\sigma_{i+1}, \varphi_{i+1})$ for $i \in \bs{1, \dots, n-1}$, with
$\varphi_n := \postcond$ and
$\varphi_0 := \precond \wedge \twp(\sigma_1, \varphi_1)$.

\paragraph{Constructing Finitely Generalised Automaton}
With tagged traces introduced, we can now construct the finitely generalised automaton. 
However, applying the original method directly remains infeasible due to the risk of introducing loops. 
To address this, we extend the usual Floyd-Hoare automaton by introducing the concept of \emph{order}, 
resulting in the \emph{ordered Floyd-Hoare automaton}.
Consider a standard Floyd-Hoare automaton, $(A, \lambda)$, consists of a PCFA $A$ with location set $L$.
An \emph{ordered Floyd-Hoare automaton} extends this to a triple $(A, \lambda, \rho)$, 
where $\rho : L \rightarrow \doubleN$ assigns a unique priority to each location. 
For a transition $\ell \xrightarrow{\sigma} \ell'$ to belong to the automaton, 
it must satisfy both the usual condition $\bs{\lambda(\ell)} \ \sigma \ \bs{\lambda(\ell')}$ 
and the additional requirement $\rho(\ell) < \rho(\ell')$.

Given a tagged trace 
\(
  \bs{\varphi_0} \ \sigma_1 \ \bs{\varphi_1} \ \dots \ \bs{\varphi_{n-1}} \ \sigma_n \ \bs{\varphi_n},
\)
we construct an ordered Floyd-Hoare automaton as follows. 
First, define a location set $\bs{\ell_0, \dots, \ell_n}$, whose size matches the length of the trace, 
unlike the non-violating case where locations correspond to distinct propositions. 
Next, assign $\lambda(\ell_i) := \varphi_i$ and $\rho(\ell_i) := i$ for each location $\ell_i$. 
Set $\ell_0$ as the initial location and $\ell_n$ as the final location. 
Finally, include all transitions satisfying the conditions for an ordered Floyd-Hoare automaton.

\paragraph{Refutational Completeness}
We here first present a formal account for the property.

\begin{theorem}[Refutational Completeness]
  Given a PCFA $A$, a pre- and a post-condition $\precond$ and $\postcond$ with a threshold $\beta$,
  if $\prob{\not\vdash \bs{\varphi_e} \ A \ \bs{\varphi_f}} > \beta$,
  then the process introduced above will terminate and report a valid counterexample.
\end{theorem}

Then, given the above process, the refutational completeness can be seen as follows.
When the violating probability indeed surpasses the given threshold, one may have a counterexample with finite number of traces, as discussed in~\cref{item: finite trace counterexample}.
For any potential counterexample, let the length of the longest trace in the example be $N$.
As discerned in~\cref{item: trace length finiteness}, and since the traces are not re-enumerated, there will inevitably be an iteration where the shortest possible trace in $\scriptL(A \cap \refinement \setminus \scriptA)$ exceeds $N$.
Consequently, all the violating traces that constitute the valid example will be stored in $\scriptA$.
Lastly, as discussed in~\cref{item: finite decidability}, this counterexample must have been detected through the examination of the storage $\scriptA$.

\section{Empirical Evaluation}
\label{sec: experiments}

To evaluate the effectiveness and efficiency of our proposed approach, we implemented our framework instantiated with trace abstraction.
We tested it on a diverse set of examples and compared its performance with state-of-the-art tools.

\paragraph{Implementation}
Our prototype, developed in Java and Scala, integrates high-performance libraries to tackle key challenges.
For satisfiability modulo theories (SMT), we employed \textsc{SMTInterpol}~\cite{smt_interpol},
selected for its robust interpolation capabilities vital to trace abstraction.
Automata manipulation is handled by the flexible Brics~\cite{brics_automaton} library,
while Breeze~\cite{hall2009scalanlp} facilitates computationally intensive numerical tasks,
such as MDP and MC reachability analysis.
To handle the numerically intensive examples, we also adopted an optimisation utilising value analysis as an auxiliary method to supplement the interpolator to refine the refinement automaton $\refinement$ by explicitly enumerating program states.
Notably, the analysis still fits perfectly into our general framework as the enumeration follows well-established algorithms for non-random programs,
without considering probabilities,
with the structural upper bound computed by the principle from~\Cref{theorem: refinement enabling},
as detailed in the appendix.
An input file is a program like that in~\Cref{example: motivation},
which is internally parsed and converted to PCFA.

\paragraph{Baseline Tools}
We compared our prototype against two state-of-the-art tools for violation bound analysis: cegispro2~\cite{batz2023probabilistic} and PSI~\cite{gehr2016psi}.
Cegispro2, a recent advancement based on fixed-point and pre-expectation techniques, has shown superior performance over tools such as Exist~\cite{bao2022data} and storm~\cite{hensel2022probabilistic} in prior studies.
PSI, a well-established tool, is known for its precise bounds, fast computation, and robust handling of both discrete and continuous random variables.
We assessed the time performance of these tools using identical validation objectives and benchmark examples.
Notably, tools and benchmarks from~\cite{pcegar,exponential_analysis_of_probabilistic_program},
though technically relevant,
were not included in the evaluation due to fundamental compatibility issues.
Specifically, \cite{pcegar} takes a PRISM-variant input format, hard to convert from and to probabilistic programs.
While \cite{exponential_analysis_of_probabilistic_program} requires manually provided invariants, hence unsuitable to compare with fully automated tools like ours, CegisPro2, and PSI.\@

\paragraph{Benchmarks}
We identify four challenging features when selecting benchmarks:
\begin{itemize}
  \item[\bf(P1)] Unbounded loops (i.e., loops with an unbounded number of iterations),
  \item[\bf(P2)] Bounded loops with over 100 times of iterations,
  \item[\bf(P3)] Loops with over 5 conditional and probabilistic branches in the body, and
  \item[\bf(P4)] Sequential or nested loops.
\end{itemize}

Following these criteria, we curated a diverse and representative set of examples from the literature, including:

\begin{itemize}
  \item We tested \emph{all loopy} examples from~\cite{10.1145/2491956.2462179}\footnote{
    The benchmarks from ~\cite{10.1145/2491956.2462179} can be obtained via the link \url{https://github.com/eth-sri/psi/tree/master/test/colorado}.
  }, a comprehensive benchmark set containing both simple yet intricate cases and complex examples with over a hundred lines and intricate branching structures.
  This benchmark suite remains widely used in recent tools~\cite{wang2024static,zaiser2025guaranteed}.

  \item We also tested \emph{all} assertion benchmarks from CegisPro2~\cite{batz2023probabilistic}, which
  demonstrated the capability of their methods by collecting examples
  from a wide range of literature and comparing them against multiple state-of-the-art tools.
  The testing results are summarized in~\Cref{tab:cp2_reality_amber_sar}.
  However, we found that their test suite has limited diversity,
  as most examples consists of a single loop with a simple loop body and a highly restricted number of iterations.

  \item To further enrich our test set, we incorporated additional loopy examples from a broader range of literature,
  including PSI~\cite{gehr2016psi}, Exist~\cite{bao2022data}, Amber~\cite{moosbrugger2021probabilistic},
  \cite{trace_abstraction_modulo_probability,carbin2013verifying}, \cite{exponential_analysis_of_probabilistic_program}, and Diabolo~\cite{zaiser2025guaranteed}.
  Some examples are derived from real-world applications.
  For instances, the ``Reliability'' example from~\cite{trace_abstraction_modulo_probability,carbin2013verifying}
  describes a pixel-block search algorithm from x264 video encoders with potential hardware failures,
  the property for which states whether the failing probability is bounded by the threshold;
  the ``Herman3'' example from~\cite{zaiser2025guaranteed}
  is an instance of the famous probabilistic self-stabilisation random algorithm~\cite{herman1990probabilistic} with three threads,
  for which we try to bound the tail probability of the random algorithm fails to stop within a bounded time;
  while the ``Coupon5'' example from~\cite{zaiser2025guaranteed} describes a well-known probabilistic problem extracted from the real-world situation~\cite{erdHos1961classical},
  widely applied in fileds like commercial analysis~\cite{yim2016design} and biology~\cite{chao1984nonparametric},
  for which we still inquire the tail probability,
  which is of interests in the real-world application;
  etc.

  \item Lastly, we noticed a lack of benchmarks with nested loops. To address this, we developed two additional examples ``LimitV'' and ``LimitVP'', by modifying the ``Limit'' example from~\Cref{example: motivation}. These new examples feature nested loops and an unbounded state space, further enriching the test set.
\end{itemize}

\begin{table}[t]
\renewcommand{\arraystretch}{1.2} 
\setlength{\tabcolsep}{4pt} 
\centering
\caption{Examples from~\cite{10.1145/2491956.2462179}:
An example is tested with multiple parameters, where ``UB'' denotes unbounded;  
if ``$\ddag$'' is used, it indicates the application of value-analysis optimization.
For each parameter, multiple different ``Bound'' values are tested as the threshold $\beta$,  
forming a \emph{verification task}.  
The ``Type'' column indicates which criteria the example with a specific parameter satisfies.  
In the ``Time'' columns for each method, the number represents seconds,  
``TO'' denotes a time-out (over 500 seconds),  
and ``N/A'' denotes not applicable, indicating the example cannot be handled by this implementation;
the bold font in a column denotes the fastest time compared with the other two.
In the ``Result'' column of our method, ``SAT'' indicates the safety of the given bound,  
while ``UnSAT($n$)'' indicates a violation with a found counterexample of $n$ traces.  
The ``Result'' of CegisPro2 is simpler, as it can either prove or report an error:  
when the example shows that the bound is a proper upper bound, we denote this by \ding{51},  
and for lower bounds, we invoke the ``lower bound'' functionality of the tool,  
denoting success in this mode by ``\ding{51} (Lower)''.
The ``Prob'' column in PSI, in turn, shows the exact violation probability value computed by PSI.\@}
    \label{tab:colo-exp}
    \vspace{-7pt}
\begin{adjustbox}{max width=\textwidth}
\begin{tabular}{|c|c|c|c|cc|cc|cc|}
\hline
\multirow{2}{*}{\textbf{Example Name}}     & \multirow{2}{*}{\textbf{Parameter}}    & \multirow{2}{*}{\textbf{Type}} & \multirow{2}{*}{\textbf{Bound}} & \multicolumn{2}{c|}{\textbf{Our Method}}                 & \multicolumn{2}{c|}{\textbf{CegisPro2}}                  & \multicolumn{2}{c|}{\textbf{PSI}}                       \\ \cline{5-10} 
                                           &                                        &                                &                                 & \multicolumn{1}{c|}{\textbf{Time (s)}} & \textbf{Result} & \multicolumn{1}{c|}{\textbf{Time (s)}} & \textbf{Result} & \multicolumn{1}{c|}{\textbf{Time (s)}} & \textbf{Prob} \\ \hline
\textit{BeauquierEtal3}                    & count\textless{}1                      & P1, P3                         & 0.6                             & \multicolumn{1}{c|}{\textbf{0.281}}    & SAT             & \multicolumn{1}{c|}{N/A}               & -               & \multicolumn{1}{c|}{N/A}               & -              \\ \hline
\multirow{2}{*}{\textit{Cav1}}             & x\textgreater{}250$^\ddag$                     & P1                             & 0.6                             & \multicolumn{1}{c|}{\textbf{84.971}}   & SAT             & \multicolumn{1}{c|}{TO}                & -               & \multicolumn{1}{c|}{N/A}               & -              \\ \cline{2-10} 
                                           & x\textgreater{}187.5$^\ddag$                   & P1                             & 0.1                             & \multicolumn{1}{c|}{\textbf{51.799}}   & SAT             & \multicolumn{1}{c|}{TO}                & -               & \multicolumn{1}{c|}{N/A}               & -              \\ \hline
\multirow{2}{*}{\textit{Cav2}}             & \multirow{2}{*}{h-t\textgreater{}4}    & \multirow{2}{*}{P1}            & 0.05                            & \multicolumn{1}{c|}{\textbf{1.329}}    & UnSAT (4)       & \multicolumn{1}{c|}{16.453}            & \ding{51} (Lower)       & \multicolumn{1}{c|}{N/A}               & -              \\ \cline{4-10} 
                                           &                                        &                                & 0.06                            & \multicolumn{1}{c|}{\textbf{1.891}}    & UnSAT (7)       & \multicolumn{1}{c|}{41.043}            & \ding{51} (Lower)       & \multicolumn{1}{c|}{N/A}               & -              \\ \hline
\textit{Cav3}                              & x-estX\textgreater{}10                 & P3                             & 0.5                             & \multicolumn{1}{c|}{TO}                & -               & \multicolumn{1}{c|}{TO}                & -               & \multicolumn{1}{c|}{TO}                & -              \\ \hline
\multirow{2}{*}{\textit{Cav4}}             & x\textless{}=3                         & P1                             & 0.1                             & \multicolumn{1}{c|}{0.702}             & SAT             & \multicolumn{1}{c|}{\textbf{0.65}}     & \ding{51}               & \multicolumn{1}{c|}{N/A}               & -              \\ \cline{2-10} 
                                           & x\textless{}=8                         & P1                             & 0.01                            & \multicolumn{1}{c|}{\textbf{1.285}}    & SAT             & \multicolumn{1}{c|}{1.612}             & \ding{51}               & \multicolumn{1}{c|}{N/A}               & -              \\ \hline
\multirow{2}{*}{\textit{Cav5}}             & \multirow{2}{*}{i\textgreater{}10}     & \multirow{2}{*}{P1, P3}        & 0.5$^\ddag$                             & \multicolumn{1}{c|}{\textbf{3.978}}    & SAT             & \multicolumn{1}{c|}{TO}                & -               & \multicolumn{1}{c|}{N/A}               & -              \\ \cline{4-10} 
                                           &                                        &                                & 0.1                             & \multicolumn{1}{c|}{\textbf{0.898}}    & UnSAT (8)       & \multicolumn{1}{c|}{TO}                & -               & \multicolumn{1}{c|}{N/A}               & -              \\ \hline
\multirow{3}{*}{\textit{Cav6}}             & N=3$^\ddag$                                    & P3                             & 0.58                            & \multicolumn{1}{c|}{\textbf{0.836}}    & SAT             & \multicolumn{1}{c|}{N/A}               & -               & \multicolumn{1}{c|}{TO}                & -              \\ \cline{2-10} 
                                           & N=4$^\ddag$                                    & P3                             & 0.56                            & \multicolumn{1}{c|}{\textbf{5.619}}    & SAT             & \multicolumn{1}{c|}{N/A}               & -               & \multicolumn{1}{c|}{TO}                & -              \\ \cline{2-10} 
                                           & N=5$^\ddag$                                    & P3                             & 0.55                            & \multicolumn{1}{c|}{\textbf{103.212}}  & SAT             & \multicolumn{1}{c|}{N/A}               & -               & \multicolumn{1}{c|}{TO}                & -              \\ \hline
\multirow{2}{*}{\textit{Cav7}}             & \multirow{2}{*}{count\textgreater{}10} & \multirow{2}{*}{P1}            & 0.3                             & \multicolumn{1}{c|}{\textbf{223.926}}  & UnSAT (229944)  & \multicolumn{1}{c|}{TO}                & -               & \multicolumn{1}{c|}{N/A}               & -              \\ \cline{4-10} 
                                           &                                        &                                & 0.5                             & \multicolumn{1}{c|}{TO}                & -               & \multicolumn{1}{c|}{TO}                & -               & \multicolumn{1}{c|}{N/A}               & -              \\ \hline
\multirow{2}{*}{\textit{BookMod}}          & count\textless{}=20                    & P1, P3                         & 0.001                           & \multicolumn{1}{c|}{\textbf{4.402}}    & SAT             & \multicolumn{1}{c|}{N/A}               & -               & \multicolumn{1}{c|}{N/A}               & -              \\ \cline{2-10} 
                                           & count\textless{}=40                    & P1, P3                         & 0.00025                        & \multicolumn{1}{c|}{\textbf{19.562}}   & SAT             & \multicolumn{1}{c|}{N/A}               & -               & \multicolumn{1}{c|}{N/A}               & -              \\ \hline
\textit{Cart}                              & count\textless{}5                      & P1, P3                         & 0.5                             & \multicolumn{1}{c|}{TO}                & -               & \multicolumn{1}{c|}{TO}                & -               & \multicolumn{1}{c|}{N/A}               & -              \\ \hline
\multirow{4}{*}{\textit{Carton5}}          & count\textless{}8                      & P1, P3                         & 0.2                             & \multicolumn{1}{c|}{\textbf{3.754}}    & SAT             & \multicolumn{1}{c|}{TO}                & -               & \multicolumn{1}{c|}{N/A}               & -              \\ \cline{2-10} 
                                           & count\textless{}7                      & P1, P3                         & 0.2                             & \multicolumn{1}{c|}{\textbf{33.327}}   & SAT             & \multicolumn{1}{c|}{TO}                & -               & \multicolumn{1}{c|}{N/A}               & -              \\ \cline{2-10} 
                                           & count\textless{}6                      & P1, P3                         & 0.2                             & \multicolumn{1}{c|}{\textbf{199.67}}   & UnSAT (820)     & \multicolumn{1}{c|}{TO}                & -               & \multicolumn{1}{c|}{N/A}               & -              \\ \cline{2-10} 
                                           & count\textless{}5                      & P1, P3                         & 0.2                             & \multicolumn{1}{c|}{\textbf{22.979}}   & UnSAT (205)     & \multicolumn{1}{c|}{TO}                & -               & \multicolumn{1}{c|}{N/A}               & -              \\ \hline
\multirow{2}{*}{\textit{Fig6}}             & \multirow{2}{*}{c\textgreater{}5}      & \multirow{2}{*}{P1}            & 0.5                             & \multicolumn{1}{c|}{\textbf{21.422}}   & UnSAT (572)     & \multicolumn{1}{c|}{TO}                & -               & \multicolumn{1}{c|}{N/A}               & -              \\ \cline{4-10} 
                                           &                                        &                                & 0.9                             & \multicolumn{1}{c|}{\textbf{34.11}}    & SAT             & \multicolumn{1}{c|}{TO}                & -               & \multicolumn{1}{c|}{N/A}               & -              \\ \hline
\multirow{2}{*}{\textit{Fig7}}             & \multirow{2}{*}{x\textless{}=1000}     & \multirow{2}{*}{P1}            & 0.0015                          & \multicolumn{1}{c|}{3.422}             & UnSAT (3)       & \multicolumn{1}{c|}{\textbf{0.515}}    & \ding{51} (Lower)       & \multicolumn{1}{c|}{N/A}               & -              \\ \cline{4-10} 
                                           &                                        &                                & 0.00198                         & \multicolumn{1}{c|}{\textbf{1.607}}    & SAT             & \multicolumn{1}{c|}{53.183}            & \ding{51}               & \multicolumn{1}{c|}{N/A}               & -              \\ \hline
\textit{InvPend}                           & pAng\textgreater{}100                  & P3                             & -                               & \multicolumn{1}{c|}{N/A}               & -               & \multicolumn{1}{c|}{N/A}               & -               & \multicolumn{1}{c|}{TO}                & -              \\ \hline
\textit{LoopPerf}                          & guardOK\textgreater{}2                 & P3, P4                         & 0.5                             & \multicolumn{1}{c|}{TO}                & -               & \multicolumn{1}{c|}{N/A}               & -               & \multicolumn{1}{c|}{TO}                & -              \\ \hline
\textit{Vol}                               & count\textgreater{}3                   & P2, P3                         & -                               & \multicolumn{1}{c|}{N/A}               & -               & \multicolumn{1}{c|}{N/A}               & -               & \multicolumn{1}{c|}{TO}                & -              \\ \hline
\textit{Herman}                            & count \textless 5                      & P2, P3                         & 0.7                             & \multicolumn{1}{c|}{\textbf{0.549}}    & SAT             & \multicolumn{1}{c|}{N/A}               & -               & \multicolumn{1}{c|}{TO}                & -              \\ \hline
\multirow{2}{*}{\textit{Israeli-jalfon-3}} & \multirow{2}{*}{count\textgreater{}=1} & \multirow{2}{*}{P1, P3}        & 0.3                             & \multicolumn{1}{c|}{\textbf{0.741}}    & UnSAT (3)       & \multicolumn{1}{c|}{64.455}            & \ding{51} (Lower)       & \multicolumn{1}{c|}{N/A}               & -              \\ \cline{4-10} 
                                           &                                        &                                & 0.6                             & \multicolumn{1}{c|}{\textbf{0.688}}    & SAT             & \multicolumn{1}{c|}{82.902}            & \ding{51}               & \multicolumn{1}{c|}{N/A}               & -              \\ \hline
\textit{Israeli-jalfon-5}                  & count\textgreater{}=1                  & P1, P3                         & 0.3                             & \multicolumn{1}{c|}{\textbf{0.741}}    & UnSAT (3)       & \multicolumn{1}{c|}{64.455}            & \ding{51} (Lower)       & \multicolumn{1}{c|}{N/A}               & -              \\ \hline
\end{tabular}
\end{adjustbox}
\vspace{-12pt}
\end{table}

\begin{table}[t]
\renewcommand{\arraystretch}{1.2} 
\setlength{\tabcolsep}{4pt} 
\centering
\caption{Other Examples: Columns have the same meaning as~\Cref{tab:colo-exp}}
    \label{tab:cp2_reality_amber_sar}
    \vspace{-7pt}
\begin{adjustbox}{max width=\textwidth}
\begin{tabular}{|c|c|c|c|cc|cc|cc|}
\hline
\multirow{2}{*}{\textbf{Example Name}}  & \multirow{2}{*}{\textbf{Parameter}} & \multirow{2}{*}{\textbf{Type}} & \multirow{2}{*}{\textbf{Bound}} & \multicolumn{2}{c|}{\textbf{Our Method}}                 & \multicolumn{2}{c|}{\textbf{CegisPro2}}                  & \multicolumn{2}{c|}{\textbf{PSI}}                       \\ \cline{5-10} 
                                        &                                     &                                &                                 & \multicolumn{1}{c|}{\textbf{Time (s)}} & \textbf{Result} & \multicolumn{1}{c|}{\textbf{Time (s)}} & \textbf{Result} & \multicolumn{1}{c|}{\textbf{Time (s)}} & \textbf{Prob} \\ \hline
\multirow{3}{*}{\textit{RwMultiStep}}   & \multirow{2}{*}{UB}                 & \multirow{2}{*}{P1}            & 0.45                            & \multicolumn{1}{c|}{\textbf{3.782}}    & SAT             & \multicolumn{1}{c|}{TO}                & -               & \multicolumn{1}{c|}{N/A}               & -              \\ \cline{4-10} 
                                        &                                     &                                & 0.39                            & \multicolumn{1}{c|}{\textbf{2.123}}    & UnSAT (4)       & \multicolumn{1}{c|}{TO}                & -               & \multicolumn{1}{c|}{N/A}               & -              \\ \cline{2-10} 
                                        & 200$^\ddag$                                 & P2                             & 0.3                             & \multicolumn{1}{c|}{154.691}           & SAT             & \multicolumn{1}{c|}{\textbf{10.543}}   & \ding{51}               & \multicolumn{1}{c|}{TO}                & -              \\ \hline
\multirow{3}{*}{\textit{brp}}           & \multirow{2}{*}{8000000000$^\ddag$}         & \multirow{2}{*}{P2}            & 0.001                           & \multicolumn{1}{c|}{TO}                & -               & \multicolumn{1}{c|}{\textbf{16.218}}   & \ding{51}               & \multicolumn{1}{c|}{TO}                & -              \\ \cline{4-10} 
                                        &                                     &                                & 0.00001                         & \multicolumn{1}{c|}{TO}                & -               & \multicolumn{1}{c|}{\textbf{10.095}}   & \ding{51}               & \multicolumn{1}{c|}{TO}                & -              \\ \cline{2-10} 
                                        & 200$^\ddag$                                 & P2                             & 0.0001                          & \multicolumn{1}{c|}{\textbf{2.609}}    & SAT             & \multicolumn{1}{c|}{3.314}             & \ding{51}               & \multicolumn{1}{c|}{TO}                & -              \\ \hline
\textit{Chain}                          & 10000$^\ddag$                               & P2                             & 0.4                             & \multicolumn{1}{c|}{\textbf{83.136}}   & SAT             & \multicolumn{1}{c|}{TO}                & -               & \multicolumn{1}{c|}{TO}                & -              \\ \hline
\multirow{4}{*}{\textit{ChainStepSize}} & \multirow{3}{*}{10000000$^\ddag$}           & \multirow{3}{*}{P2, P3}        & 0.7                             & \multicolumn{1}{c|}{TO}                & -               & \multicolumn{1}{c|}{\textbf{11.279}}   & \ding{51}               & \multicolumn{1}{c|}{TO}                & -              \\ \cline{4-10} 
                                        &                                     &                                & 0.6                             & \multicolumn{1}{c|}{TO}                & -               & \multicolumn{1}{c|}{TO}  & -               & \multicolumn{1}{c|}{TO}                & -              \\ \cline{4-10} 
                                        &                                     &                                & 0.55                            & \multicolumn{1}{c|}{TO}                & -               & \multicolumn{1}{c|}{TO}                & -               & \multicolumn{1}{c|}{TO}                & -              \\ \cline{2-10} 
                                        & 200$^\ddag$                                 & P2, P3                         & 0.8                             & \multicolumn{1}{c|}{\textbf{0.652}}    & SAT             & \multicolumn{1}{c|}{19.623}            & \ding{51}               & \multicolumn{1}{c|}{TO}                & -              \\ \hline
\multirow{2}{*}{\textit{EqualProbGrid}} & \multirow{2}{*}{UB}                 & \multirow{2}{*}{P1}            & 0.6                             & \multicolumn{1}{c|}{\textbf{0.272}}    & SAT             & \multicolumn{1}{c|}{TO}                & -               & \multicolumn{1}{c|}{N/A}               & -              \\ \cline{4-10} 
                                        &                                     &                                & 0.4                             & \multicolumn{1}{c|}{\textbf{1.374}}    & UnSAT (3)       & \multicolumn{1}{c|}{TO}                & -               & \multicolumn{1}{c|}{N/A}               & -              \\ \hline
\multirow{2}{*}{\textit{Geo}}           & x\textgreater{}3                    & P1                             & 0.3                             & \multicolumn{1}{c|}{\textbf{0.584}}    & UnSAT (1)       & \multicolumn{1}{c|}{1.028}             & \ding{51} (Lower)       & \multicolumn{1}{c|}{N/A}               & -              \\ \cline{2-10} 
                                        & x\textgreater{}5                    & P1                             & 0.9                             & \multicolumn{1}{c|}{\textbf{0.359}}    & SAT             & \multicolumn{1}{c|}{1.797}             & \ding{51}               & \multicolumn{1}{c|}{N/A}               & -              \\ \hline
\textit{Grid}                           & 100$^\ddag$                                 & P2                             & 0.93                            & \multicolumn{1}{c|}{\textbf{5.061}}    & SAT             & \multicolumn{1}{c|}{TO}                & -               & \multicolumn{1}{c|}{12.865}            & 0.5            \\ \hline
\multirow{3}{*}{\textit{ZeroConf}}      & \multirow{2}{*}{100000000}          & \multirow{2}{*}{P2}            & 0.53                            & \multicolumn{1}{c|}{TO}                & -               & \multicolumn{1}{c|}{\textbf{17.806}}   & \ding{51}               & \multicolumn{1}{c|}{TO}                & -              \\ \cline{4-10} 
                                        &                                     &                                & 0.526                           & \multicolumn{1}{c|}{TO}                & -               & \multicolumn{1}{c|}{TO}                & -               & \multicolumn{1}{c|}{TO}                & -              \\ \cline{2-10} 
                                        & 200                                 & P2                             & 0.6                             & \multicolumn{1}{c|}{\textbf{0.138}}    & SAT             & \multicolumn{1}{c|}{4.844}             & \ding{51}               & \multicolumn{1}{c|}{TO}                & -              \\ \hline
\textit{Coupon5}                        & count\textless{}100$^\ddag$                 & P2                             & 1.5E-9                             & \multicolumn{1}{c|}{\textbf{0.538}}    & SAT             & \multicolumn{1}{c|}{1.451}             & \ding{51}               & \multicolumn{1}{c|}{491.158}           & 1.273E-09      \\ \hline
\textit{Herman3}                        & count\textless{}100$^\ddag$                 & P2, P3                         & 0.01                            & \multicolumn{1}{c|}{\textbf{0.304}}    & SAT             & \multicolumn{1}{c|}{28.064}            & \ding{51}               & \multicolumn{1}{c|}{TO}                & -              \\ \hline
\textit{Reliability}                    & unrel\textgreater{}0$^\ddag$                & P3, P4                         & 0.01                            & \multicolumn{1}{c|}{\textbf{5.434}}    & SAT             & \multicolumn{1}{c|}{N/A}               & -               & \multicolumn{1}{c|}{TO}                & -              \\ \hline
\textit{SeqLoop}                        & y \textgreater 3$^\ddag$                    & P4                             & 0.3                             & \multicolumn{1}{c|}{\textbf{0.304}}    & SAT             & \multicolumn{1}{c|}{N/A}               & -               & \multicolumn{1}{c|}{TO}                & -              \\ \hline
\textit{NestedLoop}                     & y \textgreater 1030$^\ddag$                 & P4                             & 0.4                             & \multicolumn{1}{c|}{\textbf{194.848}}  & SAT             & \multicolumn{1}{c|}{N/A}               & -               & \multicolumn{1}{c|}{TO}                & -              \\ \hline
\textit{Limit}                          & X == 0                              & P1                             & 0.5                             & \multicolumn{1}{c|}{\textbf{0.15}}     & SAT             & \multicolumn{1}{c|}{N/A}               & -               & \multicolumn{1}{c|}{N/A}               & -              \\ \hline
\multirow{2}{*}{\textit{LimitV}}        & \multirow{2}{*}{X == 0}             & \multirow{2}{*}{P3, P4}        & 0.4                             & \multicolumn{1}{c|}{\textbf{5.052}}    & UnSAT (7)       & \multicolumn{1}{c|}{N/A}               & -               & \multicolumn{1}{c|}{N/A}               & -              \\ \cline{4-10} 
                                        &                                     &                                & 0.5                             & \multicolumn{1}{c|}{\textbf{0.182}}    & SAT             & \multicolumn{1}{c|}{N/A}               & -               & \multicolumn{1}{c|}{N/A}               & -              \\ \hline
\multirow{2}{*}{\textit{LimitVP}}       & \multirow{2}{*}{X == 0}             & \multirow{2}{*}{P3, P4}        & 0.3                             & \multicolumn{1}{c|}{\textbf{2.376}}    & UnSAT (3)       & \multicolumn{1}{c|}{N/A}               & -               & \multicolumn{1}{c|}{N/A}               & -              \\ \cline{4-10} 
                                        &                                     &                                & 0.45                            & \multicolumn{1}{c|}{\textbf{0.275}}    & SAT             & \multicolumn{1}{c|}{N/A}               & -               & \multicolumn{1}{c|}{N/A}               & -              \\ \hline
\multirow{5}{*}{\textit{Birthday}}      & \multirow{3}{*}{UB}                 & \multirow{3}{*}{P1, P3}        & 0.5                             & \multicolumn{1}{c|}{\textbf{26.574}}   & UnSAT (361)     & \multicolumn{1}{c|}{TO}                & -               & \multicolumn{1}{c|}{N/A}               & -              \\ \cline{4-10} 
                                        &                                     &                                & 0.7                             & \multicolumn{1}{c|}{\textbf{45.301}}   & UnSAT (1561)    & \multicolumn{1}{c|}{TO}                & -               & \multicolumn{1}{c|}{N/A}               & -              \\ \cline{4-10} 
                                        &                                     &                                & 0.99                            & \multicolumn{1}{c|}{\textbf{409.902}}  & UnSAT (43888)   & \multicolumn{1}{c|}{TO}                & -               & \multicolumn{1}{c|}{N/A}               & -              \\ \cline{2-10} 
                                        & 3                                   & P3                             & 0.4                             & \multicolumn{1}{c|}{17.002}            & SAT             & \multicolumn{1}{c|}{TO}                & -               & \multicolumn{1}{c|}{\textbf{3.703}}    & 0.388          \\ \cline{2-10} 
                                        & 5                                   & P3                             & 0.7                             & \multicolumn{1}{c|}{46.692}            & UnSAT (1261)    & \multicolumn{1}{c|}{TO}                & -               & \multicolumn{1}{c|}{\textbf{43.231}}   & 0.85           \\ \hline
\textit{Cards}                          & sum != 1000                         & P2                             & 0.3                             & \multicolumn{1}{c|}{\textbf{0.563}}    & SAT             & \multicolumn{1}{c|}{TO}                & -               & \multicolumn{1}{c|}{TO}                & -              \\ \hline
\end{tabular}
\end{adjustbox}
\vspace{-12pt}
\end{table}

Overall, the experimental dataset consists of 37 examples from diverse sources.
In these benchmarks, continuous distributions were adapted to the tools by discretization.
All tests were conducted on a MacBook Pro (M4 Pro, 48GB RAM).

\paragraph{Experimental Evaluation}
The experimental results for examples from~\cite{10.1145/2491956.2462179} are presented in Table~\ref{tab:colo-exp}, while the results for CegisPro2 and all other examples are summarized in~\Cref{tab:cp2_reality_amber_sar}.
For some examples, we tested multiple parameter settings, and for each parameter, we further selected different proof bounds ($\beta$).
Specifically, for the selection of proof bounds ($\beta$), we followed established practices in the literature~\cite{batz2023probabilistic,gehr2016psi}.
For examples sourced from CegisPro2's benchmark suite, we preserved the original threshold values and additionally tested tighter thresholds to evaluate robustness.
For examples where PSI could compute exact solutions, we selected $\beta$ values proximate to the computed exact bounds to assess precision.
For the remaining examples, we adopted the standard guess-and-refine approach commonly employed in the field.
Importantly, all threshold selections were consistent across tools to ensure fair comparison.
This process generated multiple \emph{verification tasks} per example, each evaluated using our method, CegisPro2, and PSI.\@

The results first of all serve as a sanity check for our tool, 
as the experimental data affirms the correctness of our results
from the consistency with the results from CegisPro2 and PSI.\@
Note, on the other hand, that CegisPro2 is solely capable of reporting whether a bound is proved,
and it will either crash or fail to terminate when a bound is not a proper upper bound,
let alone extracting a counterexample.
Hence, we usually run our tool and PSI first to decide whether to prove an upper bound or lower bound in CegisPro2 and in cases when both are TO,
we run both upper and lower bound computations.

As for the experimental statistical comparison,
in total, we have tested 68 verification tasks, as shown in both~\Cref{tab:colo-exp,tab:cp2_reality_amber_sar}.
Within these, our method is capable of handling 55 tasks,
timing out in 11 tasks,
and, due to the current limitation of our implementation — our tool only handles integers as data types —
there are two examples, namely ``InvPend'' and ``Vol'',
that are beyond the capability of our current implementation.
For exactly these two cases, as they contain non-easily resolvable negative numbers,
they also exceed the capacity of CegisPro2.
Notably, this is a limitation stemming from the theoretical aspect as mentioned in~\cite{batz2023probabilistic}.
While, as the two loop for too many times, PSI also timed out on them.
In comparison, CegisPro2 properly handles 22 tasks,
and PSI is capable of handling only 4 tasks.
Overall, our method demonstrates advantages (in terms of time) in 50 verification tasks in terms of speed and capability,
while CegisPro2 excels at 7 tasks,
and PSI wins at only two tasks.
In the cases where our tool reports results, we are usually much faster than the competitors.
In the cases like ``Israeli-jalfon-3'' in~\Cref{tab:colo-exp},
our tool is over 100$\times$  faster than CegisPro2
while in ``Coupon5'' in~\Cref{tab:cp2_reality_amber_sar}, our result is over 800$\times$ faster than PSI.\@
Not to mention they are often TO while we report results in a few seconds.

Delving deeper into the types of the tasks.
Observably, CegisPro2 performs well mostly in verification tasks falling into (P2) with relatively simple loop bodies.
However, their method is highly sensitive to the structure of the loop body, loop conditions and the distance to the real value,
and hence, for examples falling into other categories, it is easily overwhelmed.

The data also shows that these challenging cases are difficult for PSI,
which is a renowned tool for \emph{exact} solution computation.
It is hence solely capable of tackling bounded loops stemming from its theoretical foundation,
while from the experimental results, it is also not working swiftly for bounded loops with relatively large iterations.
For examples falling into (P1), the unrolling-based method struggles to handle the situation.

Furthermore, it is clear that the capabilities of CegisPro2 and PSI are largely disjoint, 
both intrinsically and practically. 
Intrinsically, CegisPro2 is limited in the forms of programs it can handle, 
particularly when dealing with (P4) and negative numbers, 
while PSI is theoretically limited in handling (P1). 
Practically, CegisPro2 often fails with examples in (P1) and (P3), 
while PSI typically times out with (P2). 
In contrast, our method demonstrates robustness across all these challenging features.

Overall,
the experimental results highlight the advantages of our method in addressing more general and versatile cases.
Specifically, the data demonstrates that our method exhibits evident superior performance in handling examples of type (P1).
It also typically resolves instances of (P2) efficiently with reasonable loop counts,
while showing greater robustness to variations in loop bodies and conditions.
This is evidenced by the fact that our method outperforms CegisPro2 in 9 tasks for (P2), compared to only 5 tasks where the opposite is true.
Moreover, our method robustly handles the more complex and general cases (P3) and (P4),
which are particularly challenging for CegisPro2 and PSI, especially when combined with (P1) and (P2).
Essentially, for (P3) and (P4), thanks to the separation of concerns offered by the refinement principle,
the complex structures, including nested and sequential loops, are seamlessly handled
by the established method of trace abstraction,
which has shown its excellence in non-random verification.

Beyond efficiency, our method is often capable of computing tight bounds.
For example, in the ``Grid'' and ``Coupon5'' cases in~\Cref{tab:cp2_reality_amber_sar},
the bounds computed by our method are very close to the exact values computed by PSI,
and these bounds are swiftly provable by our method.
In fact, our method can even prove the upper bound of ``Reliability'' in~\Cref{tab:cp2_reality_amber_sar}
by $0.0016$ as compared to the original $0.0072$ produced in~\cite{trace_abstraction_modulo_probability}
and $0.0051$ from Rely~\cite{carbin2013verifying}.
Also, the theoretical refutational completeness also proves itself in the experimental data;
e.g., when unbounded, the ``Birthday'' example should have probability $1$ of violating, then even when we set the threshold to be $0.99$,
our tool still successfully reports the counterexample.

In summary, our method demonstrates superior performance, effectively addressing a broader range of cases and excelling in handling complex program structures.
Even in scenarios where other tools are applicable, our method offers greater stability and efficiency, successfully covering a substantial portion of the combined capabilities of cegispro2 and PSI.

\section{Related Work}
\label{sec: related work}

\paragraph{Automaton and MDP}
When integrating graphical program representation via Control Flow Automata (CFA) or Control Flow Graph (CFG) with Markov Decision Processes (MDP),
the most straightforward approach is to use MDP as the program's semantics.
This is achieved by unrolling the CFA / CFG while exploring the program's concrete state space.
This principle supports probabilistic model checking~\cite{prism_2018,hensel2022probabilistic},
a successful lightweight probabilistic program analysis technique.
Practical tools like PRISM~\cite{prism} have been developed with industrial strength,
successfully deployed across diverse domains, some beyond computer science~\cite{prism_application_algorithm,prism_application_biology,prism_application_industry,prism_application_protocol1}.
Additionally, bounded model checking for probabilistic programs~\cite{bounded_model_checking_probabilistic_programs} has been developed,
analysing programs with potentially infinite valuations.
In this work, we propose a different approach to integrate PCFA with MDP.\@
This method, entirely structural / syntactic, converts a PCFA into an MDP by erasing computational semantics,
rather than unrolling the potential state space.

\paragraph{Probabilistic CEGAR}
The study conducted by Hermanns et al.~\cite{pcegar} bears a significant connection to our work.
They explored the extension of \emph{predicate abstraction} within a probabilistic setting using a CEGAR-guided algorithm~\cite{cegar}.
We drew inspiration from their work for our broad CEGAR framework and the idea of counterexamples,
as well as their method of verification through enumeration.
Despite these similarities, the theoretical foundation is the key divergent point between our projects with theirs.
Given the immense flexibility provided by the new refinement principle we developed in~\cref{theorem: refinement enabling},
we are able to take one more step towards generalising the framework to a versatile one that is capable of being straightforwardly instantiated by various non-probabilistic techniques,
including the predicate abstraction considered in their work.
Especially, we discuss the extension with trace abstraction, which could hardly cooperate with their method.

\paragraph{Trace Abstraction and 
Probabilities}
Trace abstraction~\cite{trace_abstraction,trace_abstraction_incremental} is a key technique in non-probabilistic program verification.
It views a program as a set of traces rather than a state space, leading to award-winning tools like~\cite{site_trace_abstraction}.
Beyond imperative program verification, it has applications in domains like termination analysis~\cite{trace_abstraction_termination_cav14}.
For combining trace abstraction with probabilities, pioneering work by Smith et al.~\cite{trace_abstraction_modulo_probability} addressed this via program synthesis.
Unlike our method, which exposes probabilities at the control-flow level, Smith et al.~encapsulated them at the statement level.
For probabilistic distributions on statements,
symbolic representations of violation probabilities are synthesised for candidate counterexample traces.
Due to their handling of probabilities,
traces remain probabilistically independent, akin to the non-probabilistic case.
This approach, however, places the burden of handling distributions on each trace,
limiting the use of existing non-probabilistic verification techniques.
Moreover, their use of the union bound~\cite{barthe2016program} also impedes completeness, even for finite trace examples like $\prob{\not\vdash \bs{\tTrue} \ X \sim flip; Y \sim flip \ \bs{X \wedge Y}}$.
Besides, trace abstractions are also applied in termination analysis, e.g., Chen et al.~\cite{ast_by_omega_decompose} aim to automatically prove almost sure termination (AST) by decomposing programs into certified $\omega$-regular automata.

\paragraph{Probabilistic Threshold Problem}
The problem considered in this work, that to automatically derive whether $\prob{\not\vdash \bs{\precond} \ P \ \bs{\postcond}} \le \beta$, has also been considered by other works.
To begin with, the mentioned work by Smith et al.~\cite{trace_abstraction_modulo_probability} also considers this problem.
Further, Wang et al.~\cite{exponential_analysis_of_probabilistic_program} analyse the violation probabilities by synthesising the bounds with an exponential template.
Such a problem can also be tackled by pre-expectation calculus and the associated expectation-transformer semantics~\cite{mciver2005abstraction,operational_semantics_of_pgcl},
which is an elegant probabilistic counterpart of the famous predicate-transformer semantics by Dijkstra in his seminal work~\cite{dijkstra1975guarded}.
The approach has been exploited in a recent work for cegispro2~\cite{batz2023probabilistic}.
The semantics is designed to compute expectations, which can also be used to compute violation probability elegantly via expectation~\cite{mciver2005abstraction}.
The key obstacle to its automation, however, is the same as in the predicate-transformer semantics -- the synthesis of (probabilistic) invariants,
which works like~\cite{bartocci2019automatic} has aimed at overcoming.
Indeed, we believe that incorporating such information in the interpolation process could potentially enhance the quality of the interpolants produced.
This enhancement may enable the resolution of previously unsolvable problems or contribute to the problem-solving time.

\paragraph{Static Analysis of Bayesian Programs}
Besides the topics, there is a rising trend in the analysis of the posterior distribution of Bayesian programs,
which additionally equip usual probabilistic programs with the {\tt observe} structure to express conditioning.
Various techniques have been proposed to address this novel problem,
including typing~\cite{beutner2022guaranteed},
fixed-point \& Optional Stopping Theorem~\cite{wang2024static},
and novel semantics~\cite{zaiser2025guaranteed}.

\section{Conclusion}
\label{sec: conclusion}

This work introduces structural abstraction for verifying the threshold query for probabilistic programs.
In this approach, transitions in a Probabilistic Control-Flow Automaton (PCFA) are treated as pure labels, abstracting the semantics and enabling a view of the PCFA as a Markov Decision Process (MDP).
From the theoretical establishment of this abstraction and the principles of refinement, our method shows the ability to separate probability from semantics.
We hence developed a general CEGAR framework capable of leveraging non-random techniques,
for which we demonstrated the instantiation via trace abstraction,
with a further optimised refutationally complete method.
Our implementation, benchmarked against advanced tools on diverse examples, highlights its adaptability and superiority in various scenarios.

\section*{Data Availability Statement}
\label{sec: data availability statement}

All data, source code, and compiled binaries reported in the experimental evaluation are publicly available at \url{https://zenodo.org/records/15760713}.

\bibliography{bibs}


\begin{thebibliography}{68}


\ifx \showCODEN    \undefined \def \showCODEN     #1{\unskip}     \fi
\ifx \showDOI      \undefined \def \showDOI       #1{#1}\fi
\ifx \showISBNx    \undefined \def \showISBNx     #1{\unskip}     \fi
\ifx \showISBNxiii \undefined \def \showISBNxiii  #1{\unskip}     \fi
\ifx \showISSN     \undefined \def \showISSN      #1{\unskip}     \fi
\ifx \showLCCN     \undefined \def \showLCCN      #1{\unskip}     \fi
\ifx \shownote     \undefined \def \shownote      #1{#1}          \fi
\ifx \showarticletitle \undefined \def \showarticletitle #1{#1}   \fi
\ifx \showURL      \undefined \def \showURL       {\relax}        \fi
\providecommand\bibfield[2]{#2}
\providecommand\bibinfo[2]{#2}
\providecommand\natexlab[1]{#1}
\providecommand\showeprint[2][]{arXiv:#2}

\bibitem[\protect\citeauthoryear{Baier and Katoen}{Baier and Katoen}{2008}]%
        {principles_of_model_checking}
\bibfield{author}{\bibinfo{person}{Christel Baier} {and}
  \bibinfo{person}{Joost-Pieter Katoen}.} \bibinfo{year}{2008}\natexlab{}.
\newblock \bibinfo{booktitle}{\emph{Principles of model checking}}.
\newblock \bibinfo{publisher}{MIT press}.
\newblock


\bibitem[\protect\citeauthoryear{Bao, Trivedi, Pathak, Hsu, and Roy}{Bao
  et~al\mbox{.}}{2022}]%
        {bao2022data}
\bibfield{author}{\bibinfo{person}{Jialu Bao}, \bibinfo{person}{Nitesh
  Trivedi}, \bibinfo{person}{Drashti Pathak}, \bibinfo{person}{Justin Hsu},
  {and} \bibinfo{person}{Subhajit Roy}.} \bibinfo{year}{2022}\natexlab{}.
\newblock \showarticletitle{Data-driven invariant learning for probabilistic
  programs}. In \bibinfo{booktitle}{\emph{International Conference on Computer
  Aided Verification}}. Springer, \bibinfo{pages}{33--54}.
\newblock


\bibitem[\protect\citeauthoryear{Barthe, Gaboardi, Gr{\'e}goire, Hsu, and
  Strub}{Barthe et~al\mbox{.}}{2016}]%
        {barthe2016program}
\bibfield{author}{\bibinfo{person}{Gilles Barthe}, \bibinfo{person}{Marco
  Gaboardi}, \bibinfo{person}{Benjamin Gr{\'e}goire}, \bibinfo{person}{Justin
  Hsu}, {and} \bibinfo{person}{Pierre-Yves Strub}.}
  \bibinfo{year}{2016}\natexlab{}.
\newblock \showarticletitle{A program logic for union bounds}.
\newblock \bibinfo{journal}{\emph{arXiv preprint arXiv:1602.05681}}
  (\bibinfo{year}{2016}).
\newblock


\bibitem[\protect\citeauthoryear{Bartocci, Kov{\'a}cs, and
  Stankovi{\v{c}}}{Bartocci et~al\mbox{.}}{2019}]%
        {bartocci2019automatic}
\bibfield{author}{\bibinfo{person}{Ezio Bartocci}, \bibinfo{person}{Laura
  Kov{\'a}cs}, {and} \bibinfo{person}{Miroslav Stankovi{\v{c}}}.}
  \bibinfo{year}{2019}\natexlab{}.
\newblock \showarticletitle{Automatic generation of moment-based invariants for
  prob-solvable loops}. In \bibinfo{booktitle}{\emph{International Symposium on
  Automated Technology for Verification and Analysis}}. Springer,
  \bibinfo{pages}{255--276}.
\newblock


\bibitem[\protect\citeauthoryear{Batz, Chen, Junges, Kaminski, Katoen, and
  Matheja}{Batz et~al\mbox{.}}{2023}]%
        {batz2023probabilistic}
\bibfield{author}{\bibinfo{person}{Kevin Batz}, \bibinfo{person}{Mingshuai
  Chen}, \bibinfo{person}{Sebastian Junges}, \bibinfo{person}{Benjamin~Lucien
  Kaminski}, \bibinfo{person}{Joost-Pieter Katoen}, {and}
  \bibinfo{person}{Christoph Matheja}.} \bibinfo{year}{2023}\natexlab{}.
\newblock \showarticletitle{Probabilistic program verification via inductive
  synthesis of inductive invariants}. In
  \bibinfo{booktitle}{\emph{International Conference on Tools and Algorithms
  for the Construction and Analysis of Systems}}. Springer,
  \bibinfo{pages}{410--429}.
\newblock


\bibitem[\protect\citeauthoryear{Beutner, Ong, and Zaiser}{Beutner
  et~al\mbox{.}}{2022}]%
        {beutner2022guaranteed}
\bibfield{author}{\bibinfo{person}{Raven Beutner}, \bibinfo{person}{C-H~Luke
  Ong}, {and} \bibinfo{person}{Fabian Zaiser}.}
  \bibinfo{year}{2022}\natexlab{}.
\newblock \showarticletitle{Guaranteed bounds for posterior inference in
  universal probabilistic programming}. In
  \bibinfo{booktitle}{\emph{Proceedings of the 43rd ACM SIGPLAN International
  Conference on Programming Language Design and Implementation}}.
  \bibinfo{pages}{536--551}.
\newblock


\bibitem[\protect\citeauthoryear{Beutner and Ong}{Beutner and Ong}{2021}]%
        {beutner2021probabilistic}
\bibfield{author}{\bibinfo{person}{Raven Beutner} {and} \bibinfo{person}{Luke
  Ong}.} \bibinfo{year}{2021}\natexlab{}.
\newblock \showarticletitle{On probabilistic termination of functional programs
  with continuous distributions}. In \bibinfo{booktitle}{\emph{Proceedings of
  the 42nd ACM SIGPLAN International Conference on Programming Language Design
  and Implementation}}. \bibinfo{pages}{1312--1326}.
\newblock


\bibitem[\protect\citeauthoryear{Beyer, Lingsch~Rosenfeld, and Spiessl}{Beyer
  et~al\mbox{.}}{2022}]%
        {beyer2022unifying}
\bibfield{author}{\bibinfo{person}{Dirk Beyer}, \bibinfo{person}{Marian
  Lingsch~Rosenfeld}, {and} \bibinfo{person}{Martin Spiessl}.}
  \bibinfo{year}{2022}\natexlab{}.
\newblock \showarticletitle{A unifying approach for control-flow-based loop
  abstraction}. In \bibinfo{booktitle}{\emph{International Conference on
  Software Engineering and Formal Methods}}. Springer, \bibinfo{pages}{3--19}.
\newblock


\bibitem[\protect\citeauthoryear{Bj{\o}rner and Phan}{Bj{\o}rner and
  Phan}{2014}]%
        {maxsmt}
\bibfield{author}{\bibinfo{person}{Nikolaj~S Bj{\o}rner} {and}
  \bibinfo{person}{Anh-Dung Phan}.} \bibinfo{year}{2014}\natexlab{}.
\newblock \showarticletitle{$\nu$Z-Maximal Satisfaction with Z3.}
\newblock \bibinfo{journal}{\emph{Scss}}  \bibinfo{volume}{30}
  (\bibinfo{year}{2014}), \bibinfo{pages}{1--9}.
\newblock


\bibitem[\protect\citeauthoryear{Carbin, Misailovic, and Rinard}{Carbin
  et~al\mbox{.}}{2013}]%
        {carbin2013verifying}
\bibfield{author}{\bibinfo{person}{Michael Carbin}, \bibinfo{person}{Sasa
  Misailovic}, {and} \bibinfo{person}{Martin~C Rinard}.}
  \bibinfo{year}{2013}\natexlab{}.
\newblock \showarticletitle{Verifying quantitative reliability for programs
  that execute on unreliable hardware}.
\newblock \bibinfo{journal}{\emph{ACM SIGPLAN Notices}} \bibinfo{volume}{48},
  \bibinfo{number}{10} (\bibinfo{year}{2013}), \bibinfo{pages}{33--52}.
\newblock


\bibitem[\protect\citeauthoryear{Carpenter, Gelman, Hoffman, Lee, Goodrich,
  Betancourt, Brubaker, Guo, Li, and Riddell}{Carpenter et~al\mbox{.}}{2017}]%
        {ppl_stan}
\bibfield{author}{\bibinfo{person}{Bob Carpenter}, \bibinfo{person}{Andrew
  Gelman}, \bibinfo{person}{Matthew~D Hoffman}, \bibinfo{person}{Daniel Lee},
  \bibinfo{person}{Ben Goodrich}, \bibinfo{person}{Michael Betancourt},
  \bibinfo{person}{Marcus Brubaker}, \bibinfo{person}{Jiqiang Guo},
  \bibinfo{person}{Peter Li}, {and} \bibinfo{person}{Allen Riddell}.}
  \bibinfo{year}{2017}\natexlab{}.
\newblock \showarticletitle{Stan: A probabilistic programming language}.
\newblock \bibinfo{journal}{\emph{Journal of statistical software}}
  \bibinfo{volume}{76}, \bibinfo{number}{1} (\bibinfo{year}{2017}),
  \bibinfo{pages}{1--32}.
\newblock


\bibitem[\protect\citeauthoryear{Chakarov and Sankaranarayanan}{Chakarov and
  Sankaranarayanan}{2013}]%
        {chakarov2013probabilistic}
\bibfield{author}{\bibinfo{person}{Aleksandar Chakarov} {and}
  \bibinfo{person}{Sriram Sankaranarayanan}.} \bibinfo{year}{2013}\natexlab{}.
\newblock \showarticletitle{Probabilistic program analysis with martingales}.
  In \bibinfo{booktitle}{\emph{International Conference on Computer Aided
  Verification}}. Springer, \bibinfo{pages}{511--526}.
\newblock


\bibitem[\protect\citeauthoryear{Chao}{Chao}{1984}]%
        {chao1984nonparametric}
\bibfield{author}{\bibinfo{person}{Anne Chao}.}
  \bibinfo{year}{1984}\natexlab{}.
\newblock \showarticletitle{Nonparametric estimation of the number of classes
  in a population}.
\newblock \bibinfo{journal}{\emph{Scandinavian Journal of statistics}}
  (\bibinfo{year}{1984}), \bibinfo{pages}{265--270}.
\newblock


\bibitem[\protect\citeauthoryear{Chen and He}{Chen and He}{2020}]%
        {ast_by_omega_decompose}
\bibfield{author}{\bibinfo{person}{Jianhui Chen} {and} \bibinfo{person}{Fei
  He}.} \bibinfo{year}{2020}\natexlab{}.
\newblock \showarticletitle{Proving almost-sure termination by omega-regular
  decomposition}. In \bibinfo{booktitle}{\emph{Proceedings of the 41st ACM
  SIGPLAN Conference on Programming Language Design and Implementation}}.
  \bibinfo{pages}{869--882}.
\newblock


\bibitem[\protect\citeauthoryear{Christ, Hoenicke, and Nutz}{Christ
  et~al\mbox{.}}{2012}]%
        {smt_interpol}
\bibfield{author}{\bibinfo{person}{J{\"u}rgen Christ}, \bibinfo{person}{Jochen
  Hoenicke}, {and} \bibinfo{person}{Alexander Nutz}.}
  \bibinfo{year}{2012}\natexlab{}.
\newblock \showarticletitle{SMTInterpol: An interpolating SMT solver}. In
  \bibinfo{booktitle}{\emph{International SPIN Workshop on Model Checking of
  Software}}. Springer, \bibinfo{pages}{248--254}.
\newblock


\bibitem[\protect\citeauthoryear{Clarke, Grumberg, Jha, Lu, and Veith}{Clarke
  et~al\mbox{.}}{2000}]%
        {cegar}
\bibfield{author}{\bibinfo{person}{Edmund Clarke}, \bibinfo{person}{Orna
  Grumberg}, \bibinfo{person}{Somesh Jha}, \bibinfo{person}{Yuan Lu}, {and}
  \bibinfo{person}{Helmut Veith}.} \bibinfo{year}{2000}\natexlab{}.
\newblock \showarticletitle{Counterexample-guided abstraction refinement}. In
  \bibinfo{booktitle}{\emph{International Conference on Computer Aided
  Verification}}. Springer, \bibinfo{pages}{154--169}.
\newblock


\bibitem[\protect\citeauthoryear{Clarke, Kroening, Sharygina, and Yorav}{Clarke
  et~al\mbox{.}}{2005}]%
        {clarke2005satabs}
\bibfield{author}{\bibinfo{person}{Edmund Clarke}, \bibinfo{person}{Daniel
  Kroening}, \bibinfo{person}{Natasha Sharygina}, {and} \bibinfo{person}{Karen
  Yorav}.} \bibinfo{year}{2005}\natexlab{}.
\newblock \showarticletitle{SATABS: SAT-Based Predicate Abstraction for
  ANSI-C}. In \bibinfo{booktitle}{\emph{Tools and Algorithms for the
  Construction and Analysis of Systems}},
  \bibfield{editor}{\bibinfo{person}{Nicolas Halbwachs} {and}
  \bibinfo{person}{Lenore~D. Zuck}} (Eds.). \bibinfo{publisher}{Springer Berlin
  Heidelberg}, \bibinfo{address}{Berlin, Heidelberg},
  \bibinfo{pages}{570--574}.
\newblock
\showISBNx{978-3-540-31980-1}


\bibitem[\protect\citeauthoryear{Cooper, Harvey, and Kennedy}{Cooper
  et~al\mbox{.}}{2004}]%
        {cooper2004iterative}
\bibfield{author}{\bibinfo{person}{Keith~D Cooper}, \bibinfo{person}{Timothy~J
  Harvey}, {and} \bibinfo{person}{Ken Kennedy}.}
  \bibinfo{year}{2004}\natexlab{}.
\newblock \bibinfo{booktitle}{\emph{Iterative data-flow analysis, revisited}}.
\newblock \bibinfo{type}{{T}echnical {R}eport}.
\newblock


\bibitem[\protect\citeauthoryear{Cousot and Monerau}{Cousot and
  Monerau}{2012}]%
        {prob_abstract_interpreation}
\bibfield{author}{\bibinfo{person}{Patrick Cousot} {and}
  \bibinfo{person}{Michael Monerau}.} \bibinfo{year}{2012}\natexlab{}.
\newblock \showarticletitle{Probabilistic abstract interpretation}. In
  \bibinfo{booktitle}{\emph{European Symposium on Programming}}. Springer,
  \bibinfo{pages}{169--193}.
\newblock


\bibitem[\protect\citeauthoryear{De~Moura and Bj{\o}rner}{De~Moura and
  Bj{\o}rner}{2008}]%
        {de2008z3}
\bibfield{author}{\bibinfo{person}{Leonardo De~Moura} {and}
  \bibinfo{person}{Nikolaj Bj{\o}rner}.} \bibinfo{year}{2008}\natexlab{}.
\newblock \showarticletitle{Z3: An efficient SMT solver}. In
  \bibinfo{booktitle}{\emph{International conference on Tools and Algorithms
  for the Construction and Analysis of Systems}}. Springer,
  \bibinfo{pages}{337--340}.
\newblock


\bibitem[\protect\citeauthoryear{Dijkstra}{Dijkstra}{1975}]%
        {dijkstra1975guarded}
\bibfield{author}{\bibinfo{person}{Edsger~W Dijkstra}.}
  \bibinfo{year}{1975}\natexlab{}.
\newblock \showarticletitle{Guarded commands, nondeterminacy and formal
  derivation of programs}.
\newblock \bibinfo{journal}{\emph{Commun. ACM}} \bibinfo{volume}{18},
  \bibinfo{number}{8} (\bibinfo{year}{1975}), \bibinfo{pages}{453--457}.
\newblock


\bibitem[\protect\citeauthoryear{Duflot, Kwiatkowska, Norman, and
  Parker}{Duflot et~al\mbox{.}}{2004}]%
        {prism_application_protocol1}
\bibfield{author}{\bibinfo{person}{M. Duflot}, \bibinfo{person}{M.
  Kwiatkowska}, \bibinfo{person}{G. Norman}, {and} \bibinfo{person}{D.
  Parker}.} \bibinfo{year}{2004}\natexlab{}.
\newblock \showarticletitle{A Formal Analysis of {Bluetooth} Device Discovery}.
  In \bibinfo{booktitle}{\emph{Proc. 1st International Symposium on Leveraging
  Applications of Formal Methods (ISOLA'04)}}.
\newblock


\bibitem[\protect\citeauthoryear{Dwork, McSherry, Nissim, and Smith}{Dwork
  et~al\mbox{.}}{2006}]%
        {dwork2006calibrating}
\bibfield{author}{\bibinfo{person}{Cynthia Dwork}, \bibinfo{person}{Frank
  McSherry}, \bibinfo{person}{Kobbi Nissim}, {and} \bibinfo{person}{Adam
  Smith}.} \bibinfo{year}{2006}\natexlab{}.
\newblock \showarticletitle{Calibrating noise to sensitivity in private data
  analysis}. In \bibinfo{booktitle}{\emph{Theory of Cryptography: Third Theory
  of Cryptography Conference, TCC 2006, New York, NY, USA, March 4-7, 2006.
  Proceedings 3}}. Springer, \bibinfo{pages}{265--284}.
\newblock


\bibitem[\protect\citeauthoryear{Erd{\H{o}}s and R{\'e}nyi}{Erd{\H{o}}s and
  R{\'e}nyi}{1961}]%
        {erdHos1961classical}
\bibfield{author}{\bibinfo{person}{P{\'a}l Erd{\H{o}}s} {and}
  \bibinfo{person}{Alfr{\'e}d R{\'e}nyi}.} \bibinfo{year}{1961}\natexlab{}.
\newblock \showarticletitle{On a classical problem of probability theory}.
\newblock \bibinfo{journal}{\emph{A Magyar Tudom{\'a}nyos Akad{\'e}mia
  Matematikai Kutat{\'o} Int{\'e}zet{\'e}nek K{\"o}zlem{\'e}nyei}}
  \bibinfo{volume}{6}, \bibinfo{number}{1-2} (\bibinfo{year}{1961}),
  \bibinfo{pages}{215--220}.
\newblock


\bibitem[\protect\citeauthoryear{Gehr, Misailovic, and Vechev}{Gehr
  et~al\mbox{.}}{2016}]%
        {gehr2016psi}
\bibfield{author}{\bibinfo{person}{Timon Gehr}, \bibinfo{person}{Sasa
  Misailovic}, {and} \bibinfo{person}{Martin Vechev}.}
  \bibinfo{year}{2016}\natexlab{}.
\newblock \showarticletitle{PSI: Exact symbolic inference for probabilistic
  programs}. In \bibinfo{booktitle}{\emph{Computer Aided Verification: 28th
  International Conference, CAV 2016, Toronto, ON, Canada, July 17-23, 2016,
  Proceedings, Part I 28}}. Springer, \bibinfo{pages}{62--83}.
\newblock


\bibitem[\protect\citeauthoryear{Goldwasser and Micali}{Goldwasser and
  Micali}{1984}]%
        {encrption_using_probability}
\bibfield{author}{\bibinfo{person}{Shafi Goldwasser} {and}
  \bibinfo{person}{Silvio Micali}.} \bibinfo{year}{1984}\natexlab{}.
\newblock \showarticletitle{Probabilistic encryption}.
\newblock \bibinfo{journal}{\emph{Journal of computer and system sciences}}
  \bibinfo{volume}{28}, \bibinfo{number}{2} (\bibinfo{year}{1984}),
  \bibinfo{pages}{270--299}.
\newblock


\bibitem[\protect\citeauthoryear{Goodman, Mansinghka, Roy, Bonawitz, and
  Tenenbaum}{Goodman et~al\mbox{.}}{2012}]%
        {ppl_church}
\bibfield{author}{\bibinfo{person}{Noah Goodman}, \bibinfo{person}{Vikash
  Mansinghka}, \bibinfo{person}{Daniel~M Roy}, \bibinfo{person}{Keith
  Bonawitz}, {and} \bibinfo{person}{Joshua~B Tenenbaum}.}
  \bibinfo{year}{2012}\natexlab{}.
\newblock \showarticletitle{Church: a language for generative models}.
\newblock \bibinfo{journal}{\emph{arXiv preprint arXiv:1206.3255}}
  (\bibinfo{year}{2012}).
\newblock


\bibitem[\protect\citeauthoryear{Grellois, Lago, and Kobayashi}{Grellois
  et~al\mbox{.}}{2020}]%
        {phors_long}
\bibfield{author}{\bibinfo{person}{Charles Grellois}, \bibinfo{person}{Ugo~Dal
  Lago}, {and} \bibinfo{person}{Naoki Kobayashi}.}
  \bibinfo{year}{2020}\natexlab{}.
\newblock \showarticletitle{On the Termination Problem for Probabilistic
  Higher-Order Recursive Programs}.
\newblock \bibinfo{journal}{\emph{Logical Methods in Computer Science}}
  \bibinfo{volume}{16} (\bibinfo{year}{2020}).
\newblock


\bibitem[\protect\citeauthoryear{Gretz, Katoen, and McIver}{Gretz
  et~al\mbox{.}}{2014}]%
        {operational_semantics_of_pgcl}
\bibfield{author}{\bibinfo{person}{Friedrich Gretz},
  \bibinfo{person}{Joost-Pieter Katoen}, {and} \bibinfo{person}{Annabelle
  McIver}.} \bibinfo{year}{2014}\natexlab{}.
\newblock \showarticletitle{Operational versus weakest pre-expectation
  semantics for the probabilistic guarded command language}.
\newblock \bibinfo{journal}{\emph{Performance Evaluation}}
  \bibinfo{volume}{73} (\bibinfo{year}{2014}), \bibinfo{pages}{110--132}.
\newblock


\bibitem[\protect\citeauthoryear{Hajdu and Micskei}{Hajdu and Micskei}{2020}]%
        {hajdu2020efficient}
\bibfield{author}{\bibinfo{person}{{\'A}kos Hajdu} {and}
  \bibinfo{person}{Zolt{\'a}n Micskei}.} \bibinfo{year}{2020}\natexlab{}.
\newblock \showarticletitle{Efficient strategies for CEGAR-based model
  checking}.
\newblock \bibinfo{journal}{\emph{Journal of Automated Reasoning}}
  \bibinfo{volume}{64}, \bibinfo{number}{6} (\bibinfo{year}{2020}),
  \bibinfo{pages}{1051--1091}.
\newblock


\bibitem[\protect\citeauthoryear{Hall, Ramage, Zaugg, Lehmann, Merritt,
  Stevens, Baldridge, Hunter, DeCaprio, Duckworth, et~al\mbox{.}}{Hall
  et~al\mbox{.}}{2009}]%
        {hall2009scalanlp}
\bibfield{author}{\bibinfo{person}{David Hall}, \bibinfo{person}{Daniel
  Ramage}, \bibinfo{person}{Jason Zaugg}, \bibinfo{person}{Alexander Lehmann},
  \bibinfo{person}{Jonathan Merritt}, \bibinfo{person}{Keith Stevens},
  \bibinfo{person}{Jason Baldridge}, \bibinfo{person}{Timothy Hunter},
  \bibinfo{person}{Dave DeCaprio}, \bibinfo{person}{Daniel Duckworth},
  {et~al\mbox{.}}} \bibinfo{year}{2009}\natexlab{}.
\newblock \bibinfo{title}{ScalaNLP: Breeze}.
\newblock
\newblock
\urldef\tempurl%
\url{https://github.com/scalanlp/breeze}
\showURL{%
\tempurl}


\bibitem[\protect\citeauthoryear{Han, Katoen, and Berteun}{Han
  et~al\mbox{.}}{2009}]%
        {trace_enumeration}
\bibfield{author}{\bibinfo{person}{Tingting Han}, \bibinfo{person}{Joost-Pieter
  Katoen}, {and} \bibinfo{person}{Damman Berteun}.}
  \bibinfo{year}{2009}\natexlab{}.
\newblock \showarticletitle{Counterexample generation in probabilistic model
  checking}.
\newblock \bibinfo{journal}{\emph{IEEE transactions on software engineering}}
  \bibinfo{volume}{35}, \bibinfo{number}{2} (\bibinfo{year}{2009}),
  \bibinfo{pages}{241--257}.
\newblock


\bibitem[\protect\citeauthoryear{Heath, Kwiatkowska, Norman, Parker, and
  Tymchyshyn}{Heath et~al\mbox{.}}{2008}]%
        {prism_application_biology}
\bibfield{author}{\bibinfo{person}{J. Heath}, \bibinfo{person}{M. Kwiatkowska},
  \bibinfo{person}{G. Norman}, \bibinfo{person}{D. Parker}, {and}
  \bibinfo{person}{O. Tymchyshyn}.} \bibinfo{year}{2008}\natexlab{}.
\newblock \showarticletitle{Probabilistic model checking of complex biological
  pathways}.
\newblock \bibinfo{journal}{\emph{Theoretical Computer Science}}
  \bibinfo{volume}{319}, \bibinfo{number}{3} (\bibinfo{year}{2008}),
  \bibinfo{pages}{239--257}.
\newblock


\bibitem[\protect\citeauthoryear{Heizmann}{Heizmann}{[n.d.]}]%
        {site_trace_abstraction}
\bibfield{author}{\bibinfo{person}{Matthias Heizmann}.}
  \bibinfo{year}{[n.d.]}\natexlab{}.
\newblock \bibinfo{title}{Uni-Freiburg : SWT - Ultimate}.
\newblock
\newblock
\urldef\tempurl%
\url{https://monteverdi.informatik.uni-freiburg.de/tomcat/Website/?ui=tool&tool=automizer}
\showURL{%
\tempurl}
\newblock
\shownote{November 02, 2021.}


\bibitem[\protect\citeauthoryear{Heizmann, Hoenicke, and Podelski}{Heizmann
  et~al\mbox{.}}{2009}]%
        {trace_abstraction}
\bibfield{author}{\bibinfo{person}{Matthias Heizmann}, \bibinfo{person}{Jochen
  Hoenicke}, {and} \bibinfo{person}{Andreas Podelski}.}
  \bibinfo{year}{2009}\natexlab{}.
\newblock \showarticletitle{Refinement of trace abstraction}. In
  \bibinfo{booktitle}{\emph{International Static Analysis Symposium}}.
  Springer, \bibinfo{pages}{69--85}.
\newblock


\bibitem[\protect\citeauthoryear{Heizmann, Hoenicke, and Podelski}{Heizmann
  et~al\mbox{.}}{2014}]%
        {trace_abstraction_termination_cav14}
\bibfield{author}{\bibinfo{person}{Matthias Heizmann}, \bibinfo{person}{Jochen
  Hoenicke}, {and} \bibinfo{person}{Andreas Podelski}.}
  \bibinfo{year}{2014}\natexlab{}.
\newblock \showarticletitle{Termination Analysis by Learning Terminating
  Programs}. In \bibinfo{booktitle}{\emph{Computer Aided Verification}},
  \bibfield{editor}{\bibinfo{person}{Armin Biere} {and}
  \bibinfo{person}{Roderick Bloem}} (Eds.). \bibinfo{publisher}{Springer
  International Publishing}, \bibinfo{address}{Cham},
  \bibinfo{pages}{797--813}.
\newblock
\showISBNx{978-3-319-08867-9}


\bibitem[\protect\citeauthoryear{Hensel, Junges, Katoen, Quatmann, and
  Volk}{Hensel et~al\mbox{.}}{2022}]%
        {hensel2022probabilistic}
\bibfield{author}{\bibinfo{person}{Christian Hensel},
  \bibinfo{person}{Sebastian Junges}, \bibinfo{person}{Joost-Pieter Katoen},
  \bibinfo{person}{Tim Quatmann}, {and} \bibinfo{person}{Matthias Volk}.}
  \bibinfo{year}{2022}\natexlab{}.
\newblock \showarticletitle{The probabilistic model checker Storm}.
\newblock \bibinfo{journal}{\emph{International Journal on Software Tools for
  Technology Transfer}} (\bibinfo{year}{2022}), \bibinfo{pages}{1--22}.
\newblock


\bibitem[\protect\citeauthoryear{Henzinger, Jhala, Majumdar, and
  McMillan}{Henzinger et~al\mbox{.}}{2004}]%
        {henzinger2004abstractions}
\bibfield{author}{\bibinfo{person}{Thomas~A Henzinger}, \bibinfo{person}{Ranjit
  Jhala}, \bibinfo{person}{Rupak Majumdar}, {and} \bibinfo{person}{Kenneth~L
  McMillan}.} \bibinfo{year}{2004}\natexlab{}.
\newblock \showarticletitle{Abstractions from proofs}. In
  \bibinfo{booktitle}{\emph{Proceedings of the 31st ACM SIGPLAN-SIGACT
  symposium on Principles of programming languages}}.
  \bibinfo{pages}{232--244}.
\newblock


\bibitem[\protect\citeauthoryear{Herman}{Herman}{1990}]%
        {herman1990probabilistic}
\bibfield{author}{\bibinfo{person}{Ted Herman}.}
  \bibinfo{year}{1990}\natexlab{}.
\newblock \showarticletitle{Probabilistic self-stabilization}.
\newblock \bibinfo{journal}{\emph{Inform. Process. Lett.}}
  \bibinfo{volume}{35}, \bibinfo{number}{2} (\bibinfo{year}{1990}),
  \bibinfo{pages}{63--67}.
\newblock


\bibitem[\protect\citeauthoryear{Hermanns, Wachter, and Zhang}{Hermanns
  et~al\mbox{.}}{2008}]%
        {pcegar}
\bibfield{author}{\bibinfo{person}{Holger Hermanns}, \bibinfo{person}{Bj{\"o}rn
  Wachter}, {and} \bibinfo{person}{Lijun Zhang}.}
  \bibinfo{year}{2008}\natexlab{}.
\newblock \showarticletitle{Probabilistic cegar}. In
  \bibinfo{booktitle}{\emph{International Conference on Computer Aided
  Verification}}. Springer, \bibinfo{pages}{162--175}.
\newblock


\bibitem[\protect\citeauthoryear{Hoare}{Hoare}{1969}]%
        {hoare1969axiomatic}
\bibfield{author}{\bibinfo{person}{Charles Antony~Richard Hoare}.}
  \bibinfo{year}{1969}\natexlab{}.
\newblock \showarticletitle{An axiomatic basis for computer programming}.
\newblock \bibinfo{journal}{\emph{Commun. ACM}} \bibinfo{volume}{12},
  \bibinfo{number}{10} (\bibinfo{year}{1969}), \bibinfo{pages}{576--580}.
\newblock


\bibitem[\protect\citeauthoryear{Jansen, Dehnert, Kaminski, Katoen, and
  Westhofen}{Jansen et~al\mbox{.}}{2016}]%
        {bounded_model_checking_probabilistic_programs}
\bibfield{author}{\bibinfo{person}{Nils Jansen}, \bibinfo{person}{Christian
  Dehnert}, \bibinfo{person}{Benjamin~Lucien Kaminski},
  \bibinfo{person}{Joost-Pieter Katoen}, {and} \bibinfo{person}{Lukas
  Westhofen}.} \bibinfo{year}{2016}\natexlab{}.
\newblock \showarticletitle{Bounded Model Checking for Probabilistic Programs}.
  In \bibinfo{booktitle}{\emph{Automated Technology for Verification and
  Analysis}}, \bibfield{editor}{\bibinfo{person}{Cyrille Artho},
  \bibinfo{person}{Axel Legay}, {and} \bibinfo{person}{Doron Peled}} (Eds.).
  \bibinfo{publisher}{Springer International Publishing},
  \bibinfo{address}{Cham}, \bibinfo{pages}{68--85}.
\newblock
\showISBNx{978-3-319-46520-3}


\bibitem[\protect\citeauthoryear{Kallmeyer and Maier}{Kallmeyer and
  Maier}{2013}]%
        {plcfrs}
\bibfield{author}{\bibinfo{person}{Laura Kallmeyer} {and}
  \bibinfo{person}{Wolfgang Maier}.} \bibinfo{year}{2013}\natexlab{}.
\newblock \showarticletitle{Data-driven parsing using probabilistic linear
  context-free rewriting systems}.
\newblock \bibinfo{journal}{\emph{Computational Linguistics}}
  \bibinfo{volume}{39}, \bibinfo{number}{1} (\bibinfo{year}{2013}),
  \bibinfo{pages}{87--119}.
\newblock


\bibitem[\protect\citeauthoryear{Kam and Ullman}{Kam and Ullman}{1976}]%
        {kam1976global}
\bibfield{author}{\bibinfo{person}{John~B Kam} {and} \bibinfo{person}{Jeffrey~D
  Ullman}.} \bibinfo{year}{1976}\natexlab{}.
\newblock \showarticletitle{Global data flow analysis and iterative
  algorithms}.
\newblock \bibinfo{journal}{\emph{Journal of the ACM (JACM)}}
  \bibinfo{volume}{23}, \bibinfo{number}{1} (\bibinfo{year}{1976}),
  \bibinfo{pages}{158--171}.
\newblock


\bibitem[\protect\citeauthoryear{Kobayashi, Dal~Lago, and Grellois}{Kobayashi
  et~al\mbox{.}}{2019}]%
        {phors}
\bibfield{author}{\bibinfo{person}{Naoki Kobayashi}, \bibinfo{person}{Ugo
  Dal~Lago}, {and} \bibinfo{person}{Charles Grellois}.}
  \bibinfo{year}{2019}\natexlab{}.
\newblock \showarticletitle{On the termination problem for probabilistic
  higher-order recursive programs}. In \bibinfo{booktitle}{\emph{2019 34th
  Annual ACM/IEEE Symposium on Logic in Computer Science (LICS)}}. IEEE,
  \bibinfo{pages}{1--14}.
\newblock


\bibitem[\protect\citeauthoryear{Kwiatkowska, Norman, and Parker}{Kwiatkowska
  et~al\mbox{.}}{2002}]%
        {prism}
\bibfield{author}{\bibinfo{person}{Marta Kwiatkowska}, \bibinfo{person}{Gethin
  Norman}, {and} \bibinfo{person}{David Parker}.}
  \bibinfo{year}{2002}\natexlab{}.
\newblock \showarticletitle{PRISM: Probabilistic symbolic model checker}. In
  \bibinfo{booktitle}{\emph{International Conference on Modelling Techniques
  and Tools for Computer Performance Evaluation}}. Springer,
  \bibinfo{pages}{200--204}.
\newblock


\bibitem[\protect\citeauthoryear{Kwiatkowska, Norman, and Parker}{Kwiatkowska
  et~al\mbox{.}}{2012}]%
        {prism_application_algorithm}
\bibfield{author}{\bibinfo{person}{M. Kwiatkowska}, \bibinfo{person}{G.
  Norman}, {and} \bibinfo{person}{D. Parker}.} \bibinfo{year}{2012}\natexlab{}.
\newblock \showarticletitle{Probabilistic Verification of Herman's
  Self-Stabilisation Algorithm}.
\newblock \bibinfo{journal}{\emph{Formal Aspects of Computing}}
  \bibinfo{volume}{24}, \bibinfo{number}{4} (\bibinfo{year}{2012}),
  \bibinfo{pages}{661--670}.
\newblock


\bibitem[\protect\citeauthoryear{Kwiatkowska, Norman, and Parker}{Kwiatkowska
  et~al\mbox{.}}{2018}]%
        {prism_2018}
\bibfield{author}{\bibinfo{person}{Marta Kwiatkowska}, \bibinfo{person}{Gethin
  Norman}, {and} \bibinfo{person}{David Parker}.}
  \bibinfo{year}{2018}\natexlab{}.
\newblock \showarticletitle{Probabilistic model checking: Advances and
  applications}.
\newblock \bibinfo{journal}{\emph{Formal System Verification: State-of the-Art
  and Future Trends}} (\bibinfo{year}{2018}), \bibinfo{pages}{73--121}.
\newblock


\bibitem[\protect\citeauthoryear{Li, Murawski, and Ong}{Li
  et~al\mbox{.}}{2022}]%
        {li2022probabilistic}
\bibfield{author}{\bibinfo{person}{Guanyan Li}, \bibinfo{person}{Andrzej
  Murawski}, {and} \bibinfo{person}{Luke Ong}.}
  \bibinfo{year}{2022}\natexlab{}.
\newblock \showarticletitle{Probabilistic Verification Beyond
  Context-Freeness}. In \bibinfo{booktitle}{\emph{Proceedings of the 37th
  Annual ACM/IEEE Symposium on Logic in Computer Science}}.
  \bibinfo{pages}{1--13}.
\newblock


\bibitem[\protect\citeauthoryear{McIver, Morgan, and Morgan}{McIver
  et~al\mbox{.}}{2005}]%
        {mciver2005abstraction}
\bibfield{author}{\bibinfo{person}{Annabelle McIver}, \bibinfo{person}{Carroll
  Morgan}, {and} \bibinfo{person}{Charles~Carroll Morgan}.}
  \bibinfo{year}{2005}\natexlab{}.
\newblock \bibinfo{booktitle}{\emph{Abstraction, refinement and proof for
  probabilistic systems}}.
\newblock \bibinfo{publisher}{Springer Science \& Business Media}.
\newblock


\bibitem[\protect\citeauthoryear{McMillan}{McMillan}{2006}]%
        {craig_interpolation}
\bibfield{author}{\bibinfo{person}{Kenneth~L McMillan}.}
  \bibinfo{year}{2006}\natexlab{}.
\newblock \showarticletitle{Lazy abstraction with interpolants}. In
  \bibinfo{booktitle}{\emph{International Conference on Computer Aided
  Verification}}. Springer, \bibinfo{pages}{123--136}.
\newblock


\bibitem[\protect\citeauthoryear{Mitzenmacher and Upfal}{Mitzenmacher and
  Upfal}{2017}]%
        {efficiency_of_probabilistic_algorithm_1}
\bibfield{author}{\bibinfo{person}{Michael Mitzenmacher} {and}
  \bibinfo{person}{Eli Upfal}.} \bibinfo{year}{2017}\natexlab{}.
\newblock \bibinfo{booktitle}{\emph{Probability and computing: Randomization
  and probabilistic techniques in algorithms and data analysis}}.
\newblock \bibinfo{publisher}{Cambridge university press}.
\newblock


\bibitem[\protect\citeauthoryear{Moosbrugger, Bartocci, Katoen, and
  Kov{\'a}cs}{Moosbrugger et~al\mbox{.}}{2021}]%
        {moosbrugger2021probabilistic}
\bibfield{author}{\bibinfo{person}{Marcel Moosbrugger}, \bibinfo{person}{Ezio
  Bartocci}, \bibinfo{person}{Joost-Pieter Katoen}, {and}
  \bibinfo{person}{Laura Kov{\'a}cs}.} \bibinfo{year}{2021}\natexlab{}.
\newblock \showarticletitle{The probabilistic termination tool amber}. In
  \bibinfo{booktitle}{\emph{International Symposium on Formal Methods}}.
  Springer, \bibinfo{pages}{667--675}.
\newblock


\bibitem[\protect\citeauthoryear{Motwani and Raghavan}{Motwani and
  Raghavan}{1995}]%
        {efficiency_of_probabilistic_algorithm_2}
\bibfield{author}{\bibinfo{person}{Rajeev Motwani} {and}
  \bibinfo{person}{Prabhakar Raghavan}.} \bibinfo{year}{1995}\natexlab{}.
\newblock \bibinfo{booktitle}{\emph{Randomized algorithms}}.
\newblock \bibinfo{publisher}{Cambridge university press}.
\newblock


\bibitem[\protect\citeauthoryear{Møller}{Møller}{2012}]%
        {brics_automaton}
\bibfield{author}{\bibinfo{person}{Anders Møller}.}
  \bibinfo{year}{2012}\natexlab{}.
\newblock \bibinfo{title}{dk.brics.automaton}.
\newblock
\newblock
\urldef\tempurl%
\url{https://www.brics.dk/automaton/}
\showURL{%
\tempurl}


\bibitem[\protect\citeauthoryear{Norman, Parker, Kwiatkowska, Shukla, and
  Gupta}{Norman et~al\mbox{.}}{2003}]%
        {prism_application_industry}
\bibfield{author}{\bibinfo{person}{G. Norman}, \bibinfo{person}{D. Parker},
  \bibinfo{person}{M. Kwiatkowska}, \bibinfo{person}{S. Shukla}, {and}
  \bibinfo{person}{R. Gupta}.} \bibinfo{year}{2003}\natexlab{}.
\newblock \showarticletitle{Using Probabilistic Model Checking for Dynamic
  Power Management}. In \bibinfo{booktitle}{\emph{Proc. 3rd Workshop on
  Automated Verification of Critical Systems (AVoCS'03)}}
  \emph{(\bibinfo{series}{Technical Report DSSE-TR-2003-2, University of
  Southampton})}, \bibfield{editor}{\bibinfo{person}{M.~Leuschel},
  \bibinfo{person}{S.~Gruner}, {and} \bibinfo{person}{S.~Lo Presti}} (Eds.).
  \bibinfo{pages}{202--215}.
\newblock


\bibitem[\protect\citeauthoryear{Pierce, de~Amorim, Casinghino, Gaboardi,
  Greenberg, Hriţcu, Sjöberg, and Yorgey}{Pierce et~al\mbox{.}}{2022}]%
        {software_foundations_1}
\bibfield{author}{\bibinfo{person}{Benjamin~C. Pierce},
  \bibinfo{person}{Arthur~Azevedo de Amorim}, \bibinfo{person}{Chris
  Casinghino}, \bibinfo{person}{Marco Gaboardi}, \bibinfo{person}{Michael
  Greenberg}, \bibinfo{person}{Cătălin Hriţcu}, \bibinfo{person}{Vilhelm
  Sjöberg}, {and} \bibinfo{person}{Brent Yorgey}.}
  \bibinfo{year}{2022}\natexlab{}.
\newblock \bibinfo{booktitle}{\emph{Logical Foundations}}.
  \bibinfo{series}{Software Foundations}, Vol.~\bibinfo{volume}{1}.
\newblock \bibinfo{publisher}{Electronic textbook}.
\newblock


\bibitem[\protect\citeauthoryear{Rabin}{Rabin}{1976}]%
        {efficiency_of_probabilistic_algorithm_3}
\bibfield{author}{\bibinfo{person}{Michael~O Rabin}.}
  \bibinfo{year}{1976}\natexlab{}.
\newblock \bibinfo{booktitle}{\emph{Probabilistic Algorithms Algorithms and
  Complexity: New Directions and Recent Results}}.
\newblock \bibinfo{publisher}{Academic Press New York}.
\newblock


\bibitem[\protect\citeauthoryear{Rothenberg, Dietsch, and Heizmann}{Rothenberg
  et~al\mbox{.}}{2018}]%
        {trace_abstraction_incremental}
\bibfield{author}{\bibinfo{person}{Bat-Chen Rothenberg},
  \bibinfo{person}{Daniel Dietsch}, {and} \bibinfo{person}{Matthias Heizmann}.}
  \bibinfo{year}{2018}\natexlab{}.
\newblock \showarticletitle{Incremental Verification Using Trace Abstraction}.
  In \bibinfo{booktitle}{\emph{Static Analysis}},
  \bibfield{editor}{\bibinfo{person}{Andreas Podelski}} (Ed.).
  \bibinfo{publisher}{Springer International Publishing},
  \bibinfo{address}{Cham}, \bibinfo{pages}{364--382}.
\newblock
\showISBNx{978-3-319-99725-4}


\bibitem[\protect\citeauthoryear{Sankaranarayanan, Chakarov, and
  Gulwani}{Sankaranarayanan et~al\mbox{.}}{2013}]%
        {10.1145/2491956.2462179}
\bibfield{author}{\bibinfo{person}{Sriram Sankaranarayanan},
  \bibinfo{person}{Aleksandar Chakarov}, {and} \bibinfo{person}{Sumit
  Gulwani}.} \bibinfo{year}{2013}\natexlab{}.
\newblock \showarticletitle{Static analysis for probabilistic programs:
  inferring whole program properties from finitely many paths}. In
  \bibinfo{booktitle}{\emph{Proceedings of the 34th ACM SIGPLAN Conference on
  Programming Language Design and Implementation}} (Seattle, Washington, USA)
  \emph{(\bibinfo{series}{PLDI '13})}. \bibinfo{publisher}{Association for
  Computing Machinery}, \bibinfo{address}{New York, NY, USA},
  \bibinfo{pages}{447–458}.
\newblock
\showISBNx{9781450320146}
\urldef\tempurl%
\url{https://doi.org/10.1145/2491956.2462179}
\showDOI{\tempurl}


\bibitem[\protect\citeauthoryear{Santos}{Santos}{1969}]%
        {prob_turing_complete}
\bibfield{author}{\bibinfo{person}{Eugene~S Santos}.}
  \bibinfo{year}{1969}\natexlab{}.
\newblock \showarticletitle{Probabilistic Turing machines and computability}.
\newblock \bibinfo{journal}{\emph{Proceedings of the American mathematical
  Society}} \bibinfo{volume}{22}, \bibinfo{number}{3} (\bibinfo{year}{1969}),
  \bibinfo{pages}{704--710}.
\newblock


\bibitem[\protect\citeauthoryear{Smith, Hsu, and Albarghouthi}{Smith
  et~al\mbox{.}}{2019}]%
        {trace_abstraction_modulo_probability}
\bibfield{author}{\bibinfo{person}{Calvin Smith}, \bibinfo{person}{Justin Hsu},
  {and} \bibinfo{person}{Aws Albarghouthi}.} \bibinfo{year}{2019}\natexlab{}.
\newblock \showarticletitle{Trace abstraction modulo probability}.
\newblock \bibinfo{journal}{\emph{Proceedings of the ACM on Programming
  Languages}} \bibinfo{volume}{3}, \bibinfo{number}{POPL}
  (\bibinfo{year}{2019}), \bibinfo{pages}{1--31}.
\newblock


\bibitem[\protect\citeauthoryear{Tolpin, van~de Meent, Yang, and Wood}{Tolpin
  et~al\mbox{.}}{2016}]%
        {ppl_anglican}
\bibfield{author}{\bibinfo{person}{David Tolpin}, \bibinfo{person}{Jan~Willem
  van~de Meent}, \bibinfo{person}{Hongseok Yang}, {and} \bibinfo{person}{Frank
  Wood}.} \bibinfo{year}{2016}\natexlab{}.
\newblock \showarticletitle{Design and Implementation of Probabilistic
  Programming Language Anglican}.
\newblock \bibinfo{journal}{\emph{arXiv preprint arXiv:1608.05263}}
  (\bibinfo{year}{2016}).
\newblock


\bibitem[\protect\citeauthoryear{Wang, Hoffmann, and Reps}{Wang
  et~al\mbox{.}}{2018}]%
        {pmaf}
\bibfield{author}{\bibinfo{person}{Di Wang}, \bibinfo{person}{Jan Hoffmann},
  {and} \bibinfo{person}{Thomas Reps}.} \bibinfo{year}{2018}\natexlab{}.
\newblock \showarticletitle{PMAF: an algebraic framework for static analysis of
  probabilistic programs}.
\newblock \bibinfo{journal}{\emph{ACM SIGPLAN Notices}} \bibinfo{volume}{53},
  \bibinfo{number}{4} (\bibinfo{year}{2018}), \bibinfo{pages}{513--528}.
\newblock


\bibitem[\protect\citeauthoryear{Wang, Sun, Fu, Chatterjee, and
  Goharshady}{Wang et~al\mbox{.}}{2021}]%
        {exponential_analysis_of_probabilistic_program}
\bibfield{author}{\bibinfo{person}{Jinyi Wang}, \bibinfo{person}{Yican Sun},
  \bibinfo{person}{Hongfei Fu}, \bibinfo{person}{Krishnendu Chatterjee}, {and}
  \bibinfo{person}{Amir~Kafshdar Goharshady}.} \bibinfo{year}{2021}\natexlab{}.
\newblock \showarticletitle{Quantitative analysis of assertion violations in
  probabilistic programs}. In \bibinfo{booktitle}{\emph{Proceedings of the 42nd
  ACM SIGPLAN International Conference on Programming Language Design and
  Implementation}}. \bibinfo{pages}{1171--1186}.
\newblock


\bibitem[\protect\citeauthoryear{Wang, Yang, Fu, Li, and Ong}{Wang
  et~al\mbox{.}}{2024}]%
        {wang2024static}
\bibfield{author}{\bibinfo{person}{Peixin Wang}, \bibinfo{person}{Tengshun
  Yang}, \bibinfo{person}{Hongfei Fu}, \bibinfo{person}{Guanyan Li}, {and}
  \bibinfo{person}{C-H~Luke Ong}.} \bibinfo{year}{2024}\natexlab{}.
\newblock \showarticletitle{Static posterior inference of Bayesian
  probabilistic programming via polynomial solving}.
\newblock \bibinfo{journal}{\emph{Proceedings of the ACM on Programming
  Languages}} \bibinfo{volume}{8}, \bibinfo{number}{PLDI}
  (\bibinfo{year}{2024}), \bibinfo{pages}{1361--1386}.
\newblock


\bibitem[\protect\citeauthoryear{Yim}{Yim}{2016}]%
        {yim2016design}
\bibfield{author}{\bibinfo{person}{Jaegeol Yim}.}
  \bibinfo{year}{2016}\natexlab{}.
\newblock \showarticletitle{Design of a smart coupon system}.
\newblock \bibinfo{journal}{\emph{International Journal of Multimedia and
  Ubiquitous Engineering}} \bibinfo{volume}{11}, \bibinfo{number}{3}
  (\bibinfo{year}{2016}), \bibinfo{pages}{187--198}.
\newblock


\bibitem[\protect\citeauthoryear{Zaiser, Murawski, and Ong}{Zaiser
  et~al\mbox{.}}{2025}]%
        {zaiser2025guaranteed}
\bibfield{author}{\bibinfo{person}{Fabian Zaiser}, \bibinfo{person}{Andrzej~S
  Murawski}, {and} \bibinfo{person}{C-H~Luke Ong}.}
  \bibinfo{year}{2025}\natexlab{}.
\newblock \showarticletitle{Guaranteed Bounds on Posterior Distributions of
  Discrete Probabilistic Programs with Loops}.
\newblock \bibinfo{journal}{\emph{Proceedings of the ACM on Programming
  Languages}} \bibinfo{volume}{9}, \bibinfo{number}{POPL}
  (\bibinfo{year}{2025}), \bibinfo{pages}{1104--1135}.
\newblock


\end{thebibliography}


\appendix

\section*{Supplementary Materials}

\section{More on Programs and PCFA}

\subsection{Program}

The program $P$ studied in our paper is essentially standard, 
which is recursively given by the following production rules:
\begin{align*}
 P := \ & X \opAssn E \\
	\ | \ & P \oplus P \\
	\ | \ & P \circledast P \\
	\ | \ & \stskip \\
	\ | \ & P; P \\
	\ | \ & \texttt{if} \ B \ \texttt{then} \ P \ \texttt{else} \ P \\
	\ | \ & \texttt{while} \ B \ \texttt{do} \ P \ \texttt{done}
\end{align*}

\subsection{Evaluation of Statements}

The definition to the evaluation of a statement $\sigma$ over a valuation $v$, 
denoted by $\interpret{\sigma}_v$, 
is given by that:
for $v = \Undefined$, for all $\sigma$, $\interpret{\sigma}_\Undefined := \Undefined$,
while for $v \neq \Undefined$:
\[
  \interpret{\stskip}_v := v 
  \qquad 
  \interpret{\dprobbranch}_v := v
  \qquad
  \interpret{*_i}_v := v
\]
where $d$ is either $\texttt{L}$ or $\texttt{R}$, and:
\[
  \interpret{X \opAssn E}_v := v[X \mapsto \interpret{E}_v]
  \quad
  \interpret{\assume~B}_v := 
  \begin{cases}
    v & \interpret{B}_v = \tTrue \\
    \Undefined & \interpret{B}_v = \tFalse
  \end{cases}
\]

\subsection{Conversion from Program to PCFA}

The conversion from program to PCFA is also standard.
Here we present the formal method.
For simplicity, we assume the tags for distribution tags and non-deterministic statements are all natural numbers.
One may easily injectively map the natural numbers to distinct elements of any set of tags.

\def\conv{\textsf{Conv}}
We first define a support function $\conv(\ell_e, P, i_d, i_n)$, which returns a triple $(A, i_d', i_n')$ where:
\begin{itemize}
  \item $\ell_0$ is a given location to serve as the initial location for the return PCFA.\@
  \item $\ell_e$ is a given location to serve as the ending location for the return PCFA.\@
  \item $P$ is the program to convert.
  \item $i_d$ is the next number to serve as the distribution tag.
  \item $i_n$ is the next number to serve as the non-deterministic tag.
  \item $i_d'$ is the new next number to serve as the distribution tag after converting $P$.
  \item $i_d'$ is the new next number to serve as the non-deterministic tag after converting $P$.
\end{itemize}

\def\newLoc{\textsf{NewLoc}}
Before giving the formal definition, we present some supportive functions:
the function $\newLoc()$ returns a distinct new location, 
we denote a creation of a new location $\ell$ by:
$\ell \leftarrow \newLoc()$.

The definition of $\conv(\ell_0, \ell_e, P, i_d, i_n)$ is recursively given by a case analysis on the kind of $P$:
\begin{itemize}
  \item When $ P = \stskip $, 
  returns a PCFA $(L, \Sigma, \Delta, \ell_0, \ell_e)$ with the original counters $i_d$ and $i_n$ where:
  \begin{align*}
    L & := \bs{\ell_0, \ell_e} \\
    \Sigma & := \bs{\stskip} \\
    \Delta & := \bs{\ell_0 \xrightarrow{\stskip} \ell_e}
  \end{align*}

  \item When $ P = X \opAssn E $, 
  returns a PCFA $(L, \Sigma, \Delta, \ell_0, \ell_e)$ with the original counters $i_d$ and $i_n$ where:
  \begin{align*}
    L & := \bs{\ell_0, \ell_e} \\
    \Sigma & := \bs{X \opAssn E} \\
    \Delta & := \bs{\ell_0 \xrightarrow{X \opAssn E} \ell_e}
  \end{align*}

  \item When $ P = P_1 \oplus P_2 $,
  create two new locations $\ell_0^1 \leftarrow \newLoc()$ and $\ell_0^2 \leftarrow \newLoc()$,
  let the results from $\conv(\ell_0^1, \ell_e, P_1, i_d, i_n)$ be $(A_1 (L_1, \Sigma_1, \Delta_1, \ell_0^1, \ell_e), \colorblock{blue}{i_d'}, \colorblock{blue}{i_n'})$
  and the results from $\conv(\ell_0^2, \ell_e, P_2, \colorblock{blue}{i_d'}, \colorblock{blue}{i_n'})$ be $(A_2 (L_2, \Sigma_2, \Delta_2, \ell_0^2, \ell_e), \colorblock{teal}{i_d''}, \colorblock{teal}{i_n''})$,
  returns a PCFA $(L, \Sigma, \Delta, \ell_0, \ell_e)$ along with the new counters $(\colorblock{teal}{i_d''} + 1)$ and $\colorblock{teal}{i_n''}$, where:
  \begin{align*}
    L & := L_1 \cup L_2 \uplus \bs{\ell_0} \\
    \Sigma & := \Sigma_1 \uplus \Sigma_2 \uplus \bs{\lprobbranch[\colorblock{teal}{i_d''}], \rprobbranch[\colorblock{teal}{i_d''}]} \\
    \Delta & := \Delta_1 \uplus \Delta_2 \uplus \bs{\begin{matrix}
      \ell_0 \xrightarrow{\lprobbranch[\colorblock{teal}{i_d''}]} \ell_0^1
     \\
      \ell_0 \xrightarrow{\rprobbranch[\colorblock{teal}{i_d''}]} \ell_0^2
    \end{matrix}}
  \end{align*}

  \item When $ P = P_1 \circledast P_2 $, 
  create two new locations $\ell_0^1 \leftarrow \newLoc()$ and $\ell_0^2 \leftarrow \newLoc()$,
  let the results from $\conv(\ell_0^1, \ell_e, P_1, i_d, i_n)$ be $(A_1 (L_1, \Sigma_1, \Delta_1, \ell_0^1, \ell_e), \colorblock{blue}{i_d'}, \colorblock{blue}{i_n'})$
  and the results from $\conv(\ell_0^2, \ell_e, P_2, \colorblock{blue}{i_d'}, \colorblock{blue}{i_n'})$ be $(A_2 (L_2, \Sigma_2, \Delta_2, \ell_0^2, \ell_e), \colorblock{teal}{i_d''}, \colorblock{teal}{i_n''})$,
  returns a PCFA $(L, \Sigma, \Delta, \ell_0, \ell_e)$ along with the new counters $\colorblock{teal}{i_d''}$ and $(\colorblock{teal}{i_n''} + 2)$, where:
  \begin{align*}
    L & := L_1 \cup L_2 \uplus \bs{\ell_0} \\
    \Sigma & := \Sigma_1 \cup \Sigma_2 \uplus \bs{*_{\colorblock{teal}{i_n''}}, *_{(\colorblock{teal}{i_d''} + 1)}} \\
    \Delta & := \Delta_1 \uplus \Delta_2 \uplus \bs{\begin{matrix}
      \ell_0 \xrightarrow{*_{\colorblock{teal}{i_n''}}} \ell_0^1
     \\
      \ell_0 \xrightarrow{*_{(\colorblock{teal}{i_d''} + 1)}} \ell_0^2
    \end{matrix}}
  \end{align*}

  \item When $ P = P_1; P_2 $, 
  create a new location $\colorblock{violet}{\ell_0'} \leftarrow \newLoc()$,
  let the results from $\conv(\ell_0, \ell_0', P_1, i_d, i_n)$ be $(A_1 (L_1, \Sigma_1, \Delta_1, \ell_0, \colorblock{violet}{\ell_0'}), \colorblock{blue}{i_d'}, \colorblock{blue}{i_n'})$
  and the results from $\conv(\colorblock{violet}{\ell_0'}, \ell_e, P_2, \colorblock{blue}{i_d'}, \colorblock{blue}{i_n'})$ be $(A_2 (L_2, \Sigma_2, \Delta_2, \ell_0', \ell_e), \colorblock{teal}{i_d''}, \colorblock{teal}{i_n''})$,
  returns a PCFA $(L, \Sigma, \Delta, \ell_0, \ell_e)$ along with the new counters $\colorblock{teal}{i_d''}$ and $\colorblock{teal}{i_n''}$, where:
  \begin{align*}
    L & := L_1 \cup L_2 \\
    \Sigma & := \Sigma_1 \cup \Sigma_2 \\
    \Delta & := \Delta_1 \uplus \Delta_2
  \end{align*}

  \item When $ P = \texttt{if} \ B \ \texttt{then} \ P_1 \ \texttt{else} \ P_2 $, 
  create two new locations $\ell_0^1 \leftarrow \newLoc()$ and $\ell_0^2 \leftarrow \newLoc()$,
  let the results from $\conv(\ell_0^1, \ell_e, P_1, i_d, i_n)$ be $(A_1 (L_1, \Sigma_1, \Delta_1, \ell_0^1, \ell_e), \colorblock{blue}{i_d'}, \colorblock{blue}{i_n'})$
  and the results from $\conv(\ell_0^2, \ell_e, P_2, \colorblock{blue}{i_d'}, \colorblock{blue}{i_n'})$ be $(A_2 (L_2, \Sigma_2, \Delta_2, \ell_0^2, \ell_e), \colorblock{teal}{i_d''}, \colorblock{teal}{i_n''})$,
  returns a PCFA $(L, \Sigma, \Delta, \ell_0, \ell_e)$ along with the new counters $\colorblock{teal}{i_d''}$ and $\colorblock{teal}{i_n''}$, where:
  \begin{align*}
    L & := L_1 \cup L_2 \uplus \bs{\ell_0} \\
    \Sigma & := \Sigma_1 \cup \Sigma_2 \cup \bs{\assume~B, \assume~\neg B} \\
    \Delta & := \Delta_1 \uplus \Delta_2 \uplus \bs{\begin{matrix}
      \ell_0 \xrightarrow{\assume~B} \ell_0^1
     \\
      \ell_0 \xrightarrow{\assume~\neg B} \ell_0^2
    \end{matrix}}
  \end{align*}
  
  \item When $ P = \texttt{while} \ B \ \texttt{do} \ P' \ \texttt{done} $, 
  create a new location $\colorblock{olive}{\ell_0'} \leftarrow \newLoc()$,
  let the results from $\conv(\colorblock{olive}{\ell_0'}, \colorblock{orange}{\ell_0}, P', i_d, i_n)$ be
  $(A (L', \Sigma', \Delta', \colorblock{olive}{\ell_0'}, \colorblock{orange}{\ell_0}), i_d', i_n')$,
  then returns a PCFA $(L, \Sigma, \Delta, \ell_0, \ell_e)$ with the new counters $i_d'$ and $i_n'$, where:
  \begin{align*}
    L & := L' \uplus \bs{\ell_e} \\
    \Sigma & := \Sigma_1 \cup \Sigma_2 \cup \bs{\assume~B, \assume~\neg B} \\
    \Delta & := \Delta_1 \uplus \Delta_2 \uplus \bs{\begin{matrix}
      \ell_0 \xrightarrow{\assume~B} \colorblock{olive}{\ell_0'}
     \\
      \ell_0 \xrightarrow{\assume~\neg B} \ell_e
    \end{matrix}}
  \end{align*}
\end{itemize}

So, given a program $P$, 
create two new locations 
$\ell_0 \leftarrow \newLoc()$ and 
$\ell_e \leftarrow \newLoc()$,
let the result from $\conv(\ell_0, \ell_e, P, 0, 0)$ be $(A, i_d, i_n)$,
the PCFA converted from $P$ is $A$.

\section{Proof to Theorem~\ref{theorem: two defs equal}}

\begin{proof}[\cref{eq: two defs equal}]
  Expand the equations in both defintions, we obtain:
  \begin{align*}
    & \sup_{v_\mathinit \models \varphi_e} 
    \sup_{\xi \in \Xi_A} 
    \sum_{\setdef{\pi \in \Pi_{v_\mathinit}^\xi}{\last(\pi) = (\ell_e, v), v \models \neg \postcond, \ell_e \in L_e}} wt(\pi) \\
    = &
    \sup_{v \models \varphi_e} 
    \sup_{\psi \in \scriptS(A)} 
    \sum_{\tau \in \scriptL(\applyPolicy{A}{\psi})} 
    \wt(\tau) \cdot \indicator{\interpret{\tau}(v) \models \neg \varphi_f}
  \end{align*}

  We then fix a specific initial valuation $v_\mathinit$ to prove that for all $v_\mathinit$, we have:
  \begin{align*}
    & \sup_{\xi \in \Xi_A} 
    \sum_{\setdef{\pi \in \Pi_{v_\mathinit}^\xi}{\last(\pi) = (\ell_e, v), v \models \neg \postcond, \ell_e \in L_e}} wt(\pi) \\
    =
    & \sup_{\psi \in \scriptS(A)} 
    \sum_{\tau \in \scriptL(\applyPolicy{A}{\psi})} 
    \wt(\tau) \cdot \indicator{\interpret{\tau}(v) \models \neg \varphi_f}
  \end{align*}

  \def\lhs{\mathit{LHS}}
  \def\rhs{\mathit{RHS}}

  We prove by firstly proving $\lhs \le \rhs$ and then $\rhs \le \lhs$.

  For $\lhs \le \rhs$, this can be given by that: 
  for every scheduler $\xi$, there exists a policy $\psi$ that can ``simulate'' it.
  This is seen by that:
  we enumerate the states involved in the run $\Pi_{v_\mathinit}^\xi$, and then select the action by the scheduler -- 
  note that the supportive domain of distribution selected by scheduler is the same as the actions.

  For $\rhs \le \lhs$, we go by showing that for every policy $\psi$, there is a sheduler $\xi$ such that the sum of the violating traces of the run guided by $\xi$ is \emph{equal} to the sum of violating traces in the applied structured by $\psi$.
  This can be shown by firstly observing that: for every policy $\psi$, there is a ``skeleton'' sub-policy $\psi_s$ (with policy states a subset of the original policy and the transition function a subset of the original policy) such that all traces in $\scriptL(\applyPolicy{A}{\psi_s})$ by initial valuation $v_\mathinit$ will not evaluate to $\Undefined$.
  Clearly, such a skeleton sub-policy induces exactly the same result as the previous one -- all the traces will be filtered out by the semantic check.
  Finally, observe that one can always construct a scheduler $\xi$ that simulates exactly such a skeleton sub-policy -- 
  by simply selects the distribution that is hinted by the application of the skeleton sub-policy.
\end{proof}

\section{Proof to Theorem~\ref{theorem: cross-structural equality}}

We in this section consider the general PCFA rather than the pure PCFA defined in~\cref{def: pcfa}.
Hence, in the following of this section, all words ``PCFA'' refers to ``general PCFA''.
As now PCFA is not necessarily deterministic, we call abbreviate deterministic (general) PCFA to be DPCFA.\@
In the following, we fix a pre- and a post-condition $\precond$ and $\postcond$.
The two are sometimes called a \emph{specification}, denoted by $\spec$.

\subsection{Normalisation}
\label{subsec: normalisation}

We first present an important concept of normalisation.

\begin{definition}[Normalised PCFA]
  A normalised PCFA is a DPCFA that additionally satisfies the condition that:
  every two probabilistic branches with the same distribution tag $i$, 
    that $\lprobbranch$ and $\rprobbranch$ in the same location \emph{point to different locations}.
\end{definition}

To reach this, a normalisation process $\fnorm{-}$ of a PCFA can be given by two steps, 
which is a usual minimalisation to automata followed by a step that consists of 
a loop of the following two components traversing all places violating the second normalisation condition
until a fixed point is reached:

\begin{enumerate}
	\item If the two branches all point to the their origin location, 
				say $\ell$, whose edge set except the current target two edges are called $S$, 
				we create a new location $\ell'$ with the modified transitions to be:
				$\ell \xrightarrow{\lprobbranch} \ell'$,
				$\ell \xrightarrow{\rprobbranch} \ell$ and
				$\ell' \xrightarrow{\rprobbranch} \ell$,
				$\ell' \xrightarrow{\lprobbranch} \ell'$.
				For other out edges, $S \setminus \bs{\lprobbranch, \rprobbranch}$,
				$\ell'$ just copy them from $\ell$;
	\item If the two branches all point to another location other than the origin, 
				we duplicate the target location and let one branch to instead point to that, 
        while copying other \emph{out} transitions.
\end{enumerate}

The loop of these two processes will eventually reach a fixed point, 
as procedure 1 will first erase all self loops
and procedure 2 will erase the non-self loops.
Also note that this process will not hamper both 
  the accepting trace set and 
  the determinism of the minimised automaton.
Hence this is a valid normalisation process -- as 
the first step guarantees the first two conditions of normalisation 
and the second step guarantees the final one.

A proof to the termination of such process can be proceeded by 
tagging each of the originally existed violations to the conditions with a unique tag, 
and during the copy of the nodes, the tags are copied.
Then, one can see that such tags are gradually erased without 
re-addition of erased ones or introduction of new ones.

\subsection{Probability from Trace View}

Another important supportive definition is the \emph{probability from trace view}.
The definition gives probability to a set of traces without referring to the host automaton,
which then may serve as the bridge between different structures.

The definition is given by the following two observations:

\begin{itemize}
  \item The key part of the definitions is the Markov chain to be extracted -- 
  a policy application involving the shape of the PCFA is simply for extracting a Markov chain.
  \item From the trace view, a Markov chain is just a special set of traces.
\end{itemize}

Hence, for a trace set, we just need to reach such special kind of traces as Markov chain directly.
In order to reach that, 
we first need to define the notion of \emph{mergeable}.
A set of paths $\Theta$ is called mergeable iff 
there exists a PCFA $\scriptM$ whose underlying MDP is a 
Markov chain such that $\Theta = \scriptL(\scriptM)$.
We call all mergeable subsets of a given path set $\Theta$ to be $\mergeable(\Theta)$.

So, given a set of paths $\Theta$
we define the violation probability of this set to be:
\begin{align}
	\traceProb[(\precond, \postcond)]{\Theta} := 
  \sup_{s \models \precond} 
  \sup_{\Pi \in \mergeable(\Theta)} 
  \sum_{\tau \in \Pi} 
  wt(\tau) \cdot [\interpret{\tau}(s) \models \neg \postcond]
	\label{eq: prob of trace set}
\end{align}

\subsection{Probability from Trace View Equals Probability of Normalised PCFA}

Then, we will need to prove that the probability from trace view equals the normalised PCFA.\@
So to start chaining the probabilities.
That is:
for all PCFA $A$:
\begin{align}
  \traceProb[(\precond, \postcond)]{\scriptL(A)}
  =
  \mdpProb{\fnorm{A}}{\precond}{\postcond}
  \label{eq: trace prob eq norm}
\end{align}

By definition, obviously, we have that:
\begin{align}
  \mdpProb{\fnorm{A}}{\precond}{\postcond}
  \le
  \traceProb[(\precond, \postcond)]{\scriptL(A)}
  \label{eq: norm le trace prob}
\end{align}

This is because every policy $\psi$ induces $\applyPolicy{fnorm{A}}{\psi}$, the set is clearly mergeable.

We then need to prove that the other direction also holds.
And we will have the following theorem to build the inequation from the opposite direction.
	
\begin{theorem}
	\label{theorem: subset CFMC implies policy}
	For any normalised PCFA $A$, 
  and any PCFA $\scriptM$ whose underlying MDP is a Markov chain, 
  if $\scriptL(\scriptM) \subseteq \scriptL(A)$, 
  then there exists a policy $\psi$ 
  such that $\scriptL(A^\psi) = \scriptL(\scriptM)$.
\end{theorem}

\begin{proof}
	Let $A = (L_A, \Sigma, \delta_A, \ell_0^A, L_e^A)$ 
  and $\scriptM = (L_\scriptM, \Sigma, \delta_\scriptM, \ell_0^\scriptM,L_e^\scriptM)$,
	where for simplicity, we assume $A$ and $\scriptM$ share the same label set.
  Also, as $\scriptM$ is deterministic, 
  we write 
  $\delta_\scriptM(\ell) = (\sigma, \ell')$ or 
  $\delta_\scriptM(\ell) = 
  (\lprobbranch[i], \ell') 
  \wedge 
  (\rprobbranch[i], \ell'')$ 
  to denote set elements.

	So we construct 
  \[ 
    \psi_\scriptM := 
    (
      L_\scriptM \uplus 
      \setdef{\ell_{(T, \ell, \ell')}}
      {T \in L_A \uplus \bs{\bot}, \ell, \ell' \in L_\scriptM \uplus \bs{\bot}}, 
      \delta, \ell_0^\scriptM
    )
  \]
  where:
  $\bot$ is a dummy symbol that is in neither $L_A$ nor $L_\scriptM$ 
  and $\delta$ is defined as the following.

  We first define for $\ell \in L_\scriptM$ a supporting function $\delta^-$ as:
	\[
    \delta^-(\ell_A, \ell) := 
    \begin{cases}
      (\sigma, \ell') & 
      \delta_\scriptM(\ell) = (\sigma, \ell'), \sigma \in \Sigma^- \\
      (i, \ell_{(T, \ell_2, \ell_3)}) & 
      \begin{matrix}
        \delta_\scriptM(\ell) = 
        (\lprobbranch[i], \ell_2) 
        \wedge 
        (\rprobbranch[i], \ell_3), 
        \\ 
        \ell_A \xrightarrow{\lprobbranch[i]} T \in \delta_A
      \end{matrix}
      \\
      (i, \ell_{(T, \ell', \bot)}) & 
      \begin{matrix}
        \delta_\scriptM(\ell) = (\lprobbranch[i], \ell'), \\
        \ell_A \xrightarrow{\lprobbranch[i]} T \in \delta_A
      \end{matrix} 
      \\
      (i, \ell_{(\bot, \bot, \ell')}) & 
      \delta_\scriptM(\ell) = (\rprobbranch[i], \ell')
    \end{cases}
	\]
	
	Then we define $\delta$ as:
	\[
    \delta(\ell_A, \ell) := 
    \begin{cases}
      \delta^-(\ell_A, \ell) & 
      \ell \in L_\scriptM 
      \\
      \delta^-(\ell_A, \ell') & 
      \ell = \ell_{(T, \ell', \_)}, \ell_A = T, \ell' \neq \bot 
      \\
      \delta^-(\ell_A, \ell') & 
      \ell = \ell_{(T, \_, \ell')}, \ell_A \neq T, \ell' \neq \bot
    \end{cases}
	\]
	
	Note that we use the next target location of $A$ 
  to help determine the next choice for the policy when encountering $\probbranchText$,
  which is where the normalisation assumption works -- that any $i$ from a single location, 
  $\lprobbranch$ and $\rprobbranch$ will point to different locations.

  \medskip

  \def\mc{{\scriptM}}
  \def\doubleT{{\mathbb{T}}}
  \def\resultStructure{{\applyPolicy{A}{\finiteMemoryPolicy[\mc]}}}
  \def\fms{\psi} 

  Then, we will need to verify the effectiveness of the construction.
	
	Finally, an induction on the length of the valid traces can be performed to show that 
  the induced CFMC of $A^{\psi_\scriptM}$ has the same accepting traces as $\scriptM$.
  Then, for any valid trace, if it is accepting in $\scriptM$,
  it must also be reaching an ending location in $\scriptA$ given the determinism,
  and it will also be in 
	Notice especially that, 
  the location $(T, \ell_{(T, \ell_1^\scriptM, \ell)})$ in 
  $A^{\psi_\scriptM}$ is effectively just $(T, \ell_1^\scriptM)$, 
  and that $(T, \ell_{(T', \ell, \ell_2^\scriptM)})$, where $T \neq T'$, 
  is effectively just $(T, \ell_2^\scriptM)$.
  The method is formally presented by:
  Let $A (L_A, \Sigma, \delta_A, \ell_0^A, L_e^A)$, $\mc (L_\mc, \Sigma, \delta_\mc, \ell_0^\mc, L_e^\mc)$,
and the resulting $\finiteMemoryPolicy[\mc]$ be:
\[
  \finiteMemoryPolicy[\scriptM] := 
  (
    L_\scriptM \uplus 
    \setdef{\ell_{(T, \ell, \ell')}}
    {T \in L_A \uplus \bs{\bot}, \ell, \ell' \in L_\scriptM \uplus \bs{\bot}}, 
    \delta_{\finiteMemoryPolicy}, \ell_0^\mc
  )
\] 

Let the set within be $\doubleT$, that is:
\[ 
  \doubleT := \setdef{\ell_{(T, \ell, \ell')}}
  {T \in L_A \uplus \bs{\bot}, \ell, \ell' \in L_\scriptM \uplus \bs{\bot}} 
\]

The the constructed $\applyPolicy{A}{\finiteMemoryPolicy[\mc]}$ is then given by:
\[ 
  (L_A \times (L_\mc \uplus \doubleT), 
  \Sigma, 
  \delta^\resultStructure, 
  (\ell_0^A, \ell_0^\mc), 
  L_e^A \times (L_\mc \uplus \doubleT)) 
\]

We will firstly need to show a lemma that: 

\begin{lemma}
  \label{theorem: trace label structure}
  For any valid trace $\tau$ of $\resultStructure$,
  for the reaching location $(\ell^A, \ell^\bullet)$
  that $(\ell_0^A, \ell_0^\mc) \xRightarrow{\tau} (\ell^A, \ell^\bullet)$,
  we have that $\ell^\bullet$ is one of the following:
  \begin{itemize}
    \item $\ell^\mc$, iff the trace $\tau$ is empty or the last statement is $\sigma \in \Sigma^-$.
    \item $\ell_{(\ell^A, \ell^\mc, \_)}$, iff $\tau$ is non-empty and the last statement is $\lprobbranch$ for some $i$.
    \item $\ell_{(\ell', \_, \ell_\mc)}$ with $\ell' \neq \ell^A$, iff $\tau$ is non-empty and the last statement is $\rprobbranch$ for some $i$.
  \end{itemize}
  We then call $\ell^\mc$ in the previous case the \emph{key location} of 
  $\ell^\bullet$ given the trace.
  And that, $\delta_\fms(\ell^A, \ell^\bullet) = \delta_\fms^-(\ell^A, \ell^\mc)$.
\end{lemma}
This can be done by a simple induction on the length of the valid traces of $\resultStructure$.

So we can show for the elements that any valid trace running at $\mc$ can be \emph{effectively}
run in $\resultStructure$.
This effectiveness says that:
for any valid trace $\tau$ in $\mc$ that $\ell_0^\mc \xRightarrow{\tau} \ell^\mc$,
we know that it must also be a valid trace for $A$ that,
$\ell_0^A \xRightarrow{\tau} \ell^A$,
then,
there it should also be a valid trace for $\applyPolicy{A}{\finiteMemoryPolicy[\mc]}$ that:
$(\ell_0^A, \ell_0^\mc) \xRightarrow{\tau} (\ell^A, \ell^\bullet)$
where 
$\ell^\bullet$ is one of the following:
\begin{itemize}
  \item $\ell^\mc$, if the trace $\tau$ is empty or the last statement is $\sigma \in \Sigma^-$.
  \item $\ell_{(\ell^A, \ell^\mc, \_)}$, if $\tau$ is non-empty and the last statement is $\lprobbranch$ for some $i$.
  \item $\ell_{(\ell', \_, \ell_\mc)}$ with $\ell' \neq \ell^A$, if $\tau$ is non-empty and the last statement is $\rprobbranch$ for some $i$.
\end{itemize}

We prove this property formally by induction on the length of $\tau$.
In the beginning, as the trace is empty, so the first case holds.
Then, in the inductive case, when $n$ holds, we prove for $n + 1$,
So, the previous segment can reach the place by the inductive condition.
What to cope is simply the last statement to enter the trace.

Assume the latest location from applying the inductive condition is: $(\ell_A, \ell^\bullet)$,
where the involving location of $\mc$ be $\ell_\mc$ by the whole trace.
And let the new statement to be $\sigma$, then there is a $\ell_A'$ such that:
$\ell_A \xrightarrow{\sigma} \ell_A'$.
So, we analyse the next statement to be appended to the trace $\sigma$,
let $\ell_\mc'$ be such that $\ell_\mc \xrightarrow{\sigma} \ell_\mc' \in \delta_\mc$.
In this case, by the proved lemma, we have that:
\[ \delta_\fms(\ell_A, \ell^\bullet) = \delta_\fms^-(\ell_A, \ell_\mc) \]
This can be easily obtained by a case analysis on the $n$ length trace (the prefix of $\tau$).
\begin{itemize}
  \item When it is a non-probabilistic statement $\sigma \in \Sigma^-$:
  $\delta_\fms(\ell_A, \ell^\bullet) = \delta_\fms^-(\ell_A, \ell_\mc) = (\sigma, \ell_\mc')$.
  \item When it is $\lprobbranch$ for some $i$:
  We have:
  \[ 
      \delta_\fms(\ell_A, \ell^\bullet) 
    = \delta_\fms^-(\ell_A, \ell_\mc)
    = (i, \ell_{(\ell_A', \ell_\mc', \_)}) 
  \]
  \item When it is $\rprobbranch$ for some $i$:
  Likewise, we have:
  \[ 
      \delta_\fms(\ell_A, \ell^\bullet) 
    = \delta_\fms^-(\ell_A, \ell_\mc)
    = (i, \ell_{(x, \ell_\mc', \_)}) 
  \]
  with $x \neq \ell_A'$.
\end{itemize}

Then, we will need to show that the backward also holds that:
given a valid trace $\tau$ of $\resultStructure$,
that:
$(\ell_0^A, \ell_0^\mc) \xRightarrow{\tau} (\ell^A, \ell^\bullet)$,
and the key location of $\ell^\bullet$ given the trace is $\ell^\mc$
it is also a valid trace in $A$ and $\scriptM$ that:
$\ell_0^A \xRightarrow{\tau} \ell^A$
and
$\ell_0^\mc \xRightarrow{\tau} \ell^M$.

We again prove the theorem by
induction on the length of $\tau$.
By \cref{theorem: trace label structure}, this is not hard.
\end{proof}

That is to say:

\begin{theorem}
	\label{theorem: subset CFMC and policy isomorphism}
	For any normalised PCFA $A$, and any Markov chain $\scriptM$, 
  if $\scriptL(\scriptM) \subseteq \scriptL(A)$, 
  then there exists a policy $\psi$ 
  such that $\scriptL(A^\psi) = \scriptL(\scriptM)$.
\end{theorem}

\subsection{Equality Between Normalised PCFA}

We then show a restricted version of the cross-structural theorem that:

\begin{lemma}
  For any two PCFA $A_1$ and $A_2$ with the same set of violating traces with respect to the given pre- and post-conditions $\precond$ and $\postcond$, we have:
  \rm
  \begin{align*}
    \mdpProb{\fnorm{A_1}}{\precond}{\postcond}
    =
    \mdpProb{\fnorm{A_2}}{\precond}{\postcond}
  \end{align*}
\end{lemma}

\begin{proof}
  We first prove an additional theorem as the last piece to prove the theorem.
  We formalise the observation about 
  the trivial fact that only the violating traces
  will contribute to the probability from trace view.
  We denote the violating trace set of a set $\Theta$ wrt a specification $\spec = (\precond, \postcond)$
  to be:
  
  \[
    \filterViolating\Theta\spec := 
    \setdef
      {\tau \in \Theta}
      {\exists s \models \precond.\interpret{\tau}_s \models \neg \postcond} 
  \]

  \begin{lemma}
    \label{theorem: only violating contributes to probability}
    Given a set of traces $\Theta$ and a specification $\spec = (\precond, \postcond)$, we have:
    \[
      \traceProb[\spec]{\Theta} = 
      \traceProb[\spec]{\filterViolating{\Theta}{\spec}} 
    \]
  \end{lemma}
  \begin{proof}
    Straightforward from definition.
  \end{proof}

  Then, we have the following equation chain:
  \begin{align*}
      &\ \mdpProb{\fnorm{A_1}}{\precond}{\postcond} & \\
    = &\ \traceProb[\spec]{\traceSet{A_1}} & (\text{\cref{eq: trace prob eq norm}}) \\
    = &\ \traceProb[\spec]{\filterViolating{\traceSet{A_1}}{\spec}} & 
      (\text{\cref{theorem: only violating contributes to probability}}) \\
    = &\ \traceProb[\spec]{\filterViolating{\traceSet{A_2}}{\spec}} & (\text{the given condition}) \\
    = &\ \traceProb[\spec]{\traceSet{A_2}} & 
      (\text{\cref{theorem: only violating contributes to probability}}) \\
    = &\ \mdpProb{\fnorm{A_2}}{\precond}{\postcond} & (\text{\cref{eq: trace prob eq norm}})
  \end{align*}
\end{proof}

\subsection{Normalised PCFA Equals Deterministic PCFA}

We then need to show that:
\begin{align}
  \mdpProb{\fnorm{A}}{\precond}{\postcond}
  =
  \mdpProb{\determinise{A}}{\precond}{\postcond}
  \label{eq: normalise equals determinise}
\end{align}

We will firstly need to show that all kinds of normalisation method will lead to the same structural value.

\begin{lemma}
  \label{theorem: normalisation has same value}
  All normalised PCFA that have the same set of traces will have the same structural probability.
\end{lemma}
\begin{proof}
  This can be done directly via the \cref{eq: trace prob eq norm},
  as the trace set probability is well-defined and unique.
\end{proof}

The above lemma means that for any kinds of normalisation method,
we will have the same result,
as long as the final one is normalised.
Hence, all we need to do is just to prove for 
\emph{one} specific normalisation method.
We then fix the method introduced previously in 
\cref{subsec: normalisation}.

And then we prove that:

\begin{lemma}
  \label{theorem: norm one step does not change}
  Every step of the normalisation procedure adotped will not change the structural probability.
\end{lemma}
\begin{proof}
This is going by adopting the \emph{equation system} for characterising the 
maximum reachability probability of an MDP,
which is exactly the structural probability.
It is a system for each of the nodes.
Given a DPCFA $A (L, \Sigma, \delta, \ell_0, L_e)$.
We will have the equation system for each location $\ell$ that:
\[
  \ell = \max 
  \setdef{\ell'}
  {\ell \xrightarrow{\sigma} \ell', \sigma \in \Sigma^-} 
  \uplus
  \setdef{\frac12 \ell'}{
    \ell \xrightarrow{\dprobbranch} \ell' \in \delta 
    \\ 
    \forall \ell''.\ell \xrightarrow{\probbranch{i}{\overline{d}}} \ell'' \notin \delta
  }
  \uplus
  \setdef{\frac12 \ell_1 + \frac12 \ell_2}{
    \ell \xrightarrow{\lprobbranch} \ell_1 \in \delta
    \\
    \ell \xrightarrow{\rprobbranch} \ell_2 \in \delta
  }
\]

Details of the original equation system can be found in~\cite{principles_of_model_checking}.

Then, we can see directly from such equation system that the split / duplication of the nodes
does not change the value of the final results at all.
To take a closer look, we anlyse the cases 
by the different situation.
For the former case in the normalisation in \cref{subsec: normalisation},
that the nodes are self-loop.
Let the node be $x$ and its new duplicating one be $x'$,
so we the equation system like:
\[
  \begin{matrix}
    x' & = & \max \bs{\frac12 x + \frac12 x', \dots} 
    \\
    x & = & \max \bs{\frac12 x + \frac12 x', \dots} 
    \\
    \dots
  \end{matrix}
\]
where by definition in \cref{subsec: normalisation}, 
the other parts of $x$ and $x'$ is not going to change.
So, we have that $x = x'$.
By applying this conclusion to the system,
we are able to safely remove the new equation for $x'$ and modify that of $x$,
which gives us the system like:
\[
  \begin{matrix}
    x & = & \max \bs{\frac12 x + \frac12 x, \dots} 
    \\
    \dots
  \end{matrix}
\]
This is exactly that of the original one 
before performing the normalisation step.
This guarantees the unchange of the result for the first step.

Then, for the second step,
let what the node to be duplicated be $y$ 
and the duplicated one be $y'$
and the source node that requires $y$ to be duplicated be $x$.
Note that, after the erasure phase,
only $x$ will be able to access $y'$.
Hence the other parts are unchanged.
Then, we have the equation system like:
\[
  \begin{matrix}
    x & = & \max \bs{\frac12 y + \frac12 y', \dots}
    \\
    y & = & \dots
    \\
    y' & = & \dots
    \\
    \dots
  \end{matrix}
\]
as the transitions are totally copied from $y$ to $y'$,
we have that $y = y'$,
which let us safely remove $y'$ and obtain a new
equation system like:
\[
  \begin{matrix}
    x & = & \max \bs{\frac12 y + \frac12 y, \dots}
    \\
    y & = & \dots
    \\
    \dots
  \end{matrix}
\]
In this new system, all other parts are not changed,
given the above discussion.
And it gives us \emph{exactly} the original one system before the normalisation step was performed.

Hence the property is shown.
\end{proof}

Then, we show a more aggressive one that:

\begin{lemma}
  \label{theorem: norm dpcfa share the same value}
  All DPCFA has the same structural probability as its normalised result by any means of normalisation.
\end{lemma}
\begin{proof}
We here fix a specific normalisation technique to DPCFA as adopting the construct in \cref{subsec: normalisation}.
Then, we try to proceed our proof by 
induction on the number of violation of the normalisation conditions.
Note that as the input is already deterministic, the violation can only happen for the second case.

In the base case, where no violation happen,
it means the DPCFA is itself normalised.
Then, by applying \cref{theorem: normalisation has same value},
we know that the property holds.
In the inductive case,
we then need to proceed one time of the normalisation to erase the violation number -- 
note that given the termination of the process,
although one time of application may not lead to decrease of the number of violation,
it will decrease eventually.
By \cref{theorem: norm one step does not change},
we know that the original one and the one that is ready to call the induction hypothesis are the same,
which, according to the induction hypothesis,
is again the same as the normalised one.

Hence the property holds.
\end{proof}

Then we have:

\begin{lemma}
  \label{theorem: determinisation share the same structural value}
  All deterministic PCFA that share the same set of traces have the same structural probability.
\end{lemma}
\begin{proof}
  Follows immediately by applying 
  \cref{theorem: norm dpcfa share the same value} 
  and 
  \cref{theorem: normalisation has same value}.
\end{proof}

Then the proof to 
\cref{eq: normalise equals determinise}
is given by:

\begin{proof}[Proof (\cref{eq: normalise equals determinise})]
  As the normalised one on the RHS of 
  \cref{eq: normalise equals determinise} 
  is also deterministic,
  it naturally equals to the LHS,
  by \cref{theorem: determinisation share the same structural value}.
\end{proof}

\subsection{Proof to \cref{theorem: cross-structural equality}}

By the above theorem for general PCFA,
all deterministic PCFA with the same set of violating traces will have the same probability.
Given the determinism assumption implicitly given for PCFA in~\cref{def: pcfa},
the~\cref{theorem: cross-structural equality} follows as a corollary.

\section{More Related Discussion}

\paragraph{Other Probabilistic Aspects}
Cousot et al.~\cite{prob_abstract_interpreation} introduced an abstract interpretation framework to the probabilistic context.
Compared with their framework, when it comes to the resolution with non-determinism, however,
our method agrees more with the \textsc{pmaf} framework by Wang et al.~\cite{pmaf},
that we both resolve probability after non-determinism.
Besides predicated verification, other aspects of probabilistic programs are also extensively researched recently.
Raven et al.~\cite{beutner2021probabilistic} considered the AST problem for functional programs;
Kobayashi et al.~\cite{phors,phors_long} showed the undecidability of the AST problem for a class of expressive probabilistic models, while Li et al.~\cite{li2022probabilistic} showed the decidability of a subclass of it yet with expressiveness over context-free.

\section{More on Proof to Lemma~\ref{thm:min(intersection)-is-pcfa}}

The only detail missing is to show properties for (**): every accepting trace is not a prefix of another accepting trace.
Recall the properties that all PCFAs should satisfy:
\hypertarget{cond-a-app}{(a)} it is deterministic, \hypertarget{cond-b-app}{(b)} with a single ending location, and \hypertarget{cond-c-app}{(c)} the ending location has no out-transition.

\begin{lemma}
  All PCFAs satisfy (**).
\end{lemma}
\begin{proof}
  Given any terminating trace $\tau$ of a PCFA $A$,
  as it is deterministic, the only location it can reach is the ending location.
  As there is no out-transition from the ending location,
  there cannot be another trace $\tau'$ that has $\tau$ as a prefix.
  This coincides with the intuition that when a computation can either be terminating or not, there cannot be a computation that is terminated but allows some more computation to follow.
\end{proof}

\begin{lemma}
  Any deterministically minimal automaton satisfies Property \hyperlink{cond-b-app}{(b)} and \hyperlink{cond-c-app}{(c)} iff it satisfies (**).
\end{lemma}
\begin{proof}
  For the forward direction,
  since any deterministically minimal automaton that satisfies Property \hyperlink{cond-b-app}{(b)} and \hyperlink{cond-c-app}{(c)} aligns with the definition of PCFA,
  we apply the former lemma directly.

  For the reverse direction,
  we first show that it satisfies Property \hyperlink{cond-c-app}{(c)}.
  This can be seen by noting that if there is an out-transition from an ending location $\ell$,
  since the automaton is minimal, this transition must have a way to reach an ending location $\ell'$ (otherwise, by minimisation, it should be removed).
  Then, given an accepting trace $\tau$ that ends at $\ell$,
  we can append the transition to $\ell'$ to $\tau$ to obtain a new trace $\tau'$ that is also accepting.
  So $\tau$ is a prefix of $\tau'$, which contradicts the assumption.
  Therefore, Property \hyperlink{cond-c-app}{(c)} must hold.

  Then, we show that it satisfies Property \hyperlink{cond-b-app}{(b)}.
  Since all the ending locations do not have out-transitions,
  they will surely be merged into one single ending location by the minimisation process.
\end{proof}

Finally, we show that:

\begin{lemma}
  If a set of traces $\Theta$ satisfies (**), then any of its subsets also satisfies (**).
\end{lemma}
\begin{proof}
  If there is a subset $\Theta' \subseteq \Theta$ that does not satisfy (**),
  then there must be two traces $\tau_1, \tau_2 \in \Theta'$ such that $\tau_1$ is a prefix of $\tau_2$.
  As $\Theta'$ is a subset of $\Theta$,
  we have that $\tau_1, \tau_2 \in \Theta$.
  This contradicts the assumption that $\Theta$ satisfies (**).
\end{proof}

\section{Exact Refinement by Value Analysis}

\subsection{Algorithmic Details}

We firstly present the whole algorithm in detail.

The first phase of the algorithm is, as mentioned, to perform a procedure to obtain the program valuations in each location.
We adapt the round-robin data-flow analysis algorithm from~\cite{kam1976global,cooper2004iterative} to fit our need.
The algorith is presented in~\cref{alg: data-flow analysis}.

\begin{algorithm}[t]
  \SetKwInOut{Input}{input}
  \SetKwInOut{Output}{output}
  \SetKwInOut{Variables}{variables}
  \Input{A PCFA $A (L, \Sigma, \Delta, \ell_0, \ell_e)$ and an initial valuation $v_0$}
  \Output{For each location $\ell$ in $L$, a valuation set $S_\ell$ containing all possible valuations
  at the location $\ell$}
  \BlankLine

  $S_{\ell_0} \leftarrow \bs{v_0}$ \;
  \For{$\ell \in L \setminus \bs{\ell_0}$}{
    $S_\ell \leftarrow \emptyset$ \;
  }

  \While{$\exists \ell.S_\ell \neq S_\ell \cup \setdef{\interpret{\sigma}_v}{
    \ell' \xrightarrow{\sigma} \ell \in \Delta, v \in S_{\ell'} \;
  }$}{
    \For{$\ell \in L$}{
      $S_\ell \leftarrow S_\ell \cup \setdef{\interpret{\sigma}_v}{
      \ell' \xrightarrow{\sigma} \ell \in \Delta, v \in S_{\ell'} \;
      }$
    }
  }

  \Return{$\setdef{S_\ell}{\ell \in L}$} \;

  \caption{Round-Robin Data-Flow Analysis Algorithm}
  \label[algorithm]{alg: data-flow analysis}
\end{algorithm}

Given the sets of valuations from the above algorithm, one may construct a refinement automaton $\refinement (L_\refinement, \Sigma, \Delta_\refinement, \ell_0^\refinement, L_e^\refinement)$ where:
\begin{itemize}
  \item The location set $L_\refinement$ to be all the possible valuations that appeared, that is:
  $L_\refinement := \bigcup_{\ell \in L} S_\ell$.
  \item The initial location $\ell_0^\refinement$ is the given initial valuation $v_0$.
  \item For the transitions $\Delta_\refinement$, there is a transition $v \xrightarrow{\sigma} v'$, iff there exists such two locations $\ell$ and $\ell'$ and a transition $\ell \xrightarrow{\sigma} \ell'$ in $\Delta$ that $v \in S_\ell$ and $v' \in S_{\ell'}$.
  \item The set of ending locations is then given by testing the reverse of $\postcond$, $L_e^\refinement := \setdef{v \in L_\refinement}{v \models \neg \postcond}$.
\end{itemize}

\subsection{The Analysis Is Exact For Finite-State Programs}

We then show that this process is in deed exact.
To be concrete, we have:

\begin{lemma}
    \label{theorem: data flow analysis is exact}
    For a PCFA $A$ with pre- and post-condition $\precond$ and $\postcond$, 
    and a refinement automaton $\refinement$ produced by 
    the lightweight data-flow analysis above: \rm
    \[
      \mdpProb{A}{\precond}{\postcond} = \structBound{\minimise(A \cap \refinement)}
    \]
\end{lemma}

\begin{proof}
  Firstly, observe that All traces within $\scriptL(A \cap \refinement)$ are violating and also compatible -- they at least share the common program valuation $v_0$.
  Hence, we canshow that:
  \begin{align*}
    \mdpProb{\minimise(A\cap\refinement)}{\precond}{\postcond}
    =
    \structBound{\minimise(A\cap\refinement)}
  \end{align*}
  This can be observed by that every policy $\psi$ applied will be violating and also compatible, hence the semantic check will always pass.
  So:
  \begin{align*}
    & \mdpProb{A}{\precond}{\postcond} \\
    =
    & \mdpProb{\minimise(A\cap\refinement)}{\precond}{\postcond}
    =
    \structBound{\minimise(A\cap\refinement)}
  \end{align*}
\end{proof}

\subsection{Value Analysis As Refinement For Infinite-State Programs}

We have shown that for finite-state programs, when the refinement automaton can be completely constructed from~\Cref{alg: data-flow analysis}, the obtained structural upper bound is exact.
But for programs where the state spaces are infinite, the~\Cref{alg: data-flow analysis} will not terminate,
value analysis still performs as a potential refinement procedure by extending $\refinement$ to some extent (by limiting to some finite number of states)
while leaving the rest of the automaton to be refined by other methods, e.g., predicate abstraction.

\section{Instantiating the General CEGAR Framework with Predicate Abstraction}

In this part, we demonstrate a direct instantiation of our general CEGAR framework with predicate abstraction.
Predicate abstraction~\cite{cegar}, a form of state abstraction, utilises a finite number of abstract states to represent a potentially infinite set of states.
More precisely, it categorises the concrete program states into certain abstract states while preserving the transition relation between them.
As will be seen, the instantiation below is rather straightforward.

\paragraph{Non-Probabilistic Predicate Abstraction}
Predicate abstraction commences with an initial set of predicates, denoted as $\Phi$, from which each assignment (of all the predicates) forms an abstract program state.
These states abstract the concrete states by establishing a transition between two (possibly identical) states if and only if at least one transition exists between the concrete states they represent.
Formally, a possible construction can be introduced as follows.
Given a set of predicates $\Phi := \bs{\varphi_1, \dots, \varphi_n}$,
an abstract program state for $\Phi$ can be represented by an $n$ dimensional vector: $(b_1, \dots, b_n)$, where $b_i$ is either $\varphi_i$ or $\neg \varphi_i$, denoting an assignment to $\Phi$.
Given the abstract states, a transition
\(
  (b_1, \dots, b_n) \xrightarrow{\sigma} (b_1', \dots, b_n')
\)
is added to the abstract (general) CFA, iff there \emph{exists} a state $v$ such that $v \models b_1 \wedge \dots \wedge b_n$ and that $\interpret{\sigma}_v \models b_1' \wedge \dots b_n'$.

Further, in the non-probabilistic scenario, one picks a trace from the abstract control-flow graph (CFG) or control-flow automaton (CFA) leading to the violating state(s), and verifies its feasibility in the actual program.
If feasible, it is reported as a valid counterexample;
otherwise, the set of predicates is refined based on the spurious counterexample, when new predicate(s) is (are) added,
for which there has been decades of research~\cite{cegar,clarke2005satabs,craig_interpolation,henzinger2004abstractions}.

\paragraph{Instantiating General CEGAR Framework}
Notably, various methods in the literature construct CFAs from predicates, leveraging CFAs and CFGs as standard representations in predicate abstraction~\cite{hajdu2020efficient,cegar,trace_abstraction,beyer2022unifying}.
These methods apply directly to PCFA, as probabilistic statements are semantically equivalent to the $\stskip$ statement.
The framework can thus be instantiated as follows:
\begin{itemize}
  \item Starting with an initial set of predicates, the initial refinement automaton $\refinement$ is constructed using standard techniques as outlined earlier.
  \item To refine a spurious counterexample, non-violating traces are used to generate new predicates, augmenting the existing set. 
  The updated refinement automaton $\refinement$ is then derived from the refined predicates.
\end{itemize}

\section{More on Experiments}

We also conducted an experiment to compare with all the examples from~\cite{exponential_analysis_of_probabilistic_program}.
The results are presented in~\cref{tab: exp bound cmp}.

Interestingly, we found that the result for the example ``Prspeed'' is strange that,
as the bound we computed is arguably exactly, the upper bounds given in~\cite{exponential_analysis_of_probabilistic_program} for this example is many times smaller than our bound.
After santitising our results with an independent tool,
we suppose there be some errors happened in their results, which are marked red in the table.

\begin{table}[t]
  \setlength{\tabcolsep}{0.5em}
\centering
{\tabulinesep=1.1mm
\resizebox{\textwidth}{!}{
\begin{tabu}{|c|c|c|c|c|c|}
\hline
\textbf{Example} & \textbf{Parameters} & \textbf{Our Results} & \textbf{Time(s)} & \textbf{Previous Results} & \textbf{rate ($\frac{\text{Previous Result}}{\text{Our Result}}$)} \\ 
\hline \hline
\multirow{3}{*}{RdAdder} & $\prob{X - E[X] \ge 25}$ & $ 1.42 \cdot 10^{-2} $ & 75 & $ 7.43 \cdot 10^{-2} $ & $ 5.23 $ \\ \cline{2-6}
& $\prob{X - E[X] \ge 50}$ & $ 4.47 \cdot 10^{-6} $ & 77 & $ 3.54 \cdot 10^{-5} $ & $ 7.92 $ \\ \cline{2-6}
& $\prob{X - E[X] \ge 75}$ & $ 9.51 \cdot 10^{-12} $ & 66 & $ 9.17 \cdot 10^{-11} $ & $ 9.64 $ \\ \hline
\multirow{3}{*}{Coupon} & $\prob{T > 100}$ & $ 1.02 \cdot 10^{-9} $ & 0.4 & $ 7.01 \cdot 10^{-5} $ & $ 6.87 \cdot 10^{4} $ \\ \cline{2-6}
& $\prob{T > 300}$ & $ 4.23 \cdot 10^{-29} $ & 1.1 & $ 7.44 \cdot 10^{-22} $ & $ 1.76 \cdot 10^{7} $ \\ \cline{2-6}
& $\prob{T > 500}$ & $ 1.75 \cdot 10^{-48} $ & 1.9 & $ 4.01 \cdot 10^{-40} $ & $ 2.29 \cdot 10^{8} $ \\ \hline
\multirow{3}{*}{RdWalk} & $\prob{T > 400}$ & $ 9.43 \cdot 10^{-9} $ & 21 & $ 2.12 \cdot 10^{-7} $ & $ 2.25 \cdot 10^{1} $ \\ \cline{2-6}
& $\prob{T > 500}$ & $ 4.46 \cdot 10^{-14} $ & 41 & $ 1.57 \cdot 10^{-12} $ & $ 3.52 \cdot 10^{1} $ \\ \cline{2-6}
& $\prob{T > 600}$ & $ 9.92 \cdot 10^{-20} $ & 81 & $ 4.81 \cdot 10^{-18} $ & $ 4.85 \cdot 10^{1} $ \\ \hline
\multirow{3}{*}{Prspeed} & $\prob{T > 150}$ & $ 1.89 \cdot 10^{-18} $ & 3.7 & \colorblock{red}{$ 7.43 \cdot 10^{-23} $} & \colorblock{red}{$ 3.93 \cdot 10^{-5} $} \\ \cline{2-6}
& $\prob{T > 200}$ & $ 1.11 \cdot 10^{-30} $ & 5.3 & \colorblock{red}{$ 8.03 \cdot 10^{-36} $} & \colorblock{red}{$ 7.23 \cdot 10^{-6} $} \\ \cline{2-6}
& $\prob{T > 250}$ & $ 1.27 \cdot 10^{-43} $ & 8.3 & \colorblock{red}{$ 2.71 \cdot 10^{-49} $} & \colorblock{red}{$ 2.13 \cdot 10^{-6} $} \\ \hline
\multirow{3}{*}{Race} & $X = 40$ & $ 2.63 \cdot 10^{-8} $ & 0.4 & $ 1.52 \cdot 10^{-7} $ & $ 5.78 $ \\ \cline{2-6}
& $X = 35$ & $ 3.53 \cdot 10^{-6} $ & 0.6 & $ 2.16 \cdot 10^{-5} $ & $ 6.12 $ \\ \cline{2-6}
& $X = 45$ & $ 1.90 \cdot 10^{-11} $ & 0.3 & $ 8.65 \cdot 10^{-11} $ & $ 4.55 $ \\ \hline
\multirow{3}{*}{1DWalk$^*$} & $X = 10$ & $ 7.79 \cdot 10^{-208} $ & 1.7 & $ 7.82 \cdot 10^{-208} $ & $ 1.00 $ \\ \cline{2-6}
& $X = 50$ & $ 1.79 \cdot 10^{-199} $ & 1.8 & $ 1.79 \cdot 10^{-199} $ & $ 1.00 $ \\ \cline{2-6}
& $X = 100$ & $ 5.04 \cdot 10^{-189} $ & 1.8 & $ 5.03 \cdot 10^{-189} $ & $ 1.00 $ \\ \hline
\end{tabu}}}
\flushleft
* : We specially adjusted the zero-tolerance for this example as the results are small.

\medskip

  \caption{Experimental Results Compared to~\cite{exponential_analysis_of_probabilistic_program}}
  \label{tab: exp bound cmp}
\end{table}


\end{document}